\def\MPL #1 #2 #3 {Mod.~Phys.~Lett.~{\bf#1},\  #2 (#3)}
\def\NPB #1 #2 #3 {Nucl.~Phys.~{\bf#1},\  #2 (#3)}
\def\PLB #1 #2 #3 {Phys.~Lett.~{\bf#1},\  #2 (#3)}
\def\PR #1 #2 #3 {Phys.~Rep.~{\bf#1},\ #2 (#3)}
\def\PRD #1 #2 #3 {Phys.~Rev.~{\bf#1},\  #2 (#3)}
\def\PRL #1 #2 #3 {Phys.~Rev.~Lett.~{\bf#1},\  #2 (#3)}
\def\RMP #1 #2 #3 {Rev.~Mod.~Phys.~{\bf#1},\  #2 (#3)}
\def\ZP #1 #2 #3 {Z.~Phys.~{\bf#1},\  #2 (#3)}
\def\IJMP #1 #2 #3 {Int.~J.~Mod.~Phys.~{\bf#1},\  #2 (#3)}

\def\sigrts{\sigma_{\tiny\rts}^{}}
\def\sighbar{\overline \sigma_{\h}}
\def\sighsmbar{\overline \sigma_{\hsm}}

\def\sighhbar{\overline \sigma_{\hh}}
\def\sighabar{\overline \sigma_{\ha}}
\def\gamha{\Gamma_{\ha}^{\rm tot}}
\def\gamhh{\Gamma_{\hh}^{\rm tot}}

\def\delmhsm{\Delta\mhsm}
\def\rtszhsm{\rts_{Z\hsm}}

\def\mgamgam{M_{\gam\gam}}
\def\call{{\cal L}}

\def\DM{D$^-$}
\def\DP{D$^+$}
\def\NSM{NS$^-$}
\def\NSP{NS$^+$}
\def\HSM{HS$^-$}
\def\HSP{HS$^+$}
\def\etc{{\it etc.}}
\def\leff{L_{\rm eff}}
\def\sign{{\rm sign}}

\def\anti{\overline}

\def\hi{\h_1}
\def\hii{\h_2}
\def\hiii{\h_3}
\def\mhi{m_{\hi}}
\def\mhii{m_{\hii}}
\def\mhiii{m_{\hiii}}

\def\mtau{m_{\tau}}
\def\etc{{\em etc.}}
\def\chisq{\chi^2}
\def\ltot{L_{\rm total}}

\def\sigmazh{\sigma_T(Z\h)}
\def\thdm{2HDM}
\def\mtau{m_{\tau}}
\def\gamhsm{\Gamma_{\hsm}^{\rm tot}}
\def\tauptaum{\tau^+\tau^-}
\def\gamgam{\gam\gam}

\def\wstar{W^{\star}}

\def\zstar{Z^{\star}}

\def\sighbar{\overline \sigma_{\h}}
\def\br{BR}

\def\rts{\sqrt s}
\def\eps{\epsilon}
\def\h{h}
\def\a{a}
\def\mh{m_{\h}}
\def\ma{m_{\a}}
\def\gamh{\Gamma_{\h}}
 
\def\lam{\lambda}

\def\eg{{\it e.g.}}
\def\etal{{\it et al.}}
\def\epem{e^+e^-}
\def\mupmum{\mu^+\mu^-}

\def\taup{\tau^+}
\def\taum{\tau^-}

\def\lsim{\mathrel{\raise.3ex\hbox{$<$\kern-.75em\lower1ex\hbox{$\sim$}}}}
\def\gsim{\mathrel{\raise.3ex\hbox{$>$\kern-.75em\lower1ex\hbox{$\sim$}}}}
\def\@versim#1#2{\vcenter{\offinterlineskip
        \ialign{$\m@th#1\hfil##\hfil$\crcr#2\crcr\sim\crcr } }}
\def\zstar{Z^\star}
\def\wstar{W^\star}
\def\slash#1{#1\hskip-8pt/\hskip4pt}

\def\ie{{\it i.e.}}

\def\gam{\gamma}

\def\anti{\overline}
\def\pbi{~{\rm pb}^{-1}}
\def\fbi{~{\rm fb}^{-1}}
\def\fb{~{\rm fb}}
\def\pb{~{\rm pb}}

\def\mev{\,{\rm MeV}}
\def\gev{\,{\rm GeV}}
\def\tev{\,{\rm TeV}}

\def\wt{\widetilde}

\def\rta{\rightarrow}
\def\mhalf{m_{1/2}}

\def\stop{\wt t}

\def\mstop{m_{\stop}}

\def\hsm{h_{\rm SM}}
\def\mhsm{m_{\hsm}}
\def\hl{h^0}
\def\hh{H^0}
\def\ha{A^0}
\def\hp{H^+}
\def\hm{H^-}
\def\hpm{H^{\pm}}
\def\mhl{m_{\hl}}
\def\mhh{m_{\hh}}
\def\mha{m_{\ha}}
\def\mhp{m_{\hp}}
\def\mhpm{m_{\hpm}}
\def\tanb{\tan\beta}

\def\mt{m_t}
\def\mb{m_b}
\def\mz{m_Z}
\def\mw{m_W}

\def\wp{W^+}
\def\wm{W^-}

\def\h{h}
\def\mh{m_{\h}}

\def\cpmone{\wt \chi^{\pm}_1}

\def\mcpmone{m_{\cpmone}}

\def\emem{e^-e^-}
\def\dmm{\Delta^{--}}
\def\mdmm{m_{\dmm}}

\def\dpp{\Delta^{++}}

\def\delm{\Delta^{-}}
\def\mdelm{m_{\delm}}

\documentclass[smus]{snow2e}
\usepackage{epsf}
\input psfig.sty
\begin{document}
\title{
{\large
             \hspace*{\fill}\phantom{help}
             \hspace*{\fill}\phantom{help}
             \hspace*{\fill} UCD-97-5  \\
             \hspace*{\fill} March, 1997 \\
             \hspace*{\fill}  \\
             \hspace*{\fill}  \\
             \hspace*{\fill}  \\
}
Higgs Boson Discovery and Properties
\thanks{
To appear in {\it Proceedings of the 1996 DPF/DPB Summer Study
on ``New Directions in High Energy Physics'' (Snowmass, 96)}, June 25 - July 12,
1996, Snowmass, Colorado.}
}

\author{ 
{\bf Authors and Conveners}
\\
J.F. Gunion$^{1,3,5}$ (U.C. Davis), L. Poggioli$^{2,3,5}$ (CERN),
R. Van Kooten$^{2,3,5}$ (Indiana) 
\\
C. Kao$^{4,5}$ (Wisconsin), 
P. Rowson$^{4}$ (SLAC) 
\\ 
\ 
\\
{\bf Working Group Members}
\\
\begin{minipage}{6.8in}
S. Abdullin$^5$ (ITEP),
V. Barger$^5$ (Wisconsin),
M. Berger$^5$ (Indiana),
D. Bauer$^5$ (U.C. Santa Barbara), 
M. Carena (CERN),
C. Damerell$^5$ (SLAC),
M. Fortner (FNAL),
R. Frey (Oregon),
H.E. Haber$^5$ (U.C. Santa Cruz),
T. Han$^5$ (U.C. Davis),
X.-G. He$^5$ (Melbourne), 
D. Hedin (FNAL), 
C. Heusch (U.C. Santa Cruz),
D. Jackson$^5$ (SLAC),
R. Jesik$^5$ (Indiana),
J. Kelly$^5$ (U.C. Davis/Wisconsin), 
S. Kim$^5$ (Tsukuba),
S. Kuhlmann$^5$ (FNAL),
C. Loomis$^5$ (Rutgers),
P. Martin$^5$ (U.C. Davis), 
T. Moroi$^5$ (LBL),
K. Pitts$^5$ (FNAL),
L. Reina (BNL),
R. Sobey$^5$ (U.C. Davis),
N. Stepanov$^5$ (ITEP),
R. Szalapski (KEK),
R. Vega (SMU), 
C. Wagner (CERN),
J. Womersley (FNAL),
W.-M. Yao$^5$ (LBL), 
R.-Y. Zhu (Cal. Tech.)
\end{minipage}
\\ \ \\
{\it 1) Primary Author; 2) Contributing Author; 3) Property Subgroup Convener;}
\\
{\it 4) Discovery Subgroup Convener; 5) Author of Summarized Contribution }
\\
}
\maketitle

\thispagestyle{empty}

\begin{abstract} 
We outline issues examined and progress made by 
the Light Higgs Snowmass 1996 working
group regarding discovering Higgs bosons
and measuring their detailed properties. We focused primarily
on what could be learned at LEP2, the Tevatron (after upgrade), the LHC,
a next linear $\epem$ collider and a $\mupmum$ collider.
\end{abstract}

\section{Introduction}

The three accelerators that exist
or are certain of being constructed are:
\begin{itemize}
\item LEP2, for which we assume $\rts=192\gev$,
and total integrated luminosity during
the time before LEP2 is shut down for LHC construction
of $L=250\pbi$ at each of the four detectors
for a total of $L=1000\pbi$, assuming that
data from the four detectors can be combined;
\item the Main Injector at the Tevatron, with $\rts=2\tev$
and $L=2\fbi$ per year for CDF and D0, each, for
a total of $L=4\fbi$ per year or $L=12\fbi$ for three years; and
\item the LHC, with $\rts=14\tev$ and $L=100\fbi$ for 
each detector (ATLAS and CMS), for a total of $L=200\fbi$
per year, or $L=600\fbi$ for three years of operation.
\end{itemize}
Possible upgrades and future machines include:
\begin{itemize}
\item an upgrade of the Main Injector 
so as to enable $30\fbi$ each to be accumulated by CDF and D0;
\item
a Next Linear $\epem$ Collider (NLC) with $\rts=500\gev$ and four-year
integrated luminosity of about $200\fbi$;
\item
a First Muon Collider (FMC) with four-year integrated luminosity
of $L=200\fbi$ which could be spread out between operation at $\rts=500\gev$
and running at $\rts$ in the vicinity of the mass of an already
detected Higgs boson or in a range designed to scan for an undiscovered
Higgs boson.
\end{itemize}
{\bf Notational Convention:}
In the following discussions, we use the notation NLC for results
that could be achieved in either $\epem$ 
or $\mupmum$ collisions\footnote{At $\rts=500\gev$, $\mupmum$ collision
results will be similar to $\epem$ collision results
if new detector backgrounds are not an issue.} at $\rts=500\gev$.
The notation FMC will be reserved for $s$-channel Higgs production results.

During the Snowmass workshop, we were able to pursue only a limited
set of projects.  The results obtained by various members of the group
will be summarized and their overall impact on the study and discovery
of Higgs bosons will be noted. We shall also outline an ongoing program 
for delineating the role that the various machines mentioned above
will play in pinning down the properties of a Higgs boson
with Standard Model-like properties. Our discussion will
be confined to five models:
\begin{itemize}
\item 
the Standard Model (SM), with a single Higgs boson, $\hsm$.
\item
the minimal supersymmetric standard model (MSSM)
with exactly two Higgs doublets
resulting in five Higgs eigenstates: two CP-even bosons, $\hl$
and $\hh$ with $\mhl\leq\mhh$; one CP-odd Higgs, $\ha$; and a charged
Higgs pair, $\hpm$.
\item
the non-minimal supersymmetric standard model (NMSSM) with a single
Higgs singlet field added to the two Higgs doublet fields ---
if CP is conserved, this adds a third CP-even eigenstate and a second
CP-odd eigenstate to the spectrum.
\item
a general two-Higgs-doublet model (\thdm)
of type-II (see Ref.~\cite{dpfreport}). In such a model,
the masses and couplings of the three neutral Higgs bosons
are free parameters; the neutral
Higgs bosons can be CP-mixed states.
\item
a Higgs sector containing a doubly-charged Higgs boson ($\dmm$).
\end{itemize}
This report is not intended as a general
review of Higgs boson physics.  It is designed to be read in conjunction
with the recent review of Ref.~\cite{dpfreport}, the NLC Physics
report Ref.~\cite{nlcreport}, and the muon-collider
Higgs physics study of Ref.~\cite{bbgh}. 

\section{The SM or a SM-like Higgs boson}

Although there has been extensive study of the $\hsm$, we found
a remarkably large number of new projects to pursue. In particular,
we found that
a detailed delineation of the extent to which the fundamental properties
of the $\hsm$ could be determined at a given accelerator or
combination of accelerators was lacking. In what follows we
present results obtained during the course of the workshop
and summaries of earlier work in the following areas:
\begin{description}
\item{A)} the discovery reach of TeV33; 
\item{B)} strategies for
verifying the fundamental properties of the $\hsm$ using a combination
of LEP2, TeV33, and LHC data, including some first estimates of errors;
\item{C)} optimizing the measurements of $\sigma\br(\hsm\to b\anti b)$, 
$\sigma\br(\hsm\to c\anti c)$ and $\sigma\br(\hsm\to WW^\star)$ at the NLC
for various production modes and determining
ratios of branching ratios.
\item{D)} determining $\sigma\br(\hsm\to\gam\gam)$ at the NLC;
\item{E)} determining the $ZZ\hsm$ coupling at the NLC;
\item{F)} determining the branching ratios and $WW\hsm$ coupling at the NLC;
\item{G)} determining the total width (and thereby the partial widths)
of the $\hsm$, including relevance and current status
of measuring a) the total width by $s$-channel scanning at the FMC
and b) $\Gamma(\hsm\to\gam\gam)$ at the $\gam\gam$
collider facility at the NLC --- this section ends with a summary
of the errors/precisions with which fundamental $\hsm$
properties can be determined using NLC data alone, $s$-channel FMC
data alone and a combination of NLC and $s$-channel FMC data;
\item{H)} determining the mass of the $\hsm$ at TeV33, the LHC, the NLC
and the FMC;
\item{I)} verifying the spin, parity and CP of the $\hsm$.
\end{description}
Many new results are contained in these summaries.

Throughout our discussions, the branching ratios of the $\hsm$
will play a major role, especially those for the $b\anti b$,
$W\wstar$  and $Z\zstar$ channels.  These three branching ratios 
are tabulated for $\mhsm\leq 170\gev$ in Table~\ref{hsmbrs} for later reference.
(For a full figure, see Ref.~\cite{dpfreport}.)
Note, in particular, that the $W\wstar$ mode only really begins
to be competitive with the $b\anti b$ mode when $\mhsm\gsim 130\gev$,
and that it causes a precipitous decline in the $b\anti b$ branching ratio
by $\mhsm\gsim 150\gev$. In some of the error estimates to be
presented, we have extrapolated simulations performed at only a few masses
to a larger range of masses using the mass dependence
of the $b\anti b$ and $W\wstar$ branching ratios. 
Another important point is also immediately apparent. For $\mhsm\leq 170\gev$, 
$\br(\hsm\to Z\zstar)$ is always much smaller than $\br(\hsm\to W\wstar)$.
At the NLC, where backgrounds in the $W\wstar$ channel
are not a particular problem (since a Higgs mass peak
can be reconstructed in the four-jet $W\wstar$ final state), 
this has meant that detection of the $\hsm$ in its $Z\zstar$
decay mode has received little attention. However, at the LHC
the $Z\zstar$ channel has been the preferred channel due to
the inability to detect $W\wstar$ in its four-jet mode (because
of the large jet backgrounds in $pp$ collisions) and the
lack of a clear mass peak in the purely-leptonic or mixed modes.
The much larger $W\wstar$ branching
ratio has led us to pay increased attention to the $W\wstar$
mode at the LHC in this report.

\begin{table}[hbt]
\caption[fake]{$b\anti b$, $W\wstar$ and $Z\zstar$ branching
ratios for the $\hsm$ in the $\mhsm<2\mz$ mass region.}
\begin{center}
\begin{tabular}{|c|c|c|c|c|c|c|}
\hline
 Mass (GeV) & 110 & 120 & 130 & 140 & 150 & 170 \\
\hline
 $\br(b\anti b)$ & 0.78 & 0.69 & 0.53 & 0.34 & 0.17 & 0.008 \\
 $\br(W\wstar)$ & 0.04 & 0.13 & 0.29 & 0.50 & 0.70 & 0.97 \\
 $\br(Z\zstar)$ & 0.002 & 0.01 & 0.03 & 0.06 & 0.08 & 0.02 \\
\hline
\end{tabular}
\end{center}
\label{hsmbrs}
\end{table}

\begin{figure}[h]
\leavevmode
\begin{center}
\centerline{\psfig{file=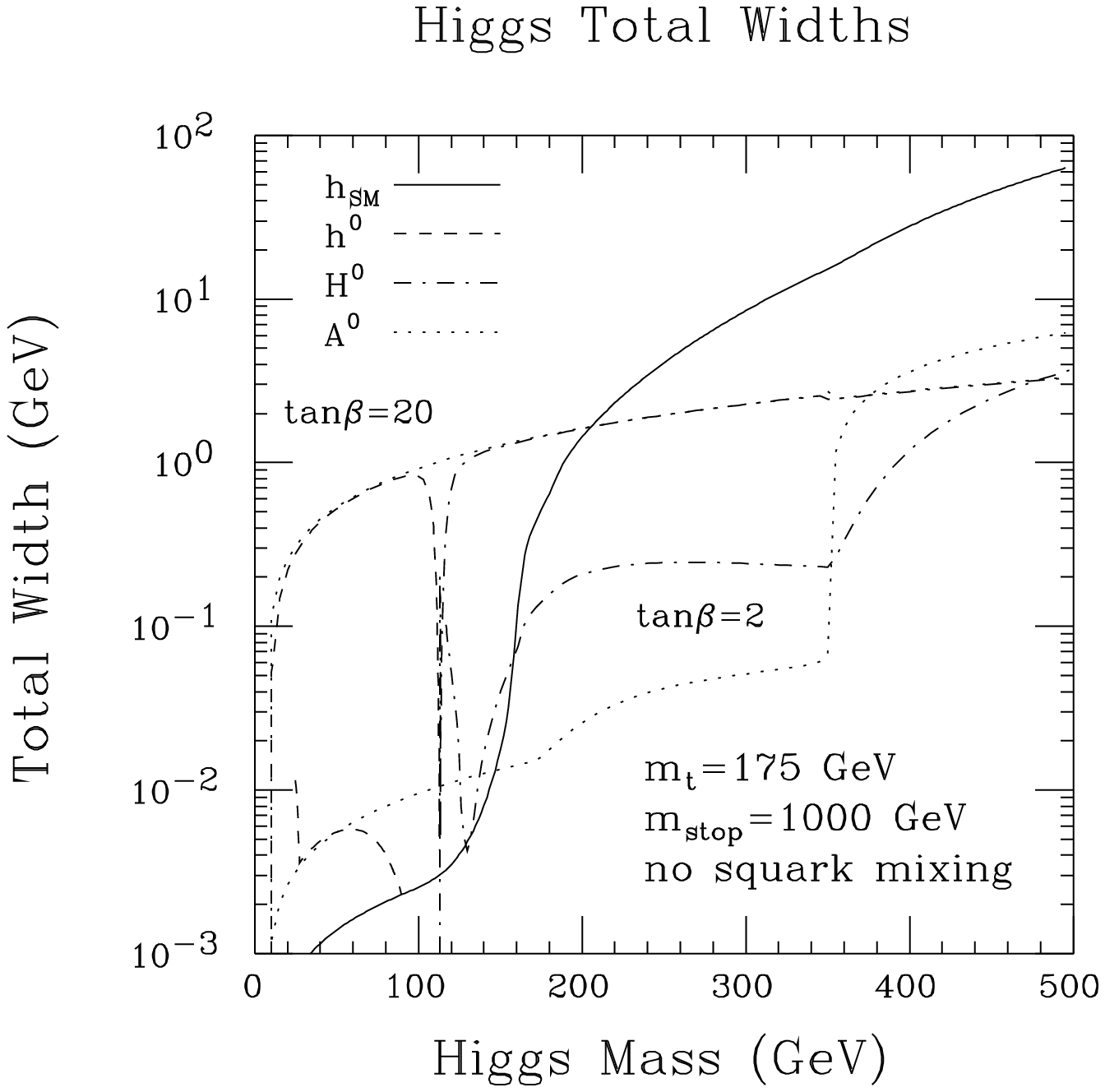,width=3.25in}}
\end{center}
\caption{
Total width versus mass of the SM and MSSM Higgs bosons
for $\mt=175\gev$.
In the case of the MSSM, we have plotted results for
$\tan\beta =2$ and 20, taking $\mstop=1\tev$ and
including two-loop/RGE-improved Higgs mass corrections and
neglecting squark mixing; SUSY decay channels are assumed to be absent.}
\label{hwidths}
\end{figure}

Also of considerable importance is the expected width of a SM-like
Higgs boson.  The predicted width, $\gamhsm$, is plotted in
Fig.~\ref{hwidths} as a function of $\mhsm$.  The main features
to take note of are:
\begin{itemize}
\item $\gamhsm$ is very small for $\mhsm\lsim 2\mw$.
Indeed, for $\mhsm\lsim 140\gev$, $\gamhsm<10\mev$.
\item $\gamhsm$ grows rapidly for $\mhsm>2\mw$ due to the
turn-on of $\hsm\to \wp\wm,ZZ$ decays.
\end{itemize}

\subsection{Discovery of $\hsm$ in the $W\hsm$ and $Z\hsm$ modes at TeV33}

A thorough assessment of the $W\hsm\to\ell\nu b\anti b$ \protect\cite{kky}
and $Z\hsm\to(\nu\anti\nu,\ell^+\ell^-)b\anti b$ \protect\cite{wmyao}
channels at TeV33 was made. In addition, a first exploration of
$Z\hsm$ detection in the $4b$ final state was initiated \cite{jesik}.

Earlier results for the $W\hsm$ mode were improved upon by:
\begin{itemize}
\item
using the CDF soft lepton $b$-tagging and loose secondary vertex $b$-tagging
in addition to the CDF secondary vertex $b$-tagging;
\item
requiring that $|\cos(\theta)|<0.8$, where $\theta$ is the scattering
angle of the Higgs in the $W$-Higgs c.m. system.
\end{itemize}
With these additional cuts, it is found that the SM Higgs with
$\mhsm= 60$, $80$, $100$, $120\gev$ could be discovered at the $\rts=2\tev$
TeV33 with integrated luminosity of $L=3.5$, $5.5$, $11$ and $24.5\fbi$,
respectively --- $30\fbi$ would probably probe up to $\mhsm=125\gev$.
If results from both CDF and D0 could be combined, one might even reach
$\mhsm=130\gev$.
The $Z\hsm$ study showed that this channel can provide support
for the $W\hsm$ discovery. For $L=30\fbi$, the statistical significances
achieved in the $Z\hsm$ channel 
by employing $b$-tagging and a series of cuts were
$S/\sqrt B=4.3$, $3.8$, $3.5$, and $2.5$ for $\mhsm=90$, $100$, $110$ and
$120\gev$, respectively. In both the $W\hsm$ and $Z\hsm$
channels, the above results require
reconstructing the mass of the two tagged $b$-jets.  Accepted $b\anti b$
mass intervals were in the range from $24\gev$ (at $\mhsm\sim 60\gev$)
to $39\gev$ (at $\mhsm\sim 120\gev$).

The above results assume increased importance in the context of the MSSM,
in which the upper bound on the probably SM-like $\hl$ is of order $130\gev$.
Indeed, values of $\mhl$ in the $80-120\gev$ range are most typical
in grand-unified (GUT) models with GUT-scale boundary conditions
that yield automatic electroweak symmetry
breaking (EWSB) via the renormalization group equations (RGE's), provided the
squark masses are small enough to avoid naturalness problems.
It seems that TeV33 has a good chance of discovering the $\hl$
of the MSSM.  However, in the NMSSM the lightest Higgs need not have
full coupling to $WW$ and $ZZ$. Even if its mass lies in the $\lsim
125\gev$ range, the other parameters of the model
can easily be chosen so that it would not be detectable at TeV33.

If a SM-like Higgs boson
is sufficiently light to be discovered at TeV33 (more
generally if a SM-like Higgs has mass below $2\mw$)
both the NLC and the FMC would be highly desirable machines
capable of measuring crucial properties of a SM-like Higgs boson.
In particular, for $\mhsm\lsim 2\mw$, a FMC optimized for $\rts=\mhsm$ running
would be a Higgs factory \cite{bbgh}
capable of directly measuring (by scanning) the total Higgs width
and coupling ratios with great accuracy.
Indeed, our final summary tables for the $\hsm$
show that if $\mhsm\lsim 150\gev$
then it would be extremely desirable to have both the
NLC (or FMC running at $\rts=500\gev$) and a FMC devoted to
$\rts=\mhsm$ measurements in the $s$-channel.
However, observation of a Higgs boson at TeV33 is unlikely to
come soon enough to guide us should a decision between the FMC
and the NLC (or a second NLC) become necessary.

Finally, a few brief remarks regarding 
the $Z\hsm\to 4b$ final state detection mode.
Final results have not yet been obtained, but progress has been made
\cite{jesik}. First, it is found that it will be possible
to trigger with about 60\% efficiency
on the $4b$ final states using a ``standard'' (\ie\
as employed in top quark studies) lepton plus jets trigger,
where the lepton comes from semi-leptonic decay of one of the $b$'s.
Somewhat higher efficiency can probably be achieved with increased
electron triggering acceptance and by employing secondary vertex triggers.
Using the various codes for the $4b$ final state backgrounds 
\cite{4bcodes} (which are in good agreement) a reasonable
signal over background is found if two pairs of $b$'s are required
to have high mass.\footnote{This means that signal to background
will possibly also be acceptable for supersymmetric model $\hl\ha$ 
Higgs pair production.} On the other hand, $gg\to \hsm b\anti b$
looks hard since the associated $b\anti b$ does not generally have
a high pair mass; in particular, it would probably be necessary
to veto charm at the $1\%$ level.\footnote{Of course, for
high $\tanb$ in the MSSM, the
$gg\to \ha b\anti b$ and either $gg\to \hh b\anti b$ (for $\mha\gsim 130\gev$)
or $gg\to \hl b\anti b$ (for $\mha\lsim 130\gev$)
rates are greatly enhanced relative to $gg\to \hsm b\anti b$ and detection
above some minimum value of $\tanb$ would become possible.}

\subsection{Strategies for verifying the properties of the $\hsm$
using LEP2, TeV33 and LHC data only 
\protect\cite{gp}}

In this continuing project, the goal is to fully
enumerate the important strategies and measurements at LEP2, the Tevatron
and the LHC that
will be required to maximize information regarding the couplings
of a Standard-Model-like Higgs boson and, thereby, our ability
to verify its SM-like nature.  Ultimately, as we shall discuss in
later subsections, experimental data 
from the NLC and/or FMC will also be available that will
vastly expand our ability to verify the properties of a SM-like
Higgs boson. However, in the next decade or so, the challenge will
be to extract maximal information from the former three operating
accelerators. Ideally, one would wish
to determine, {\it in a model-independent fashion}, all of the tree-level and
one-loop couplings of the $\hsm$, its spin, parity, and CP nature,
and its total width. Here we outline the extent to which
this will be possible using data from the three machines.

\subsubsection{Enumeration of mass regions and reactions}

The discussion divides naturally into five different mass regions:
\begin{description}
\item {M1:} $\mhsm\lsim 95\gev-100\gev$. Detection of the $\hsm$
should be possible at all three machines: LEP2, the Tevatron,
and the LHC.
\item {M2:} $95-100\gev \lsim\mhsm\lsim 130\gev$. Detection should be possible
at the Tevatron and the LHC, but not at LEP2. Note
that we are adopting the optimistic conclusions discussed above
that the mass range for which detection at TeV33 will 
be viable in the $W\hsm$, $\hsm\to b\anti b$ mode includes the region 
between 120 and 130 GeV, and that up to 130 GeV 
some information can also be extracted
at TeV33 from the $Z\hsm$ mode. 
At the LHC, modes involving $\hsm\to b\anti b$ are 
currently regarded as being quite problematic above 120 GeV.
Nonetheless, we will consider them.
Of course, $\hsm\to Z\zstar$ and $W\wstar$
decay modes will not yet be significant, and the Higgs remains very
narrow. 
\item {M3:} $130\gev\lsim\mhsm\lsim 150-155\gev$. Detection is only
possible at the LHC, $Z\zstar$ and $W\wstar$ decay modes
emerge and become highly viable, the Higgs remains narrow.
\item {M4:} $155\lsim\mhsm\lsim 2\mz$.  The real $WW$ mode turns on,
$Z\zstar$ reaches a minimum at $\mhsm\sim 170\gev$. The inclusive $\gam\gam$
mode is definitely out of the picture. The Higgs starts to get
broad, but $\gamhsm\lsim 1\gev$.
\item {M5:} $\mhsm\gsim 2\mz$. Detection will only be possible
at the LHC, $ZZ$ and $WW$ modes are dominant,
and the Higgs becomes broad enough that a {\it direct}
determination of its width becomes conceivable
by reconstructing the $ZZ\to 4\ell$ final state mass
(probable resolution being of order $1\%\times\mhsm$ at CMS
and $1.5\%\times\mhsm$ at ATLAS).
\end{description}
The possible modes of potential use for determining
the properties of the $\hsm$ at each of
the three machines are listed below.  Even very
marginal modes are included when potentially crucial
to measuring an otherwise inaccessible Higgs property.
\smallskip

\noindent\underline{LEP2}
\begin{description}
\item {LP1:} $\epem\to \zstar\to Z\hsm\to Z b\anti b$
\item {LP2:} $\epem\to \zstar\to Z\hsm\to Z \tauptaum$
\item {LP3:} $\epem\to \zstar\to Z\hsm\to Z X$
\end{description}
\underline{Tevatron/TeV33}
\begin{description}
\item {T1:} $\wstar\to W\hsm\to Wb\anti b$
\item {T2:} $\wstar\to W\hsm\to W\tauptaum$
\item {T3:} $\zstar\to Z\hsm\to Zb\anti b$
\item {T4:} $\zstar\to Z\hsm\to Z\tauptaum$
\end{description}
\underline{LHC: $\mhsm\lsim 2\mw,2\mz$}
\begin{description}
\item {L1:} $gg\to\hsm\to\gamgam$
\item {L2:} $gg\to\hsm\to Z\zstar$
\item {L3:} $gg\to\hsm\to W\wstar$
\item {L4:} $WW\to\hsm\to\gamgam$
\item {L5:} $WW\to\hsm\to Z\zstar$
\item {L6:} $WW\to\hsm\to W\wstar$
\item {L7:} $\wstar\to W\hsm\to W\gam\gam$
\item {L8:} $\wstar\to W\hsm\to W b\anti b$
\item {L9:} $\wstar\to W\hsm\to W\tauptaum$
\item {L10:} $\wstar\to W\hsm \to WZ\zstar$
\item {L11:} $\wstar\to W\hsm \to WW\wstar$
\item {L12:} $t\anti t \hsm\to t\anti t \gam\gam$
\item {L13:} $t\anti t \hsm\to t\anti t b\anti b$
\item {L14:} $t\anti t \hsm \to t\anti t \tauptaum$
\item {L15:} $t\anti t \hsm \to t\anti t Z\zstar$
\item {L16:} $t\anti t \hsm \to t\anti t W\wstar$
\end{description}
\underline{LHC: $\mhsm\gsim 2\mw,2\mz$}
\begin{description} 
\item {H1:} $gg\to\hsm\to ZZ$
\item {H2:} $gg\to\hsm\to WW$
\item {H3:} $WW\to\hsm\to ZZ$
\item {H4:} $WW\to\hsm\to WW$
\item {H5:} $\wstar\to W\hsm\to WWW$
\item {H6:} $\wstar\to W\hsm\to W ZZ$
\end{description}
For $\mhsm\gsim 2\mw,2\mz$, we ignore $b\anti b$ decays of the $\hsm$
as having much too small a branching ratio, and 
$t\anti t$ decays are not relevant for $\mhsm\lsim 2\mt$.

We now tabulate the reactions of potential use in the five
different mass regions, M1, M2, M3, M4 and M5.
\begin{description}
\item{M1:} LP1, LP2, LP3, T1, T2, T3, T4, L1, L4, L7, L8, L9, L12, L13,
L14.
\item{M2:} T1, T2, T3, T4, L1, L4, L7, L8, L9, L12, L13, L14.
\item{M3:} L1, L2, L3, L4?, L5, L6, L7, L10, L11, L12?, L15, L16.
\item{M4:} L2, L3, L5, L6, L10, L11, L15, L16.
\item{M5:} H1, H2, H3, H4, H5, H6.
\end{description}

\subsubsection{Using Observed Rates to Extract Higgs Couplings}

Again, we divide our discussion according to the five different
mass regions listed above.						

\noindent\underline{M1}

Rates for reactions LP1, LP3, T1, T3, L1, L7, L8, L12, L13
will be well measured.  Our ability
to observe reactions LP2, T2, T4, L4, L9, L14
and determine with some reasonable accuracy the ratio
of the rates for these reactions to the better measured
reactions and to each other is less certain.
Considering only the well-measured rates to begin with
we find that we should be able to determine the following
quantities.
\begin{itemize}
\item Measurement of the rate for LP3 (\ie\ $Z\hsm\to Z X$
with $Z\to \epem,\mupmum$)
determines the $ZZ\hsm$ coupling (squared). For $\mhsm\sim 90-100\gev$,
$\sigma(Z\hsm)\sim 0.5\pb$ ($\rts=192\gev$), implying for $L=1000\pbi$
an event rate of about $0.06\times 500=30$.
Taking $S/B\sim 1$\footnote{Here, and in what follows, we 
denote the signal event rate by $S$ and the background event rate by $B$.}
for $\mhsm\sim\mz$ (we cannot use $b$-tagging
for this inclusive mode) 
gives a $1\sigma$ error, $\pm\sqrt{S+B}/S$, of $\pm 26\%$
on $\sigma(Z\hsm)$, corresponding to a $\sim \pm 12\%$ error
on the $ZZ\hsm$ coupling. For $\mhsm$ significantly below $\mz$,
$B/S$ will be smaller, and $S$ larger, implying smaller errors.
\item LP1/LP3 gives $\br(\hsm\to b\anti b)$, which can
be checked against the SM prediction, but on
its own does not allow a model-independent determination
of the $\hsm\to b\anti b$ coupling. For $\mhsm\sim\mz$, using $\br(\hsm\to
b\anti b)\sim 0.89$ and a $b$-tagging efficiency of 50\%
per $b$, we get $S=500\times 0.89\times (1-[0.5]^2)\sim 334$ 
in the $Z b\anti b$ 
channel. The net efficiency associated with the use of the various $Z$
decay modes is probably not more than 70\%, implying a usable $S=233$.
Taking $S/B=1$ ($b$-tagging included) we get $\sqrt{S+B}/S\sim 0.1$
for the $1\sigma$ error on $\sigma(Z\hsm)\br(\hsm\to b\anti b)$.
The error on $\br(\hsm\to b\anti b)$ will then be dominated by
the $\sigma(Z\hsm)$ error of $\sim\pm 26\%$.
\item The ratio T1/LP1 yields the  $(WW\hsm)^2/(ZZ\hsm)^2$ coupling-squared
ratio, and multiplying by the LP3 determination of $(ZZ\hsm)^2$
we get an absolute magnitude for $(WW\hsm)^2$.
The statistical error for T1
can be estimated from the results presented in Ref.~\cite{kky}.
For $\mhsm\sim \mz$, we average the $S$ and $B$ values presented in
Table I of Ref.~\cite{kky} for $\mhsm=80\gev$ and $\mhsm=100\gev$, obtaining
$S\sim 75$ and $B\sim 324$, implying $\sqrt{S+B}/S\sim 0.26$, for $L=10\fbi$.
Going to $60\fbi$ ($L=30\fbi$ for each of the two detectors) would
reduce the fractional error to $\sim 0.11$. 
Combining with the $\sim 0.1$ error on LP1
implies an error for T1/LP1 of order $\sim\pm 15\%$.
Systematic uncertainty would probably also
be present in relating the $\sigma(W\hsm)$ factor in the T1 rate
to the $WW\hsm$ coupling, and in the exact efficiencies
for isolating the $T1$ reaction. 
It is hard to imagine that $(WW\hsm)^2/(ZZ\hsm)^2$
could be determined to better than $\sim \pm 20\%$.
\item The ratio T1/T3 gives an independent determination of
$(WW\hsm)^2/(ZZ\hsm)^2$. The T3 error can 
be estimated from the results presented
in Ref.~\cite{wmyao}, Table I. For $L=60\fbi$, at $\mhsm=90\gev$ we
find $S\sim 216$ and $B\sim 1066$, implying $\sqrt{S+B}/S\sim 0.16$.
combining this with the T1 error quoted above and
including systematics, which might not be so large
for this type of ratio, we might achieve a $\sim \pm 20\%$ determination of
$(WW\hsm)^2/(ZZ\hsm)^2$. 
If this and the previous determination can be combined, then
a net error of order $\pm 14\%$ would appear to be possible.
Given the $\sim\pm 26\%$ error in the determination of $(ZZ\hsm)^2$
from $\sigma(Z\hsm)$, we obtain an error of $\sim\pm 30\%$ 
for $(WW\hsm)^2$.
\item The ratio T1/$\br(b\anti b)$ gives $(WW\hsm)^2$ and T3/$\br(b\anti b)$
gives $(ZZ\hsm)^2$. Given the $\sim\pm 26\%$ error on $\br(b\anti b)$
from LP1/LP3 and the $\sim\pm 15\%$ and $\sim\pm 19\%$ 
errors on T1 and T3 (treated 
individually, implying that systematic errors --- we take $10\%$ ---
should be included), 
we obtain about $\pm 30\%$ and $\pm 32\%$ error on the absolute magnitudes
of the individual $(WW\hsm)^2$ and $(ZZ\hsm)^2$ couplings-squared,
respectively. Combining with the previously
discussed determinations we see that errors on $(WW\hsm)^2$ and $(ZZ\hsm)^2$
of order $\sim\pm 20\%$ and $\sim\pm 22\%$ are to be expected, respectively.
\item The ratios L7/L8 and L12/L13 yield two independent
determinations of  $\br(\gam\gam)/\br(b\anti b)$. 
Alternatively, if it is difficult to separate L7 from L12
and/or L8 from L13
(\ie\ $W\hsm$ from $t\anti t\hsm$ production), we can take
(L7+L12)/(L8+L13) to get a single determination.
Multiplying by $\br(b\anti b)$, we get $\br(\gamgam)$. 
\begin{itemize}
\item
A first estimate of the errors for L7 and L12, performed 
in Ref.~\cite{dpfreport}, gave errors of $\sim\pm 13\%$,
assuming no inefficiencies associated with separating
L7 from L12. The individual errors on L7 and L12 
were re-examined for this report, the new estimates being
$\sim\pm 15\%$. 
If we combine L7+L12, the net error
on the sum would then be of order $\sim\pm 10\%$.
\item
Remarkably, the errors associated with separating
L7 from L12 are small. For
example, misidentification of $t\anti t\hsm$ as $W\hsm$
would mean that
(a) both $b$-jets are missed 
{\it and} 
(b) in addition to one $W$ from one of the $t$'s observed
in the leptonic decay mode the second $W$ must
decay to two jets with
mass different from $\mw$ or to $\ell\nu$ where the $\ell$ is
mis-identified.\footnote{Thus, misidentification has a probability
of $(1-\eps_{b-tag})^2[\br(W\to 2j)Prob(m_{2j}\slash\sim\mw)+\br(W\to
\ell\nu)\eps_{\ell-misid}]$ estimated at 
$(0.5)^2[0.8\cdot 0.1+0.2\cdot0.1]\sim 0.025$.}
The net probability for $t\anti t\hsm$ misidentification would
then be of order 2.5\%.
\item
Modes L8 and L13 ($Wb\anti b$ and $t\anti t b\anti b$) are still under study
by CMS. The ATLAS study \cite{ATLAS}
states that isolation of L8 may be impossible
at high luminosity because of the difficulty of vetoing extra jets.
The error on the L13 mode event rate can be estimated
for $\mhsm\sim \mz$ using Table 11.8 from Ref.~\cite{ATLAS}.
Averaging $80\gev$ and $100\gev$ results yields $S\sim 1355$
and $B\sim 37850$ for $L=100\fbi$.  At $L=600\fbi$, $S/\sqrt B\sim 17$
and $\sqrt{S+B}/S\sim 0.06$.
In the next item, we assume that the amount of contamination from L8 is small.
\item
Thus, so long as the (large) backgrounds in the $b\anti b$ channels are
well-understood, extraction of $\br(\gam\gam)/\br(b\anti b)$ in
the form L12/L13 would be possible.  
Using the above estimates of $\pm 15\%$ for L12 and $\pm 6\%$ for L13,
a statistical error of $\pm 17\%$ would be found for
$\br(\gam\gam)/\br(b\anti b)$.
Combining with the $\pm 26\%$ error on $\br(b\anti b)$ from LEP2
implies error on $\br(\gam\gam)$ of $\sim\pm 31\%$.
\end{itemize}

\begin{table}[hbt]
\caption[fake]{We tabulate the approximate error in the determination
of $\sigma(gg\to\hsm)\br(\hsm\to \gam\gam)$ as a function
of $\mhsm$ (in GeV) assuming $L=300\fbi$ for the CMS 
and ATLAS detectors at the LHC.}
\begin{center}
\begin{tabular}{|c|c|c|c|c|}
\hline
 Mass & 90 & 110 & 130 & 150 \\
\hline
CMS Error & $\pm 9\%$ & $\pm 6\%$ & $\pm 5\%$ & $\pm 8\%$ \\
ATLAS Error & $\pm 23\%$ & $\pm 7\%$ & $\pm 7\%$ & $\pm 10\%$ \\
\hline
Combined Error & $\pm 8.5\%$ & $\pm 4.5\%$ & $\pm 4.0\%$ & $\pm 6.2\%$ \\
\hline
\end{tabular}
\end{center}
\label{2gamerrors}
\end{table}

\item L1/$\br(\gamgam)$ yields the magnitude of the $(gg\hsm)^2$
coupling-squared, which is primarily sensitive to the $t\anti t\hsm$
coupling. The error on the L1 rate is quite different for ATLAS
and CMS.  At $\mhsm\sim \mz$, for $L=300\fbi$ ATLAS \cite{ATLAS} expects 
$S\sim 1650$ and $B\sim 142800$ yielding an error of $\sim\pm 23\%$;
CMS (see Fig. 12.3 and associated tables in Ref.~\cite{CMS}) 
expects $S\sim 3825$ and $B\sim 115429$ yielding 
${\sqrt{S+B}/ S}\sim 0.09$.
(For later reference, we give the errors on the L1 event rate for CMS
and ATLAS with $L=300\fbi$ in Table~\ref{2gamerrors}.)
The much better CMS result derives from: i) CMS includes
a QCD correction factor of $K=1.5$ in the L1 rate, whereas
ATLAS does not; ii) for $\mhsm\sim \mz$, ATLAS reduces the $\gam$
efficiency from 80\% to 72\% (needed to reject the $Z$ continuum).
As a compromise, we adopt the approach of computing the
net L1 error by combining CMS and ATLAS as stated, thereby obtaining 
an error on the L1 rate of $\sim\pm 8\%$.
In any case, it will be much smaller than the error of $\sim\pm 31\%$ 
on $\br(\gamgam)$, which will therefore
dominate the error on $(gg\hsm)^2$.
\item L12/L7 and L13/L8 yield independent determinations of 
the $t\anti t\hsm/WW\hsm$ event rate ratio. 
By multiplying by the previously determined value
of $(WW\hsm)^2$ we get an absolute magnitude for the $(t\anti t\hsm)^2$
coupling-squared which can be checked against the $gg\hsm$ result.
As noted earlier, L12 can be efficiently separated from L7, whereas
isolation of L8 is very uncertain at high luminosity.
Since the L7 and L12 rates have errors of $\sim \pm 15\%$ (see above),
we predict an error on L12/L7 of about $\pm 21\%$, not including
any systematic uncertainty. Given the $\sim\pm 20\%$ error in $(WW\hsm)^2$,
an error of $\sim \pm 30\%$ for $(t\anti t\hsm)^2$ is anticipated,
\ie\ comparable to that coming from the $(gg\hsm)^2$ coupling-squared
determination.
\end{itemize}
What is missing from the above list is any determination
of the $(b\anti b \hsm)$, $(\tauptaum\hsm)$ and $(\gamgam\hsm)$ couplings,
any check that fermion couplings are proportional to the
fermion mass (other than the $(t\anti t\hsm)$ coupling magnitude),
and the Higgs total width. Given the $(WW\hsm)$ and $(t\anti t\hsm)$
couplings we could compute the expected value for the $(\gamgam\hsm)$
coupling, and combine this with $\br(\gamgam)$ to get
a value for $\gamhsm$.  $\br(b\anti b)\gamhsm$ then yields
$b\anti b\hsm$ and we would have a somewhat indirect check
that $b\anti b\hsm/t\anti t\hsm=\mb/\mt$. Some systematic
uncertainty in the correct values of $\mb$ and $\mt$
would enter into this check, but the propagation of 
the already rather significant statistical
errors would be the dominant uncertainty.

In the above, a very critical ingredient was the small probability
of mis-identifying a $t\anti t\hsm$ event as a $W\hsm$ event,
and vice versa. Further careful studies of this issue by the detector
groups would be useful.

Let us now ask what we would gain by adding reactions LP2, T2,
T4, L4, L9, and L14. LP2/LP1, T2/T1, T4/T3 L9/L8 and L14/L13 would all
allow different determinations of $\tauptaum\hsm/b\anti b\hsm$.
This would certainly be of significant value, but with what
accuracy could these ratios be measured? After including
efficiencies for $\tau$ identification, the rate for
$Z\hsm\to 2j+2\tau$ at LEP2 is about 8 events over a background
of 5 for $L=1\fbi$, for
$\mhsm\simeq\mz$. This makes use of the estimated mass resolution
$\sigma_m\sim 2-3 \gev$ for a $2\tau$ pair. The known $Z$ branching fractions
could then be used to extract the $\hsm\to \tauptaum$ portion
of the net rate.  At best, LP2/LP1 could be extracted
with $\sim\pm50\%$ accuracy implying (taking the square root)
that  the $(\tauptaum\hsm)/(b\anti b\hsm)$  coupling
ratio could be extracted with an error of order $\sim \pm 30\%$.

What about $W\hsm$ and $t\anti t\hsm$
production with $\hsm\to \tauptaum$ at TeV33 (T2 and T4) and the LHC (L9 and
L14)? At the time that this report
is being prepared, the status of T2 and T4 at TeV33 is still
being debated.  We will not attempt any estimates.
At the LHC, L9 and L14 are not deemed observable at $\mhsm\lsim 100\gev$
because of the very large backgrounds associated with $Z\to\tauptaum$.

Finally, the unstudied mode L4 does not provide any crucial new
information given that the $WW\hsm$ coupling cannot be very well
determined, and given that it would probably be difficult
to separate $WW\to \hsm$ fusion from $gg\to \hsm$ fusion for
the low values of $\mhsm$ appropriate to mass region M1.

We end by summarizing in Table~\ref{m1errors} the errors on fundamental
branching ratios, couplings-squared, and ratios thereof as obtained
above at $\mhsm\sim\mz$ by combining LEP2, TeV33 and LHC data.

\begin{table}[hbt]
\caption[fake]{Summary of approximate
errors for branching ratios and couplings-squared
at $\mhsm\sim \mz$ in the M1 mass region.
Where appropriate, estimated systematic errors are included.
Quantities not listed cannot be determined in a model-independent manner.
As discussed in the text, directly measured products of couplings-squared times
branching ratios can often be determined with better accuracy.}
\begin{center}
\begin{tabular}{|c|c|}
\hline
 Quantity & Error \\
\hline
 $\br(b\anti b)$ & $\pm 26\%$ \\
\hline
 $(WW\hsm)^2/(ZZ\hsm)^2$ & $\pm 14\%$ \\
\hline
 $(WW\hsm)^2$ & $\pm 20\%$ \\
\hline
 $(ZZ\hsm)^2$ & $\pm 22\%$ \\
\hline
 $(\gam\gam\hsm)^2/(b\anti b\hsm)^2$ & $\pm 17\%$ \\
\hline
 $\br(\gam\gam)$ & $\pm 31\%$ \\
\hline
 $(gg\hsm)^2$ & $\pm 31\%$ \\
\hline
 $(t\anti t\hsm)^2/(WW\hsm)^2$ & $\pm 21\%$ \\
\hline
 $(t\anti t\hsm)^2$ & $\pm 30\%$ \\
\hline
\end{tabular}
\end{center}
\label{m1errors}
\end{table}

\noindent\underline{M2}

Rates for reactions T1, T3, L1, L7, L8, L12, L13
will be well measured.  Reactions T2, T4, L4, L9, L14
are less robust. Relative to mass region M1, we suffer
the crucial loss of a measurement of the magnitude of the $(ZZ\hsm)$
coupling-constant-squared.
Considering first the well-measured rates, we should be able to
determine the following quantities.
\begin{itemize}
\item 
The ratio T1/T3 gives a determination of
$(WW\hsm)^2/(ZZ\hsm)^2$. Following a similar procedure as at 
$\mhsm\sim\mz$, the statistical error for T1
can be estimated from the results presented in Ref.~\cite{kky}.
For $\mhsm\sim 100,120\gev$,
Table I of Ref.~\cite{kky} shows 
$S\sim 52,27$ and $B\sim 257,137$, implying $\sqrt{S+B}/S\sim 0.34,0.47$, 
for $L=10\fbi$.
Going to $60\fbi$ would reduce the fractional error to $\sim 0.14,0.19$.
Table I of Ref.~\cite{wmyao} for reaction T3 implies $S\sim 184,102$ 
and $B\sim 990,756$ for $L=60\fbi$, implying fractional error of 
$\sim 0.19,0.29$ at $\mhsm=100,120\gev$.
The resulting error on the ratio of the couplings-squared,
$(WW\hsm)^2/(ZZ\hsm)^2$, would then be $\sim\pm 23\%,\pm 34\%$
at these two masses.
\item
The errors for L7 and L12
are predicted to be similar in the M2 mass range to those found in
the M1 mass range, \ie\ of order $\pm 15\%$.
\item
The utility of the $b\anti b$ final states at the LHC, modes L8 and L13,
is still being debated.  No explicit CMS results are available at
the time of writing. ATLAS states \cite{ATLAS} 
that only the $t\anti t\hsm$ process
can be extracted at high luminosity when $\hsm\to b\anti b$.
Here, we note that even if 
L8 and L13 are not viable discovery channels, it may still be possible
to get a semi-accurate measurement of important ratios of branching ratios
once the Higgs has been discovered. A rough estimate of the
accuracy with which L13 can be measured is possible from
Table 11.8 in Ref.~\cite{ATLAS}. For $L=100\fbi$,
ATLAS expects $S=870,420,283$ and $B=35100,28300,20000$
at $\mhsm=100,120,130\gev$ (where the $130\gev$
numbers are obtained by extrapolation).
Assuming that CMS studies will ultimately
yield similar results we upgrade these numbers to $L=600\fbi$, 
and find accuracies for the L13 rate of 
$\pm 9\%,\pm 16\%,\pm 21\%$ at the above respective masses.
\item 
The ratios L7/L8 and L12/L13 yield two independent
determinations of  $\br(\gam\gam)/\br(b\anti b)$.  At the moment we
can only estimate the accuracy of the L12/L13 determination
of $\br(\gam\gam)/\br(b\anti b)$: using $\pm15\%$ for the
error in L12 and the above estimates for the L13 errors we obtain
errors for L12/L13 of $\pm 17\%,\pm22\%,\pm25\%$ at $\mhsm=100,120,130\gev$.
\item L12/L7 and L13/L8 yield independent determinations of
$(t\anti t\hsm)^2/(WW\hsm)^2$.
Since L8 is dubious, we focus on L12/L7. 
Since the numerator and denominator errors are both of order $\pm 15\%$
in the M2 mass region, 
the error on this ratio is of order $\pm 21\%$, substantially better
than the TeV33 expectation of $\sim\pm 34\%$.
\end{itemize}
Thus, we will have ways of determining the
$(WW\hsm):(ZZ\hsm):(t\anti t\hsm)$  coupling ratios, but no absolute
coupling magnitudes are directly determined, and there is no test
of the fermion-Higgs coupling being proportional to fermion
mass. Once again,
an important ingredient in determining the
$(WW\hsm)^2/(t\anti t\hsm)^2$ ratio is the ability to separate
$W\hsm$ from $t\anti t\hsm$ final states in the $\gamgam$ decay mode of the
$\hsm$.

To proceed further, requires more model input.  Given that
we know (in the SM) how to compute $\br(\gamgam)$
from the $WW\hsm$ and $t\anti t\hsm$ couplings, and given
that we know the ratio of the latter, $\br(\gamgam)/\br(b\anti b)$
would yield a result for $t\anti t\hsm/b\anti b\hsm$
which could then be checked against the predicted $\mt/\mb$.

Let us now ask what we would gain by adding reactions T2,
T4, L4, L9, and L14. T2/T1, T4/T3, L9/L8 and L14/L13 would all allow
different determinations of $(\tauptaum\hsm)^2/(b\anti b\hsm)^2$.
This would allow a model independent check of the predicted
$\mtau^2/[3\mb^2(\mhsm)]$ result. A first look at the LHC L9 and L14
rates is described below; recall that $\mhsm\gsim 100\gev$, \ie\
in the M2 mass region, is required in order that the $Z\to\tauptaum$
backgrounds to L9 and L14 be manageable.
(We continue to leave aside the $\tauptaum$ modes T2 and T4
at TeV33 as being too uncertain.) 
Reaction L4 does not provide new information, and will not be considered.

We have estimated rates for L9 at the LHC.
At $\mhsm=110\gev$, $\sigma \br(W\hsm\to l\nu \tauptaum)\sim 19\fb$.
The $\Delta m_{\tauptaum}$ would be about $11\gev$ ($21\gev$)
at low (high) luminosity. 
The acceptance factor (which takes into account the kinematical
cuts, mass bin acceptance, the $\tau$ identification efficiency
and the efficiency of reconstructing the escaping neutrinos) is
only about $0.15\%$ ($0.07\%$), at low (high) $L$. At high $L$
with $600\fbi$ (3 years running), this would leave us with
$N=600\fbi\times 19\fb\times 0.0007=8$ events.  This is clearly
a very marginal rate.

An alternative approach to identifying the $\tauptaum$
final state is to use $\tauptaum\to \ell+hadron+X$,
which has an effective $\br\sim 50\%$, implying about 25 events
per detector at low luminosity ($L=30\fbi$). For $L=600\fbi$
one would have 500 events. But substantial cuts would
be need to eliminate backgrounds
from $W\tauptaum$, $t\anti t$ and $WW\to \ell\nu\ell\nu$.

The final mode is $\tauptaum\to \ell\nu\ell\nu$ with
$\br\sim 0.12$, implying about 12 events for a detector-summed
$L=60\fbi$ (low luminosity) or 120 events
for total $L=600\fbi$. However, this is before any cuts
required to eliminate backgrounds.

We are not optimistic that L9 can be measured
at a useful level of accuracy at the LHC.
It appears that any determination of the $(\tauptaum\hsm)^2/(b\anti b\hsm)^2$
coupling-squared ratio will be extremely rough.

Finally, if signal L4
proves viable, L1/L4 would give $(gg\hsm)^2/(WW\hsm)^2$, which
in the SM would yield a determination of $(t\anti t\hsm)^2/(WW\hsm)^2$
that could be checked against the L12/L7 determination.
The key question is whether the $WW$ fusion reaction can be
separated from the $gg$ fusion reaction in order to get at L1/L4.
Some work by the ATLAS collaboration \cite{wwseparation}
showed that this may be very difficult at Higgs masses in
the $100\gev$ range.

We summarize as a function of $\mhsm$ in Table~\ref{m2errors} 
the errors for the few coupling-squared ratios that
can be determined in the M2 mass region.

\begin{table}[hbt]
\caption[fake]{Summary of approximate errors for coupling-squared ratios
at $\mhsm=100,110,120,130\gev$ in the M2 mass region.
As discussed in the text, directly measured products of couplings-squared times
branching ratios can often be determined with better accuracy.}
\footnotesize
\begin{center}
\begin{tabular}{|c|c|c|c|c|}
\hline
 Quantity & \multicolumn{4}{c|}{Errors} \\
\hline
\hline
 Mass (GeV) & 100 & 110 & 120 & 130 \\
\hline
 $(WW\hsm)^2/(ZZ\hsm)^2$ & $\pm 23\%$ & $\pm 26\%$ & $\pm 34\%$ & $-$ \\
\hline
 $(\gam\gam\hsm)^2/(b\anti b\hsm)^2$ & $\pm 17\%$ & $\pm 19\%$ & $\pm 22\%$ &
 $\pm 25\%$ \\
\hline
 $(t\anti t\hsm)^2/(WW\hsm)^2$ & $\pm 21\%$ & $\pm 21\%$ & $\pm 21\%$ &
 $\pm 21\%$  \\
\hline
\end{tabular}
\end{center}
\label{m2errors}
\end{table}

\noindent\underline{M3}

Of the potential channels listed under M3, only L1 and L2
are thoroughly studied and certain to be measurable over
this mass interval. L1 should be viable for $\mhsm\lsim
150\gev$. L2 (the $gg\to \hsm\to
Z\zstar$ reaction) should be good for $\mhsm\gsim 130\gev$.
With these two modes alone, we discover the Higgs, 
and for $130\lsim \mhsm\lsim 150\gev$ we can determine
$\br(\gamgam)/\br(Z\zstar)$.  
The errors for the measurement of L2 have been estimated
from the high luminosity results presented in Table 29 of Ref.~\cite{atlas48}.
For $L=600\fbi$ we find the errors listed in Table~\ref{4lerrors}. As
expected, quite decent results are obtained for $\mhsm\gsim 130\gev$.
The errors in the $\gam\gam$ mode L1 rate obtained by
combining ATLAS and CMS results
would be $\pm 4\%-\pm 5\%$ for $\mhsm$ in the $110-130\gev$
range, rising to $\sim \pm 6\%$ at $\mhsm=150\gev$;
see Table~\ref{2gamerrors}. The errors for $(\gam\gam\hsm)^2/(ZZ\hsm)^2$
deriving from the L1/L2 ratio are tabulated in Table~\ref{m3errors}.
This ratio is interesting, but cannot be unambiguously 
interpreted.

\begin{table}[hbt]
\caption[fake]{We tabulate the error in the determination
of $\sigma(gg\to\hsm)\br(\hsm\to 4\ell)$ as a function
of $\mhsm$ (in GeV) assuming $L=600\fbi$ at the LHC.}
\begin{center}
\begin{tabular}{|c|c|c|c|c|c|}
\hline
 Mass & 120 & 130 & 150 & 170 & 180 \\
 Error & $\pm 25\%$ & $\pm 9.5\%$ & $\pm 5.3\%$ & $\pm 11\%$ & $\pm 6.1\%$ \\
\hline
 Mass & 200 & 220 & 240 & 260 & 280 \\
 Error & $\pm 7.8\%$ & $\pm 6.9\%$ & $\pm 6.2\%$ & $\pm6.2\%$ & $\pm6.2\%$ \\
\hline
 Mass & 300 & 320 & 340 & 360 & 380 \\
 Error & $\pm 6.2\%$ & $\pm6.2\%$ & $\pm 6.1\%$ & $\pm 6.0\%$  & $\pm 6.4\%$ \\
\hline
 Mass & 400 & 500 & 600 & 700 & 800 \\
 Error & $\pm 6.7\%$ & $\pm 9.4\%$ & $\pm 14\%$ & $\pm 20\%$ & $\pm28\%$ \\
\hline
\end{tabular}
\end{center}
\label{4lerrors}
\end{table}

The L3 mode was first examined in detail in Refs.~\cite{gloveretal,hanetal}.
It was found that with some cuts it might be possible to dig out 
a signal in the $\ell\nu\ell\nu$ decay mode of the $W\wstar$ final state.
A more recent study \cite{ditdr} 
focusing on the $\mhsm\gsim 155\gev$ mass region
finds that additional cuts are necessary in the context of a more
complete simulation, but that very promising $S/\sqrt B$ can
be obtained. Here
we give a rough extrapolation into the $130-150\gev$ mass region
of their results by simply using the mass dependence of $\br(\hsm\to W\wstar)$.
We do not include the rise in the cross section as $\mhsm$ decreases
since it is likely that there will be a compensating decrease in
the efficiency with which the cuts of Ref.~\cite{ditdr} accept events.
We begin with the $\mhsm=155\gev$, $L=5\fbi$ result from their Table~2
of $S=49$ and $B=92$.  We upgrade to $L=600\fbi$ and correct
for $\br(\hsm\to W\wstar)$ to obtain the statistical
errors for $\sigma(gg\to\hsm\to
W\wstar)$ listed in Table~\ref{ggwwstarerrors}; this table also includes
the $\mhsm\geq 155\gev$ results. Presumably, one must also allow for
a $\sim \pm 10\%$ systematic uncertainty in absolute normalization.
This would then be the dominant error!  However, we do not include
this systematic error in the errors quoted for the $(WW\hsm)^2/(ZZ\hsm)^2$
coupling-squared ratio as computed from L3/L2. The amount
of systematic error that should be incorporated in estimating
the error for such a ratio requires further study.
The resulting statistical $(WW\hsm)^2/(ZZ\hsm)^2$ 
errors are tabulated in Table~\ref{m3errors}. Apparently
L3/L2 will provide a decent measurement of the $(WW\hsm)^2/(ZZ\hsm)^2$ 
coupling-squared ratio, thereby allowing a check
that custodial SU(2) is operating, so long as the systematic error
is $\lsim 10\%$.

\begin{table}[hbt]
\caption[fake]{We tabulate the statistical error in the determination
of $\sigma(gg\to\hsm\to W\wstar)$ as a function
of $\mhsm$ (in GeV) assuming $L=600\fbi$ at the LHC. For $\mhsm\leq150\gev$,
the errors are based on extrapolation from $\mhsm\geq 155\gev$
results. See text.}
\begin{center}
\begin{tabular}{|c|c|c|c|c|c|}
\hline
 Mass & 120 & 130 & 140 & 150 & $155-180$ \\
 Error & $\pm 12\%$ &  $\pm 6\%$ &  $\pm 3\%$ &  $\pm 3\%$ & $\pm 2\%$ \\
\hline
\end{tabular}
\end{center}
\label{ggwwstarerrors}
\end{table}

\begin{table}[hbt]
\caption[fake]{We tabulate the statistical errors at $\mhsm=120,130,150\gev$
in the determinations
of $(\gam\gam\hsm)^2/(ZZ\hsm)^2$ and $(WW\hsm)^2/(ZZ\hsm)^2$,
assuming $L=600\fbi$ at the LHC.}
\begin{center}
\begin{tabular}{|c|c|c|c|}
\hline
 Quantity & \multicolumn{3}{c|}{Errors} \\
\hline
\hline
Mass (GeV) & 120 & 130 & 150 \\
\hline
 $(\gam\gam\hsm)^2/(ZZ\hsm)^2$
& $\pm 25\%$ & $\pm 11\%$ & $\pm 10\%$ \\
\hline
 $(WW\hsm)^2/(ZZ\hsm)^2$
 & $\pm 27\%$ & $\pm 11\%$ & $\pm 6\%$ \\
\hline
\end{tabular}
\end{center}
\label{m3errors}
\end{table}

The L4 mode could become of critical importance, since
L4/L1 yields a determination of $(WW\hsm)^2/(gg\hsm)^2$
which (assuming only SM particles in the loops)
yields a value of $(WW\hsm)^2/(t\anti t\hsm)^2$.
But, at best the L4 mode might survive for $\mhsm\lsim 140\gev$.
Further, the ability to separate $WW$ fusion from $gg$ fusion
production has not been studied at $\mhsm$ values this low.
The $WW$ fusion rate is $\sim 1/5$ of the $gg$ fusion rate;
see Fig.~15, Ref.~\cite{dpfreport}.
   
Let us now turn to other modes. Consider L10, L11, L15, and L16.
To begin, we relate L15 to L2. The maximum rate
for $gg\to\hsm\to Z\zstar\to 4\ell$ is 69 events at $\mhsm=150\gev$ for
$L=100\fbi$, implying about 410 events at combined $L=600\fbi$.
The L15 $t\anti t \hsm \to t\anti t 4\ell$ rate is about a factor
of 50 smaller at this mass 
implying 8 events.  This seems too marginal to warrant further
consideration. L10 would be still worse. The L11 and L16 
($W\ell\nu\ell\nu$ and $t\anti t\ell\nu\ell\nu$ final state channels) 
each have $\sigma\br(\hsm\to \ell\nu\ell\nu)\sim 1.3\fb$. (No
$t$ or $W$ branchings ratios are included; tagging with the two leptons
from the $\hsm$ decay is sufficient.)
The $L=600\fbi$ event rates for each channel would thus be of order
800, \ie\ larger than the L10 and L15 $4\ell$ event rates.
But the inability to reconstruct the resonance mass in this channel
would make extraction of a signal difficult. Separation
of $W\hsm$ from $t\anti t\hsm$ events could be performed as sketched
earlier, but the input event rates would be lower due to the
necessity of focusing on particular $W$ and $t$ decay final states.
Still, further work on these $\ell\nu\ell\nu$ channels is clearly warranted,
especially in light of the good results obtained in the inclusive
$\ell\nu\ell\nu$ final state.
Could the $\ell\nu 2j$ Higgs decay channel be used
in $W\hsm$ and $t\anti t\hsm$
associated production?  For the moment, 
we adopt a pessimistic attitude. Clearly, given
the importance of L11/L16 as a means of determining
the $(WW\hsm)^2/(t\anti t\hsm)^2$ coupling-squared ratio, much more effort
should be devoted in both the $\ell\nu\ell\nu$ and $\ell\nu 2j$ channels 
to determining if it will be possible to
separately measure L11 and L16.

How about L5 and L6?  Using a ratio of 1/5 for
the $WW/gg$ fusion production cross section ratio, 
we are left with about 80 events
in the (L5) $WW\to\hsm \to Z\zstar\to 4\ell$ mode at $\mhsm=150\gev$;
spectator jet tagging might allow a small background.
If we assume 20\% efficiency for double tagging adequate
to effectively remove the $gg$ fusion process (L2),
we would be left with 16 events.  While far from wonderful,
this would allow in principle a $\lsim \pm25\%$ determination
of the L5/L2 ratio implying an implicit determination of the $(t\anti
t\hsm)/(WW\hsm)$ ratio to $\sim\pm13\%$.  The L6 mode perhaps deserves a
look, since it might turn out that double
spectator tagging could keep the $gg$-fusion and other backgrounds small. 
L6/L5 would then yield $(WW\hsm)^2/(ZZ\hsm)^2$, 
which could be combined with the
L5/L2 result to give the very important set of relative weights:
$(WW\hsm):(ZZ\hsm):(t\anti t\hsm)$. If these relative weights agree with
expectations for the $\hsm$, it would be hard to imagine that the
observed Higgs boson is not a SM-like Higgs. As noted, the ability to
separate $WW$ fusion events from $gg$ fusion events with decent
efficiency down at this low mass, using spectator jet tagging, will be
critical for the above procedure.  
 
\noindent\underline{M4}

Let us now turn to the $155\lsim\mhsm\lsim 2\mz$ mass region.
The most significant variation in this region arises due to
the fact that as $\hsm\to WW$ becomes kinematically
allowed at $\mhsm\sim 160\gev$, the $\hsm\to Z\zstar$ branching ratio
dips, the dip being almost a factor of 4 at $\mhsm=170\gev$;
see Table~\ref{hsmbrs}. As a consequence, at $\mhsm=170\gev$ $S/B$
(using ATLAS numbers)
drops to 20/9.5 for $L=100\fbi$ compared to 69/10 at $\mhsm=150\gev$.
Nonetheless, these $S$ and $B$ rates show that
L2 can still be regarded as iron-clad throughout this region
provided adequate $L$ is accumulated.
For $L=600\fbi$, an accurate measurement of $(gg\hsm)^2\br(\hsm\to Z\zstar)$
is clearly possible; results were already tabulated in Table~\ref{4lerrors}.

L3 is now an on-shell $WW$ final state, and,
according to the results summarized in Table~\ref{ggwwstarerrors},
can be measured with good statistical accuracy in the $\ell\nu\ell\nu$
final state of the $\hsm\to WW$ Higgs decay.
The statistical accuracy for $(WW\hsm)^2/(ZZ\hsm)^2$ deriving
from L3/L2 is tabulated in Table~\ref{m4errors}.

The fact that L3 provides a good signal can be traced
to the large associated rates.
The cross section for L3 is about $16\pb$. Neglecting the $WW$
fusion inclusive contribution would mean that we could
just collect events inclusively.  Taking $\br(\hsm\to WW)\sim 1$,
$\br(W\to\ell\nu)\sim2/9,\br(W\to 2j)\sim 2/3$ and $L=600\fbi$, we get
$\sim 5\times 10^5$ events in the $\ell\nu\ell\nu$ channel
and $\sim 3\times 10^6$ events in the $\ell\nu 2j$ channel.
Although the continuum $WW$ and the $t\anti t$ backgrounds
are large, there is lots of room for making cuts of the type
considered in \cite{ditdr}, which achieve
$S/B=1$ and $S/\sqrt B=5-10$ in the M4 mass region
for only $L=5\fbi$.  Thus,
the error on the L3/L2 determination of $(WW\hsm)^2/(ZZ\hsm)^2$
in the M4 mass region is dominated by that for the $4\ell$
channel (tabulated in Table~\ref{4lerrors}).

Rates associated with measuring L5 are expected to be low
given the small $\br(\hsm\to Z\zstar)$ in this mass region and the 
probably low efficiency for the double spectator tagging required to isolate
the $WW$ fusion process.  We have made a rough estimate of what might be
expected as follows. We take
1/5 as the ratio for the $WW$ fusion production rate
as compared to the $gg$ fusion rate. We then assume a tagging efficiency
(associated with eliminating the $gg$ fusion signal)
for both signal and background of order 20\%. The result is that L5 errors
would be about a factor of 5 larger than the L2 errors listed 
in Table~\ref{4lerrors},
implying at least $\pm 25\%$ statistical error for measuring the L5 rate.
This in turn implies at least $\pm 25\%$ statistical error for
measuring $(WW\hsm)^2/(gg\hsm)^2$ via L5/L2.  While not particularly
wonderful, this level of error would at least be useful.
A more detailed study should be performed to see if one could do better.

L6 would now be an on-shell final state, and might be measurable.
The cross section for $WW$ fusion is about $3\pb$ in this mass region.
Assuming 20\% efficiency for double spectator tagging, 
$\br(\hsm\to WW)\sim 1$,
$\br(W\to\ell\nu)\sim2/9,\br(W\to 2j)\sim 2/3$ and $L=600\fbi$, we get
$\sim 2\times 10^4$ events in the $\ell\nu\ell\nu$ channel
and $\sim 1.2\times 10^5$ events in the $\ell\nu 2j$ channel.
It should be possible to get a decent measurement of L6 given
the background reduction that would be obtained as part
of the double-tagging procedure used to make $gg$ fusion small.
L6/L3 would determine $(WW\hsm)^2/(gg\hsm)^2$ and, thence, 
yield and implicit determination of $(WW\hsm)/(t\anti t\hsm)$.

We discard out-of-hand the L10 and L15 reactions
given that $\br(\hsm\to Z\zstar)$ is in the dip region. The L11 and L16 
reactions become on-shell decays, and probably deserve a close look,
given that their ratio would yield the vital $WW\hsm/t\anti t\hsm$
ratio. We have not performed a study for this report.
However, event rates are again encouraging. L11 has a cross section of
about $0.3\pb$ and L16 is about $0.2\pb$. Assuming 10\% efficiency
for tagging and
isolating these processes from one another, $\br(\hsm\to WW)\sim 1$,
and the standard $W$ decay branching ratios, we get 
$\sim 9\times 10^2$ and $\sim 5\times 10^3$ events in the
$\ell\nu\ell\nu$ and $\ell\nu 2j$ channels, respectively (for $L=600\fbi$).
Given that backgrounds associated with these final states 
could be small because of our ability to tag these channels,
the above event numbers might be sufficient to yield
a reasonable determination of the L16/L11 ratio that would
give a value for $(t\anti t\hsm)^2/(WW\hsm)^2$.

\begin{table}[hbt]
\caption[fake]{We tabulate the statistical errors at $\mhsm=155,170,180\gev$
in the determination of $(WW\hsm)^2/(ZZ\hsm)^2$ from L3/L2,
assuming $L=600\fbi$ at the LHC.}
\begin{center}
\begin{tabular}{|c|c|c|c|}
\hline
 Quantity & \multicolumn{3}{c|}{Errors} \\
\hline
\hline
Mass (GeV) & 155 & 170 & 180 \\
\hline
 $(WW\hsm)^2/(ZZ\hsm)^2$ & $\pm6\%$ & $\pm 11\%$ & $\pm 7\%$ \\
\hline
\end{tabular}
\end{center}
\label{m4errors}
\end{table}

\noindent\underline{M5} 

Finally we consider $\mhsm\gsim 2\mz$.
The first important remark is that $\gamhsm$ becomes
measurable in the $4\ell$ channel once $\gamhsm\gsim (1\%-1.5\%)\times\mhsm$,
which occurs starting at $\mhsm\sim 200\gev$ where $\gamhsm\sim 2\gev$.
Quantitative estimates for the precision of the $\gamhsm$ measurement
will be discussed in Section G.
At $\mhsm=210$, $250$, $300$, and $400\gev$,
rough percentage error expectations (assuming $L=600\fbi$ for ATLAS+CMS)
for $\gamhsm$ are $\pm 21\%$, $\pm 7\%$, $\pm 4\%$ and $\pm 3\%$,
respectively. 

Among the H1 to H6 modes, only H1 is gold-plated, and of course
it alone provides very limited information about the actual
Higgs properties. As described for the M4 mass region,
the mode H2 has been studied for masses close to $2\mz$ 
in the $\ell\nu\ell\nu$ final state in \cite{gloveretal,hanetal}
and in the M4 mass region in \cite{ditdr}. These
results indicate that reasonable to good accuracy for the H2/H1 ratio,
implying a reasonably accurate implicit
determination of $(WW\hsm)^2/(ZZ\hsm)^2$,
might be possible for Higgs masses not too far above $2\mz$.
One could also ask if it would be possible to separate out the $WW$
final state in the $\ell\nu jj$ mode where a mass peak could be
reconstructed (subject to the usual two-fold ambiguity procedures).
Event rates would be quite significant, and a Monte Carlo 
study should be performed. 

Processes H3 and H4 would have to be separated from H1 and H2
using spectator jet tagging to isolate the former $WW$
fusion reactions. If this were possible, then
H3/H1 and H4/H2 would both yield a
determination of $(t\anti t\hsm)^2/(WW\hsm)^2$ under
the assumption that the $t$-loop dominates the $(gg\hsm)$ coupling.
However, the mass range for which separation of H3 and H4 would be possible
is far from certain.\footnote{A recent study \cite{wwpoggioli} has shown
that forward jet tagging allows isolation of H4 in the $\ell\nu jj$ final state
for $\mhsm\gsim 600\gev$ (\ie\ beyond
the mass range being explicitly considered here), but suggests that
the $W$+jets background is difficult to surmount for lower masses.
However, strategies in the mass range down near $2\mz$ could be quite
different given the much larger signal rates.}

Isolation of H4 is of particular importance given that
$\gamhsm$ becomes directly measurable in the $4\ell$ final state
once $\mhsm\gsim 2\mz$.  This is because the rate for H4
is proportional to $(WW\hsm)^2\br(\hsm\to WW)$.  Multiplying
by $\gamhsm$ yields $(WW\hsm)^4$, which implies a very accurate
determination of the $WW\hsm$ coupling for even modest 
accuracy of the experimental inputs.  Thus, further study of H4
for all values of $\mhsm$ above $2\mz$ is a priority.
If the $(WW\hsm)^2:(ZZ\hsm)^2:(t\anti t\hsm)^2$ ratios 
could also be determined (using H1-H4 as outlined above),
then the $(WW\hsm)^2$ magnitude would yield
absolute values for $(ZZ\hsm)^2$ and $(t\anti t\hsm)^2$
and, thence, a detailed test of the SM predictions.

We have not pursued the processes H5 and H6, as they will
have lower rates.  On the other hand, the backgrounds will
be different, and one could imagine using them to confirm
some of the results obtained from H1 through H4.

\subsection{Measuring $\sigma\br(\hsm\to c\anti c, b\anti b,W\wstar)$
using NLC and $s$-channel FMC data}

We divide the discussion into:
\begin{itemize}
\item
measurements that would be performed
by running at $\rts= 500\gev$ 
at the NLC (or in NLC-like running at the FMC) ---
the production modes of interest are 
$\epem\to Z\hsm$, $\epem\to\epem\hsm$ ($ZZ$-fusion)
and $\epem\to\nu\anti\nu\hsm$ ($WW$-fusion);\footnote{In the following,
we will consistently use the notation $\epem\hsm$ and $\nu\anti\nu\hsm$
for the $ZZ$ fusion and $WW$ fusion contributions to these final state channels
only. The contributions to these same final states from $Z\hsm$ with
$Z\to\epem$ and $Z\to\nu\anti\nu$, respectively, and interference
at the amplitude level with the $ZZ$ and $WW$ fusion graphs is
presumed excluded by appropriate cuts requiring that the $\epem$
or $\nu\anti\nu$ reconstructed mass not be near $\mz$.}
\item
measurements performed in $s$-channel production at the FMC ---
the production mode being $\mupmum\to\hsm$.
\end{itemize}
In the first case, we presume that $L=200\fbi$ is available
for the measurements at $\rts=500\gev$. (Such operation at a FMC,
would only be appropriate if the NLC has not been constructed
or is not operating at expected instantaneous luminosity.)
In the second case, we implicitly presume that the NLC is
already in operation, so that a repetition 
of $\rts=500\gev$ data collection would not be useful and
devoting all the FMC luminosity to $s$-channel Higgs
production would be entirely appropriate.  The errors we quote
in this second case will be those for only $L=50\fbi$
at $\rts=\mhsm$ (exactly). This is because the crucial measurement
of $\gamhsm$ by scanning the Higgs peak in the $s$-channel requires
devoting significant luminosity to the wings of the peak (see later
discussion). 

\subsubsection{Measurements at $\protect\rts=500\gev$}

The accuracy with which cross section times branching ratio
can be measured in various channels will prove to be vitally important
in determining the branching ratios themselves and, ultimately,
the total width and partial widths of the Higgs boson, which are 
its most fundamental properties.  In addition, the ratios
\begin{equation}
{\sigma\br(\hsm\to c\anti c)\over \sigma\br(\hsm\to b\anti b)}\,,~~~
{\sigma\br(\hsm\to W\wstar)\over \sigma\br(\hsm\to b\anti b)}
\label{ratios}
\end{equation}
will themselves be a sensitive probe of deviations from SM predictions
to the extent that SM values for these branching ratios can
be reliably computed (see later discussion).
It should be noted that the $c\anti c$ and $W\wstar$ modes are
complementary in that for $\mhsm\lsim 130\gev$ only the $c\anti c$
mode will have good measurement accuracy, while for $\mhsm\gsim 130\gev$
accuracy in the $W\wstar$ mode will be best.

\begin{figure}[htb]
\leavevmode
\begin{center}
\centerline{\psfig{file=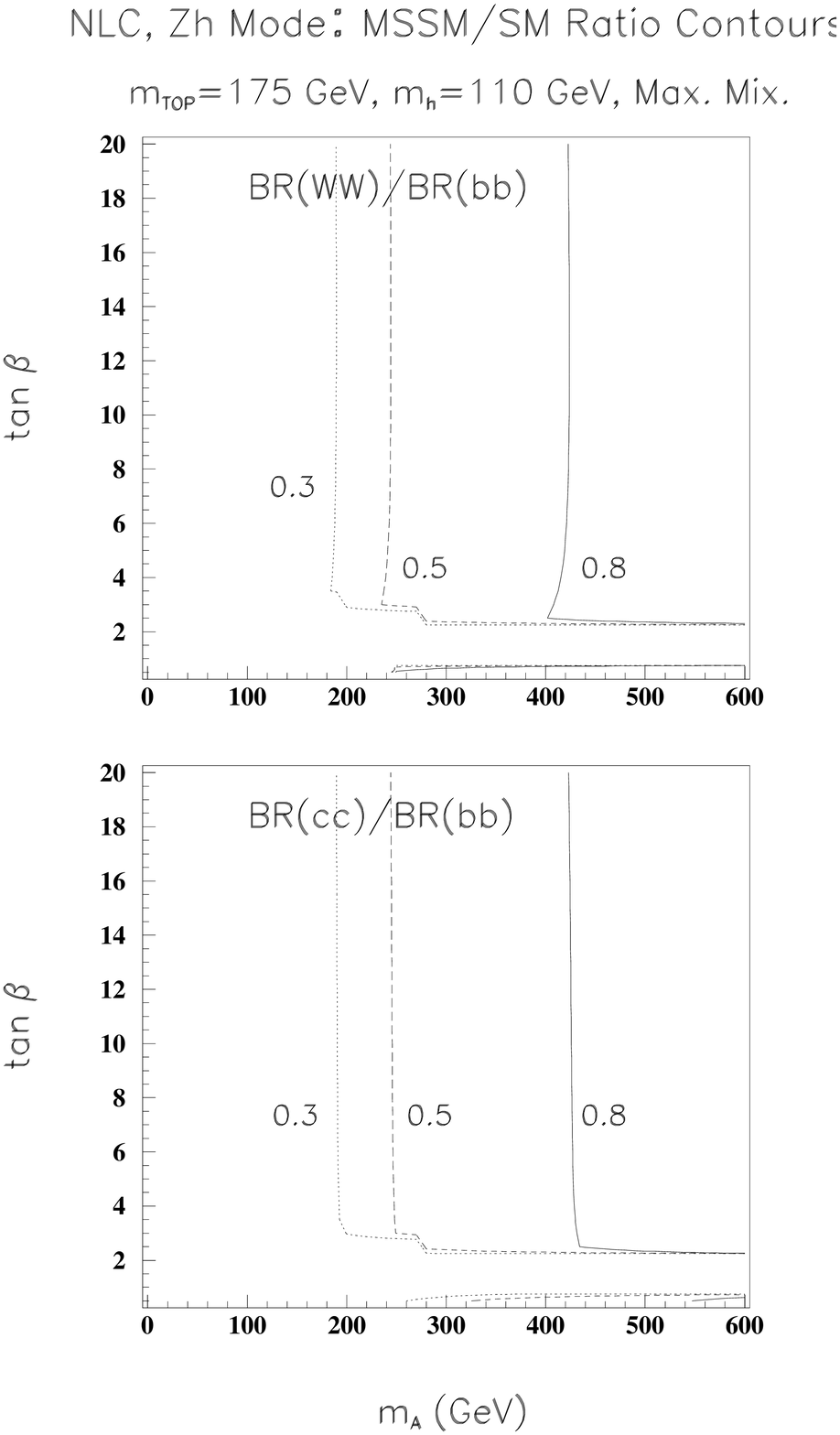,width=3.5in}}
\end{center}
\caption{Constant value contours in $(\mha,\tanb)$ parameter space
for the ratios $[W\wstar/b\anti b]_{\hl}/[W\wstar/b\anti b]_{\hsm}$  and
$[c\anti c/b\anti b]_{\hl}/[c\anti c/b\anti b]_{\hsm}$.
We assume ``maximal-mixing'' in the squark sector and present
results for the case of fixed $\mhl=110\gev$. The band extending
out to large $\mha$ at $\tanb\sim 2$ is where $\mhl=110\gev$
is theoretically disallowed in the case of maximal mixing.
For no mixing, see Ref.~\protect\cite{dpfreport}, the vertical contours
are essentially identical --- only the size of the disallowed band
changes.}
\label{figdevsm}
\end{figure}

The $\hl$ of the MSSM
provides a particularly useful testing ground for the accuracy
with which the above ratios must be determined
in order that such deviations be detectable.  As $\mha$ increases,
the $\hl$ becomes increasingly SM-like.  In typical GUT-unified versions
of the MSSM, $\mha$ values above $200\gev$ are the norm
and deviations of the $\hl$'s couplings and branching ratios from
those of the $\hsm$ will only be detectable if the branching ratios
can be determined with good accuracy.  
The survey of Ref.~\cite{dpfreport} and further work performed
for this workshop \cite{gdev} shows that the $c\anti c$, $b\anti b$
and $WW^\star$ partial widths and ratios of branching ratios
provide sensitivity to $\hl$ vs. $\hsm$ deviations out to higher
values of $\mha$ than any others.  In particular, the $c\anti c/b\anti b$
and $W\wstar/b\anti b$ ratio deviations essentially depend only upon $\mha$
and are quite insensitive to details of squark mixing and so forth.
To illustrate, we present in Fig.~\ref{figdevsm} 
the ratio of the MSSM prediction to the SM prediction
for these two ratios taking $\mhl=110\gev$ (held fixed, implying
variation of stop masses as $\mha$ and $\tanb$ are changed)
and assuming ``maximal mixing'' in the stop sector
(as defined in Ref.~\cite{dpfreport}). Results are presented using contours in
the $(\mha,\tanb)$ parameter space. Aside from an enlargement
of the allowed parameter space region, the ``no mixing'' scenario
contours are essentially the same. Results for larger $\mhl$ are very similar
in the allowed portion of parameter space. We observe that 
it is necessary to detect deviations in the ratios at the level
of 20\% in order to have sensitivity up to $\mha\sim 400\gev$.
Of course, for a Higgs mass as small as $\mhl=110\gev$, only the $c\anti c$
branching ratio has a chance of being measured with reasonable accuracy.
Indeed, the $W\wstar$ branching ratio will inevitably 
be poorly measured for the $\hl$ of the MSSM if
stop squark masses are $\lsim 1\tev$ implying $\mhl\lsim 130\gev$.
In non-minimal supersymmetric models the lightest Higgs can, however,
be heavier and the $W\wstar$ branching ratio would then prove useful.

There are both experimental and theoretical
sources of uncertainty for the branching ratio ratios
of Eq.~(\ref{ratios}). We discuss first the systematic uncertainties that
are present in the theoretical computations.  The primary uncertainty
is that associated with knowing the running $b$ and $c$ quark masses
at the Higgs mass scale. These were recently reviewed \cite{djouadi}
with rather optimistic conclusions. The values obtained
in Ref.~\cite{narison} from QCD sum rule calculations are
$m_c(m_c)=1.23^{+0.02}_{-0.04} \pm 0.06 \gev$ and 
$m_b(m_b)=4.23^{+0.03}_{-0.04} \pm 0.04 \gev$,
where the first error is that from $\alpha_s(\mz)=0.118\pm0.006$
and the second error is twice that claimed in \cite{narison}.
With these inputs, one finds for
$\mhsm\sim100\gev$ the result $m_c(\mhsm)=0.62\pm0.05\pm0.02\gev$,
the first error being that from $\alpha_s$ uncertainties, including
those deriving from the running.
The uncertainty in $\br(\hsm\to c\anti c)\propto m_c^2(\mhsm)$ is 
then $\pm 15\%$. Analogously,
the error for $\br(\hsm\to b\anti b)$ is about $\pm 4\%$.
In the 10 years between now and operation of the NLC, it is reasonable
to suppose that the $\alpha_s$ errors will be reduced to less than half 
the current value. The NLC itself will allow further improvement in 
the $\alpha_s$ determination \cite{burrows}.
Further improvement in the sum rule errors
should also be possible, and fully competitive lattice calculation errors 
should be commonplace by the end of the century.
Further, some of the uncertainties in
the running $\alpha_s$ and other components of the theoretical calculations 
are common to the $b$
and $c$ channels, and will cancel out in the $c\anti c/b\anti b$
ratio of interest. In all, we find it not unreasonable to suppose
that an accuracy of $\lsim\pm 10\%$ can be achieved for the theoretical
computations of the ratios of Eq.~(\ref{ratios}).

Early studies of the experimental accuracy with which the separate event
rates for $Z\hsm$ production with $\hsm$ decaying to $b\anti b$,
$c\anti c$ and $W\wstar$ could be measured are
summarized in Ref.~\cite{dpfreport}; accuracies for 
the latter two were not encouraging.
This was re-examined during the workshop \cite{rvdj}.

\begin{figure}[htb]
\leavevmode
\begin{center}
\centerline{\psfig{file=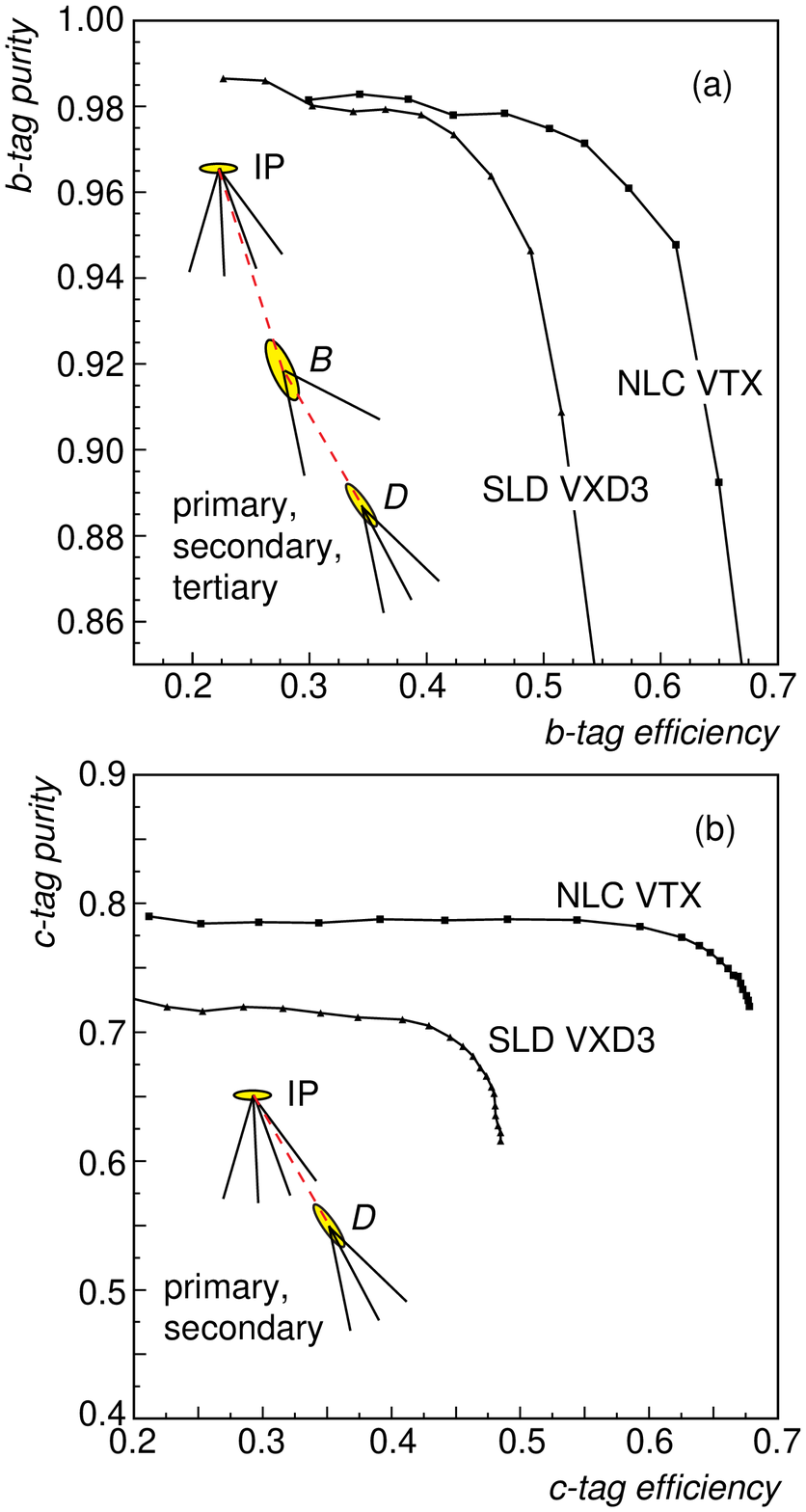,width=3.5in}}
\end{center}
\caption{Purity vs. efficiency for $b$ and $c$ single jet tagging
using the topological tagging techniques of Ref.~\protect\cite{rvdj}.}
\label{figtagging}
\end{figure}

We consider first $\hsm\to b\anti b$ and $\hsm \to c\anti c$. 
It is found that the separate $b\anti b$
and, especially, $c\anti c$ channel event rates
can be measured in $Z\hsm$ production 
with greater accuracy than previously estimated, provided one uses topological
tagging techniques (as opposed to simple impact parameter tagging).
Most importantly, the topological tagging allows a clean separation
of the $c\anti c$ Higgs decay mode from the $gg$ mode.\footnote{The $gg$
mode was simulated using the HAZA Monte Carlo generator
\cite{janotnew} followed by default JETSET fragmentation \cite{PYTHIA}.}
The purity of $b$ and $c$ topological single
jet tagging as a function of the efficiency
required is illustrated in Fig.~\ref{figtagging}, where the present 
performance of the SLD VXD3 upgrade pixel vertex detector is shown along with
that predicted for a proposed pixel detector (NLC VTX) \cite{damerell} in
a typical NLC detector. This method allows for the 
reconstruction of a primary,
secondary, and tertiary vertex to identify the presence of a $b$ quark, 
only a primary and a secondary for a $c$ quark, and tracks only coming
from a primary in the case of a jet originating from a gluon.

The resulting ability to separate $b\anti b$, $c\anti c$
and $gg,q\anti q$ decays of the $\hsm$ in $Z\hsm$ events at $\rts=500\gev$
was studied using simulations performed assuming the
performance of the NLC detector \cite{nlc} at the smeared four-vector
level, signals with $\mhsm=120\gev$ and $130\gev$, 
and considering the known Standard
Model backgrounds. For determining $\sigma(Z\hsm)\br(\hsm\to C)$ 
($C=b\anti b$ or $c\anti c$), both
$Z\to\epem,\mupmum$ and $Z\to jj$ decays
(with full kinematically constrained fitting for both) are retained.
The topological tagging works so well that $b\anti b$ events
can be identified with sufficient purity\footnote{Here, and in the numbers 
quoted below, we refer to event, or decay-channel, purity (as opposed
to single jet tagging purity as plotted in Fig.~\ref{figtagging}).
Event/channel purity is defined as the number of events selected by the tagging
procedure for a particular channel $C$ that are truly from $\hsm\to C$ 
decays divided by the total number so selected, including all $\hsm$ decays 
with relative branching ratios as predicted in the SM.}
by tagging just one (or both) of the $b$-jets. To isolate $c\anti c$ events
with adequate purity, we require that both the $c$ and the $\anti c$
be tagged.
For tagging $\hsm\to b\anti b$ events at $\mhsm=120\gev$ (for example), 
a sample operating point was chosen to give 60\%
efficiency for tagging one or both $b$-quarks with a purity
of 95.4\% (and efficiency for tagging $c$-quark decay events of 2.6\%).
For tagging $c$ quark decay events,
the operating point chosen resulted in an efficiency of 40\% for 
tagging both the $c$ and $\anti c$ quarks and 
a $c\anti c$ channel purity of 77.5\% 
(and efficiencies of 11\% for events where the $\hsm$
decays to $b$-quarks and 0.2\%
for events where the $\hsm$ decays to light quarks or gluons\footnote{Gluon
splitting to heavy quarks is included in the Monte Carlo; at LEP
energies the probabilities for $g\to c\anti c$ and $g\to b\anti b$
are of order 2.5\% and 0.5\%, respectively.}).

These results represent a very substantial improvement over earlier expectations
using impact parameter only. For Higgs masses below about $130\gev$, 
it will be possible \cite{rvdj} to measure 
$\sigma(Z\hsm)\br(\hsm\to b\anti b)$ with an error of $\sim\pm 2.5\%-\pm 3.5\%$
and $\sigma(Z\hsm)\br(\hsm\to c\anti c)$
to about $\pm 10\%$, for $L=200\fbi$.
This implies $\sim \pm 11\%$ error for 
$\br(\hsm\to c\anti c)/\br(\hsm\to b\anti b)$. 

Although not specifically studied for this report, a crude estimate \cite{ghs}
suggests that the analogous
procedure in the $\epem\hsm$ final state mode would yield 
a similar level of error for this ratio. (See 
the later $\br(\hsm\to\ b\anti b)$ discussion for comparative $Z\hsm$
and $\epem\hsm$ errors in the $b\anti b$ decay mode.)
The ratio could again be measured in the $WW$-fusion
$\nu\anti\nu\hsm$ final state. There,
the error on $\sigma(\nu\anti\nu\hsm)\br(\hsm\to b\anti b)$
is expected to be in the $\pm 2.5\%-\pm 3.5\%$ range for 
$\mhsm\lsim 140\gev$ as estimated in \cite{dpfreport} and reconfirmed
at this workshop. The $c\anti c$ final state has not been studied
yet, but it would seem that accuracies in the $\pm 10\%$
vicinity for $\sigma(\nu\anti\nu\hsm)\br(\hsm\to c\anti c)$ are
not out of the question.
Combining \cite{rickjack} just the $Z\hsm$ and $\epem\hsm$
modes, we could probably achieve $\sim\pm 7\%-\pm8\%$
error for $c\anti c/b\anti b$. Including the $\nu\anti\nu\hsm$
final state might allow us to reach $\lsim\pm 7\%$.
If we combine this error in quadrature with the earlier estimate
of $\lsim\pm 10\%$ for systematic error in the theoretical
calculation of the $c\anti c/b\anti b$ ratio, we arrive at a net 
error of $\lsim 12\%$. Fig.~\ref{figdevsm} shows that this would allow 
differentiation of the $\hl$ from the $\hsm$
at the $2\sigma$ level out to $\mha\sim 450\gev$. This is a very encouraging
result.  The dominance of the theoretical error in the above estimates
indicates the high priority of obtaining theoretical predictions for 
$c\anti c/b\anti b$ that are as precise as possible. 
Overall, precision $\hl$ measurements at $\rts=500\gev$ with $L=200\fbi$
appear to have a good chance of probing the heavier Higgs
mass scale (which is related to important SUSY-breaking parameters)
even when the heavier Higgs bosons can not be (pair) produced
without going to higher energy.

\begin{figure}[htb]
\leavevmode
\begin{center}
\centerline{\psfig{file=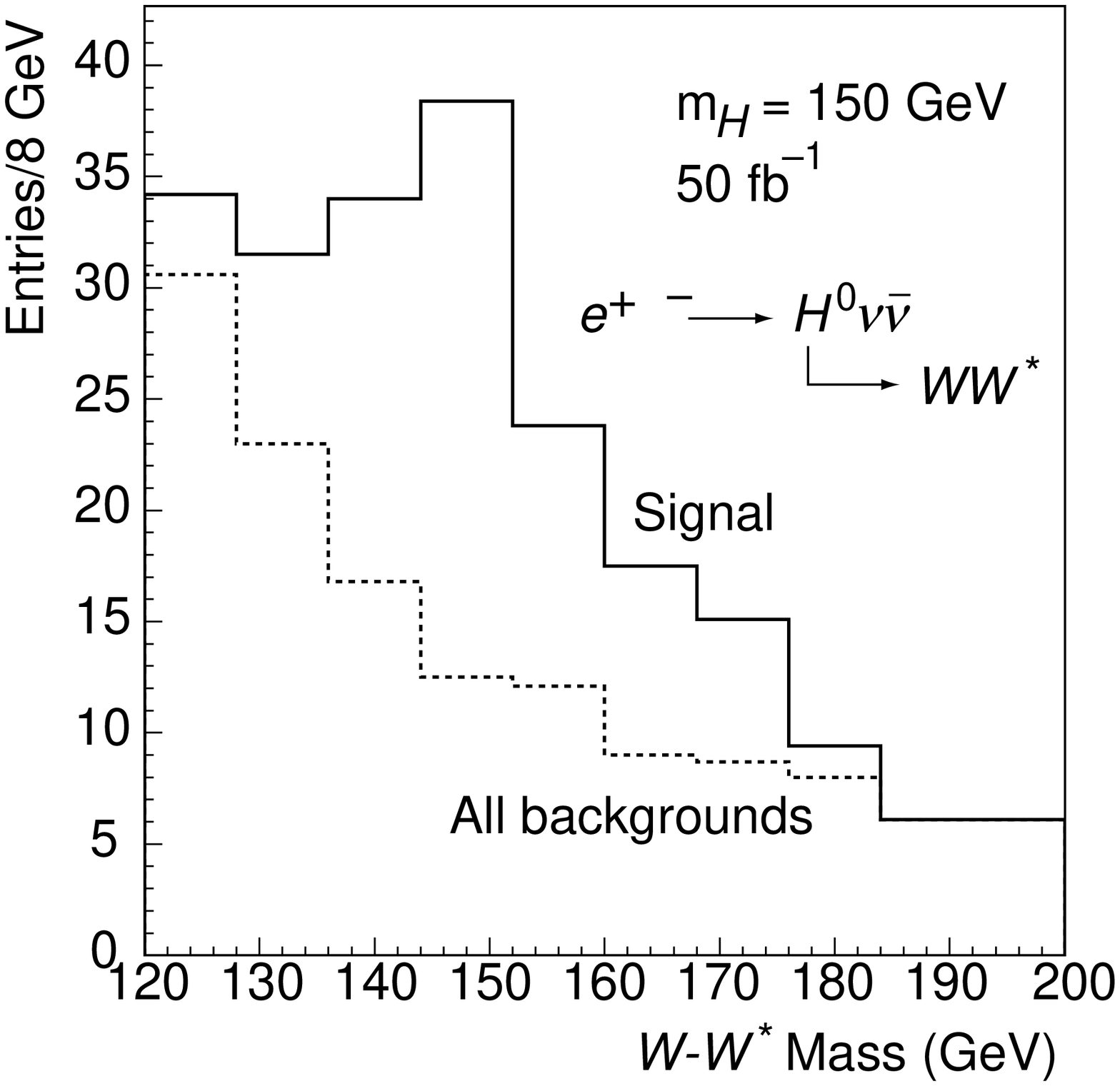,width=3.5in}}
\end{center}
\caption{Signal and background rates for $L=50\fbi$ at $\protect\rts=500\gev$
for $\epem\to \nu\anti \nu W\wstar$ as a function of $W\wstar$ mass,
taking $\mhsm=150\gev$.}
\label{figwwstar}
\end{figure}

We now consider the $W\wstar$ mode, which would be relevant for a SM-like
Higgs with mass above $130\gev$.
Both $\hsm$ production via $WW$ fusion, $\epem\to \nu\anti\nu\hsm$,
and $Z\hsm $ production followed by $\hsm$ decay into $W\wstar$ 
for heavier Higgs masses were simulated \cite{rickww}.
For $\mhsm = 150\gev$, the cross-sections for these two production modes
are roughly equal and it is advantageous to use both for more statistics.
We shall see \cite{rickjack} that the measurement of
$\sigma(WW \to \hsm)\br(\hsm \to W\wstar)$ allows a direct probe
of the $(WW \hsm)^2$ coupling and a determination
of the total $\hsm$ width. For the 2nd ratio of Eq.~(\ref{ratios}),
it will also be important to compare
rates for the $\nu\anti\nu b\anti b$ and $\nu\anti\nu W\wstar$ final states
and rates for the $Z b\anti b$ and $Z W\wstar$ final states.

For the case of $Z\hsm$ production followed by $\hsm$ decay into $W\wstar$,
two topologies were examined: the first is the final state
containing six jets --- two from hadronic decay of
the $Z$ and two jets from each of the $W$ bosons; the second final state 
considered is that with two leptons 
from the $Z$ and four jets from the $W\wstar$.
Simulations were
performed at $\rts=500\gev$ assuming the performance of the
NLC detector \cite{nlc} at the smeared four-vector level, a signal with
$\mhsm = 130-170\gev$, and considering the known Standard Model backgrounds.
After cuts demanding large visible energy, and that the event be
well contained, a kinematic constrained fit was performed taking into
account $E_{cm}$, $\mz$, and one on-shell $\mw$ after assigning 
the $Z$ mass to the quark
or lepton pair  with invariant mass closest to and within
8~GeV of $m_Z$. Requiring the fit probability to be greater than 10\%
greatly reduced the background from $WW$, $t\bar{t}$ and light quarks.
The purity is then enhanced further by employing the previously described
topological quark tags as anti-tags on the jets assigned to the $W$ bosons,
i.e. requiring the jets fail the $b$ and $c$ topological tags.
As an example, in $L=50\fbi$ of data with $\mhsm = 150\gev$,
65 signal events survive on a
background of 21 events, of which only 4.2 are from Higgs decays into
heavy quarks and gluons. Extrapolating to $L=200\fbi$,
$S=260$ with $B=84$ implies $\sqrt{S+B}/S=0.07$ for the indicated mass.
The situation deteriorates considerably for $\mhsm = 130\gev$ with
$S/B \approx 1.0$.
For $L=200\fbi$ the statistical accuracy with which $\sigma(Z\hsm)\br(\hsm\to
W\wstar)$ can be measured is about $\pm 22\%$ and $\pm 10\%$
for $\mhsm=130$ and $140\gev$, respectively.
For $\mhsm$ above $150\gev$, the accuracy of the measurement
improves over the $\mhsm=150\gev$ result, falling to a low of about $\pm 6\%$
at $\mhsm=170\gev$.

Of course, as the $W\wstar$ mode gets stronger, the $b\anti b$ mode weakens,
see Table~\ref{hsmbrs}.
Thus, the $Z\hsm\to Zb\anti b$ rate is measured with progressively poorer
accuracy as $\mhsm$ increases. At $\mhsm\sim 150\gev$, for example,
we \cite{rickjack} estimate that the earlier $\mhsm\sim 110\gev$ errors for
$\sigma(Z\hsm)\br(\hsm\to b\anti b)$ will have increased
by about a factor of two to $\sim\pm 6\%$,
rising rapidly to $\sim\pm 28\%$ at $\mhsm=170\gev$.
Combining \cite{rickjack} the $W\wstar$ and $b\anti b$ mode errors
in the $Z\hsm$ production mode, we find errors for $\br(\hsm\to
W\wstar)/\br(\hsm\to b\anti b)$ of 
roughly $\pm 22\%$, $\pm 11\%$, $\pm 9\%$ and $\sim\pm 28\%$
at $\mhsm=130$, $140$, $150$ and $170\gev$, respectively.

For the case of $\hsm$ production via $WW$ fusion followed by the Higgs decaying
into $W\wstar$, the final state is $\nu \anti{\nu} W \wstar$.  Cuts are
made demanding visible energy less than $0.5 E_{cm}$, large missing mass, no
isolated leptons, large missing transverse momentum, a large
acoplanarity angle between the reconstructed $W$ axes, and that the
missing momentum vector does not point in the forward direction.
These cuts reduce the dangerous $eeWW$, $\nu \bar{\nu} WW$, and
$e \nu W Z$ backgrounds.  The event is forced to be reconstructed into
four jets, with two required to have invariant mass close to the $W$ mass
and the remaining two jets to have invariant mass well below the $W$ mass
(from the $\wstar$).  The heavy quark topological tag is then again used
as an anti-tag to increase the purity in the $W$ sample.  The visible
mass of the entire event is then examined for peaking at the Higgs mass.
A huge peak results at lower masses due to $e \nu W$ and $Z \nu \bar{\nu}$.

A typical result is that of Fig.~\ref{figwwstar} for $\mhsm=150\gev$,
where $L=50\fbi$ is assumed. In general, 
the statistical accuracy with which $\sigma(\nu\anti\nu\hsm)\br(\hsm\to
W\wstar)$ can be measured is estimated \cite{rickjack} 
to be very similar to that
found for $\sigma(Z\hsm)\br(\hsm\to W\wstar)$ above: for $L=200\fbi$
the rough errors for the former are $\pm 22\%$, $\pm 10\%$, $\pm 8\%$ and 
$\pm 7\%$ for $\mhsm=130$, $140$, $150$ and $170\gev$, respectively.
At these same masses,
the corresponding accuracy for $\sigma(\nu\anti\nu\hsm)\br(\hsm\to b\anti b)$
is $\sim\pm 3\%$, $\sim\pm 4\%$, $\sim \pm 7\%$ and
$\gsim\pm 33\%$, respectively.
At $\mhsm=150\gev$, the $W\wstar$ branching ratio is still more
difficult to measure than $b\anti b$ because of the larger background and lower 
efficiency for isolating the final state. However, by $\mhsm=170\gev$
the $b\anti b$ branching ratio has become so small that errors in this channel
rapidly increase. The above errors imply $\nu\anti\nu \hsm$ channel
errors for $\br(\hsm\to W\wstar)/\br(\hsm\to b\anti b)$ 
of $\sim \pm22\%$, $\sim \pm11\%$, $\sim \pm10\%$ or 
$\gsim \pm 33\%$ at $\mhsm=130$, $140$, $150$ or $170\gev$, respectively.

Combining \cite{rickjack} the $Z\hsm$ and $\nu\anti\nu\hsm$ channel 
results, we obtain accuracies for $\br(W\wstar)/\br(b\anti b)$
of roughly $\pm 16\%$, $\pm 8\%$ and $\pm 7\%$ for $W\wstar/b\anti b$
for $\mhsm=130$, $140$ and $150\gev$.  
At $\mhsm=120\gev$ and $170\gev$, we \cite{rickjack} estimate the errors to be 
$\sim\pm 23\%$ and $\sim\pm 21\%$, respectively. (We have not
pursued the degree to which these errors would be further reduced by including
the $\epem\hsm$ channel determination of this ratio.)
Fig.~\ref{figdevsm} (which is fairly
independent of the actual $\mhl$ value aside from the extent
of the allowed parameter region) implies that a $\lsim 10\%$ error,
as achieved for $\mhsm$ in the $140-150\gev$ mass range, would be a very
useful level of accuracy in the MSSM should stop quark masses (contrary
to expectations based on naturalness) be sufficiently above 1 TeV
to make $\mhl=140-150\gev$ possible.
In the NMSSM, where the lightest higgs (denoted $\h_1$)
can have mass $\mhi\sim 140-150\gev$ and the second lightest ($\h_2$)
often has mass in the $\mhii\sim150-190\gev$ range, even if stop masses
are substantially below 1 TeV, deviations from SM expectations
are typically even larger.  This exemplifies
the fact that the $W\wstar/b\anti b$ ratio will
provide an extremely important probe of a non-minimal Higgs sector
when both the $W\wstar$ and $b\anti b$ decays have significant
branching ratio.

The NLC errors for the $(c\anti c\hsm)^2/(b\anti b\hsm)^2$ and
$(W\wstar\hsm)^2/(b\anti b\hsm)^2$ coupling-squared ratios 
outlined above for $L=200\fbi$ at $\rts=500\gev$
are repeated in the NLC summary table, Table~\ref{nlcerrors}.

\subsubsection{Measuring $\sigma(\mupmum\to\hsm)\br(\hsm\to b\anti
b,W\wstar,Z\zstar)$ in $s$-channel FMC production}

\begin{table}[h]
\caption[fake]{Summary of approximate errors for
$\sigma(\mupmum\to\hsm)\br(\hsm\to b\anti b, W\wstar, Z\zstar)$,
assuming $L=50\fbi$ devoted to $\protect\rts=\mhsm$ and beam energy resolution
of $R=0.01\%$.}
\footnotesize
\begin{center}
\small
\begin{tabular}{|c|c|c|c|c|c|}
\hline
 Channel & \multicolumn{5}{c|}{Errors} \\
\hline
\hline
{$\bf\mhsm$}{\bf (GeV)} & {\bf 80} & {\bf 90} & {\bf 100} & {\bf 110} & {\bf 120} \\
\hline
$b\anti b $ & 
$\pm 0.2\%$ & $\pm 1.6\%$ & $\pm 0.4\%$ & $\pm 0.3\%$ & $\pm 0.3\%$ \\
\hline
$W\wstar $ &
$-$ & $-$ & $\pm 3.5\%$ & $\pm 1.5\%$ & $\pm 0.9\%$ \\
\hline
$Z\zstar $ &
$-$ & $-$ & $-$ & $\pm 34\%$ & $\pm 6.2\%$ \\
\hline
\hline
{$\bf\mhsm$}{\bf (GeV)} & {\bf 130} & {\bf 140} & {\bf 150} & {\bf 160} & {\bf 170} \\
\hline
$b\anti b $ & 
$\pm 0.3\%$ & $\pm 0.5\%$ & $\pm 1.1\%$ & $\pm 59\%$ & $-$ \\
\hline
$W\wstar $ &
$\pm 0.7\%$ & $\pm 0.5\%$ & $\pm 0.5\%$ & $\pm 1.1\%$ & $\pm 9.4\%$ \\
\hline
$Z\zstar $ &
$\pm 2.8\%$ & $\pm 2.0\%$ & $\pm 2.1\%$ & $\pm 22\%$ & $\pm 34\%$ \\
\hline
\hline
{$\bf\mhsm$}{\bf (GeV)} & {\bf 180} & {\bf 190} & {\bf 200} & {\bf 210} & {\bf 220} \\
\hline
$W\wstar $ &
 $\pm 18\%$ & $\pm 38\%$ & $\pm 58\%$ & $\pm 79\%$ & $-$ \\
\hline
$Z\zstar $ &
$\pm 25\%$ & $\pm 27\%$ & $\pm 35\%$ & $\pm 45\%$ & $\pm 56\%$ \\
\hline
\end{tabular}
\end{center}
\label{fmcsigbrerrors}
\end{table}

The accuracies expected for these measurements were determined in
Ref.~\cite{bbgh} under the assumption
that the relevant detector challenges associated with detecting
and tagging final states in the potentially harsh FMC environment can be met.
As explained in the introduction to this
section, if $L=200\fbi$ is used so as to optimize the Higgs peak scan
determination of $\gamhsm$, then the equivalent $\rts=\mhsm$ Higgs 
peak luminosity accumulated 
for measuring $\sigma(\mupmum\to\hsm)\br(\hsm\to X)$
in various channels is of order $L=50\fbi$. The associated errors 
expected for $\sigma(\mupmum\to\hsm)\br(\hsm\to
b\anti b, W\wstar,Z\zstar)$ are summarized as a function of $\mhsm$
in Table~\ref{fmcsigbrerrors}. As is apparent from the
table, the errors are remarkably small for $\mhsm\lsim 150\gev$.
As already stated, detector performance in the FMC environment
will be critical to whether or not such small errors can be achieved
in practice. As an example, to achieve the good $b$-tagging
efficiencies and purities employed in obtaining the NLC detector errors
given in this report, a relatively clean environment is required
and it must be possible to get as close as 1.5 cm to the beam.
FMC detectors discussed to date do not allow for instrumentation this
close to the beam. More generally, in all the channels
it is quite possible that the FMC
errors will in practice be at least in the few per cent range.
This, however, would still constitute an extremely valuable
level of precision.

For later purposes, it is important to understand the relation
between $\sigma(\mupmum\to\hsm)$ and the $\Gamma(\hsm\to\mupmum)$
partial width (which is directly proportional to the $(\mupmum\hsm)^2$
coupling-squared). 
Very generally, the average cross section for production
of any Higgs boson in the $s$-channel, $\sighbar$, is obtained
by convoluting the standard Breit-Wigner shape for the Higgs
resonance with a Gaussian distribution in $\rts$ of width 
$\sigrts$. For a distribution centered at $\rts=\mh$, $\sighbar$ is given by 
$\sighbar\sim 4\pi\mh^{-2}\br(\h\to\mupmum)$ if $\sigrts\ll\gamh$
and by
$\sighbar\sim2\pi^2\mh^{-2}\Gamma(\h\to\mupmum)/(\sqrt{2\pi}\sigrts)$
if $\sigrts\gg\gamh$. To get near maximal $\sighbar$ and
to have sensitivity to $\gamh$ via scanning in $\rts$ (see later
subsection) it
is important that $\sigrts$ be no larger than $2-3\times\gamh$.
Fig.~\ref{hwidths} shows that $\gamh< 1-10\mev$ is typical of the $\hsm$
for $\mhsm\lsim 140\gev$.
Using the parameterization 
$\sigrts\simeq 7\mev\left({R\over 0.01\%}\right)\left({\rts\over 
100\gev}\right)$ for $\sigrts$ in terms of the beam energy resolution, $R$,
we see that very excellent resolution $R\sim0.01\%$ typically
yields $\sigrts\sim  2-3\times\gamhsm$ when $\mhsm\lsim 140\gev$.
In this mass region,
$\sighsmbar$ is then roughly proportional to $\Gamma(\hsm\to\mupmum)/\sigrts$
with small corrections sensitive to $\gamh$.  Thus, a measurement
of $\sighsmbar\br(hsm\to X)$ in the $\mhsm\lsim 140\gev$
mass region can be readily converted to a determination of
$\Gamma(\hsm\to\mupmum)\br(\hsm\to X)$ provided that $\gamhsm$
is measured with good accuracy (given that $\sigrts$ will be accurately known).
As reviewed in a later subsection, one finds that (with $R=0.01\%$)
very good precision for $\gamhsm$
is possible in $\mupmum\to\hsm$ collisions by employing a 
simple scan of the Higgs resonance peak.

\subsection{Measuring $\sigma\br(\hsm\to\gam\gam)$ at the NLC \protect\cite{gm}}

We will later review why a determination of $\br(\hsm\to\gam\gam)$
is the only means for extracting $\gamhsm$ in the $\mhsm\lsim 130\gev$
mass range.  Of course, $\br(\hsm\to\gam\gam)$ and especially
$\Gamma(\hsm\to\gam\gam)$ are of special interest themselves in that
the $\gam\gam\hsm$ coupling is sensitive to one-loop graphs 
involving arbitrarily heavy states (that get their mass from the $\hsm$
sector vev --- to be contrasted with, for example, heavy SUSY partner states 
which decouple since they get mass from explicit SUSY breaking).

\begin{figure}[htb]
\leavevmode
\begin{center}
\centerline{\psfig{file=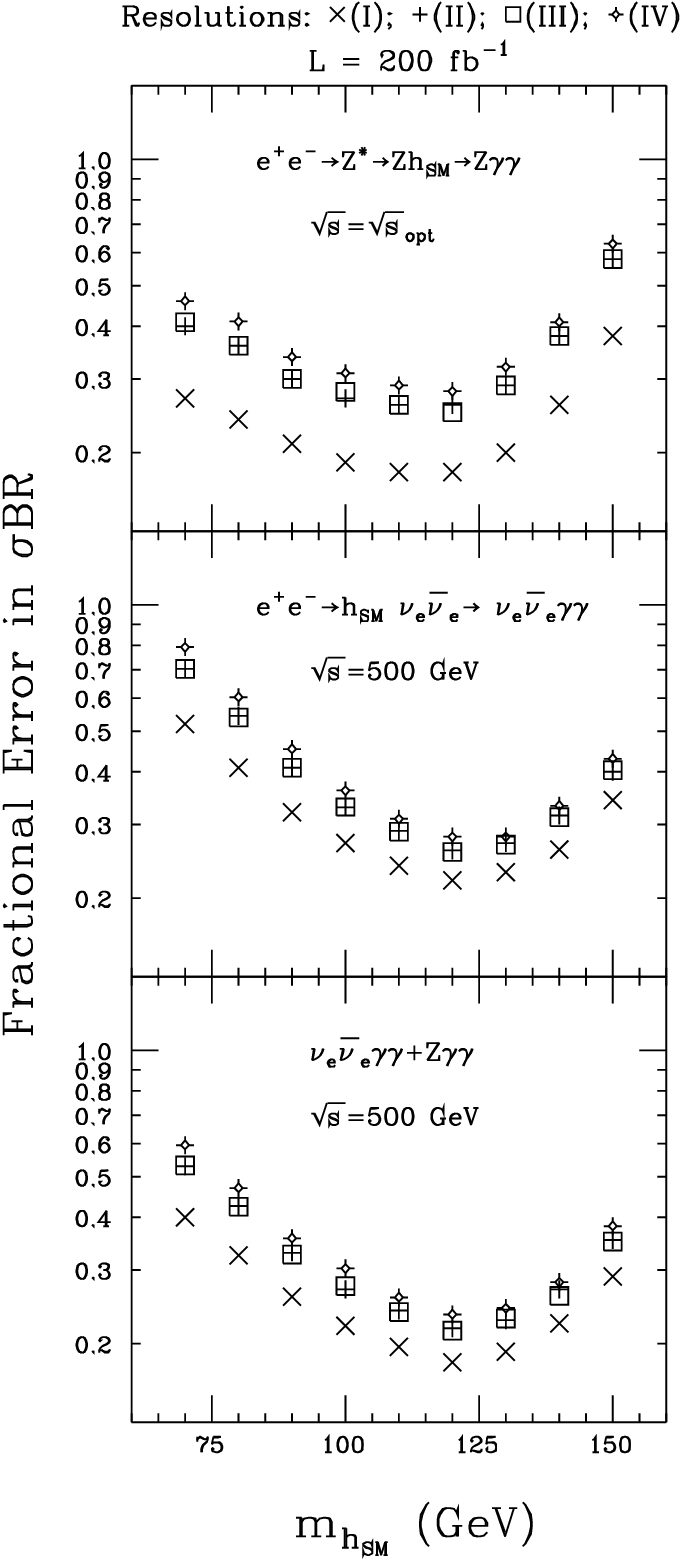,width=2.7in}}
\end{center}
\caption{The fractional error in the measurement of
$\sigma(\nu_e \bar\nu_e \hsm)\br(\hsm\to\gam\gam)$
[$\sigma(Z\hsm)\br(\hsm\to\gam\gam)$]
as a function of $\mhsm$ assuming $L=200\fbi$ is accumulated at 
$\protect\rts=500\gev$ [$\protect\rts=\protect\rts_{\rm opt}$].
Also shown is the fractional $\sigma\br(\hsm\to\gam\gam)$
error obtained by combining $Z\hsm$ and $\nu_e\anti\nu_e\hsm$
channels for $L=200\fbi$ at $\protect\rts=500\gev$.
Results for the four electromagnetic calorimeter 
resolutions described in the text are given.}
\label{figgamgamerrors}
\end{figure}

At the NLC, the only means of getting at $\br(\hsm\to\gam\gam)$
is to first measure $\sigma\br(\hsm\to\gam\gam)$ in all accessible
production modes, and then divide out by the $\sigma$'s as
computed using other data. One finds that
the errors in the $\sigma$'s are small so that
the net error for $\br(\hsm\to\gam\gam)$ is essentially that obtained
by combining the statistical errors for the available $\sigma\br$ measurements.
The $\sigma\br(\hsm\to\gam\gam)$ errors have been studied for the $Z\hsm$ and
$\nu\anti\nu\hsm$ ($WW$-fusion) production modes in Ref.~\cite{gm}.
The error for $\sigma(Z\hsm)\br(\hsm\to\gam\gam)$ is minimized
for a given total luminosity $L$ by running at a $\rts$ value that
is near the maximum of the $Z\hsm$ cross section, 
roughly $\rts\sim \mhsm+\mz+$ a few GeV; precise optimal $\rts_{\rm opt}$
values are given in Ref.~\cite{gm}.  The error for
$\sigma(\nu\anti\nu\hsm)\br(\hsm\to\gam\gam)$ at given $L$ 
is minimized by operating at the highest available $\rts$; this maximizes
the $WW$-fusion cross section. 
The errors in $\sigma(Z\hsm)\br(\hsm\to\gam\gam)$ for $L=200\fbi$ accumulated
at $\rts=\rts_{\rm opt}$ and in $\sigma(\nu\anti\nu\hsm)\br(\hsm\to\gam\gam)$
for $L=200\fbi$ accumulated at $\rts=500\gev$ are plotted as a function
of $\mhsm$ in the first two windows of Fig.~\ref{figgamgamerrors}.
The effective $\sigma\br(\hsm\to\gam\gam)$ error
obtained by combining the statistics for the $Z\hsm$ and $WW$-fusion modes
assuming $L=200\fbi$ is accumulated at $\rts=500\gev$ is plotted in
the third window of Fig.~\ref{figgamgamerrors}. 
The effective $\sigma\br(\hsm\to\gam\gam)$ error for $L=200\fbi$ at
$\rts=\rts_{\rm opt}$ is essentially the same as for the $Z\hsm$
mode alone, the $WW$-fusion contribution to the statistics being
unimportant at these low $\rts$ values.

Results are presented for four different electromagnetic 
calorimeter resolutions: 
I corresponds to the very excellent resolution of the CMS calorimeter
\cite{CMS}.
II and III are somewhat optimistic limits of the resolutions currently
planned for the NLC detectors \cite{nlc}; and IV is the resolution planned for
the JLC detector \cite{jlci}. 
(For details and references, see Ref.~\cite{gm}.)

It is important to compare the $\sigma\br(\hsm\to\gam\gam)$ 
error found using the $Z\hsm$ mode statistics
for $L=200\fbi$ at $\rts=\rts_{\rm opt}$ to the error
found by combining $WW$-fusion and $Z\hsm$ statistics
for $L=200\fbi$ at $\rts=500\gev$ (window 1 vs. window 3 of
Fig.~\ref{figgamgamerrors}). 
We find that in resolution cases II-IV (I) the $Z\hsm$, $\rts=\rts_{\rm opt}$
measurement yields smaller errors for $70\lsim\mhsm\lsim 100\gev$
($70\lsim\mhsm\lsim 120\gev$). The $Z\hsm$ mode at $\rts=\rts_{\rm opt}$
is most superior to the combined $WW$-fusion plus $Z\hsm$, $\rts=500\gev$
error if $\mhsm=70\gev$: for excellent calorimeter 
resolution case I (`standard' resolution cases II/III),
$Z\hsm$ at $\rts=\rts_{\rm opt}$ yields an error of $\pm 27\%$ ($\pm 40\%$)
vs. combined $WW$-fusion plus $Z\hsm$, $\rts=500\gev$ error of $\pm 40\%$
($\pm 53\%$). However, the above $Z\hsm|_{\rts=\rts_{\rm opt}}$ 
advantage would be lost if the instantaneous luminosity ($\call$)
at $\rts_{\rm opt}\sim 165\gev$ is more than a factor of 2.2 (1.8)
below that at $\rts=500\gev$ in resolution case I (cases II/III).
If the interaction region is designed for maximal $\call$ at $\rts=500\gev$,
$\call$ at $\rts=\rts_{\rm opt}$ would decrease by an even larger
factor since $\call \propto (\rts/500\gev)^2$ \cite{ji} as the energy
is lowered;
for any $\mhsm$ the best results would be obtained by running at $\rts=500\gev$.
Although it would not be
all that expensive to build new quads \etc\ suited to a lower $\rts$ \cite{ji},
any significant associated loss in running time 
would quickly offset the potential benefits. Further, lower energy
operation might decrease sensitivity to other types of new physics.
If $\mhsm$ is known ahead of time (from LEP2 or LHC) to
be below $100\gev$ ($120\gev$) or so, for which focusing on $Z\hsm$ production
at $\rts=\rts_{\rm opt}$ would be appropriate 
in resolution cases II-IV (I), 
then an interaction region with maximal $\call$ at $\rts=\rts_{\rm opt}$
could be included in the design from the beginning.

Clearly, the most likely situation is that
$L=200\fbi$ is accumulated at $\rts=500\gev$ and that
the calorimeter is at the
optimistic end of current plans for the NLC detector (cases II and III).
After combining the statistics for the $WW$-fusion
and $Z\hsm$ modes, the errors 
in $\sigma\br(\hsm\to\gam\gam)$ range from
$\sim\pm 22\%$ at $\mhsm=120\gev$
to $\sim\pm 35\%$ ($\sim\pm 53\%$) at $\mhsm=150\gev$ ($70\gev$).
In the $100\lsim \mhsm\lsim 140\gev$ mass 
region, the errors are smallest and lie in the $\pm22\%-\pm 27\%$ range.

We note that it is also possible to 
consider measuring $\sigma\br(\hsm\to\gam\gam)$
in the $\epem\to \epem\hsm$ ($ZZ$-fusion) production mode.
A study of this case \cite{pmartin} shows, however,
that the errors will be much worse than found for 
either $Z\hsm$ production or $WW$-fusion production. For instance,
compared to the $Z\hsm$ channel, where all $Z$ decay modes can
be included,~\footnote{This is possible 
since we can constrain the recoil mass,
constructed from $\rts$ and the momenta of the two photons
from $\hsm\to\gam\gam$ decay, to be close to $\mz$.}
the $\epem\hsm$ rate with $M_{\epem}\not\sim \mz$ is substantially smaller.

Let us now turn to the errors that can be expected for the coupling-squared
ratio $(\gam\gam\hsm)^2/(b\anti b\hsm)^2$.
We have already tabulated 
in Tables~\ref{m1errors} and \ref{m2errors} the
errors expected from LHC data for $\mhsm\leq 130\gev$;
the LHC error varies from $\pm 17\%$ to $\pm 25\%$ as $\mhsm$ goes
from $90\gev$ to $130\gev$. Above $\mhsm=130\gev$, the LHC
error for the ratio is expected to be quite large.
At the NLC, $(\gam\gam\hsm)^2/(b\anti b\hsm)^2$
can be computed in the $Z\hsm$ and $WW$-fusion production modes 
(treated separately) 
as $\sigma\br(\hsm\to \gam\gam)/\sigma\br(\hsm\to b\anti b)$;
the numerator and denominator in
this latter ratio can be obtained (assuming reasonable knowledge
of efficiencies) from measured event rates.
We will presume that all NLC measurements are performed 
by accumulating $L=200\fbi$ at $\rts=500\gev$.
The $L=200\fbi$, $\rts=500\gev$ errors for the
denominator in the $Z\hsm$ and $WW$-fusion production modes
have been given in the previous section. The $WW$-fusion numerator
errors are those given in the 2nd window of Fig.~\ref{figgamgamerrors}.
The $Z\hsm$ numerator errors have not been separately plotted,
but are those implicit in the 3rd window of Fig.~\ref{figgamgamerrors}.
The $Z\hsm$ and $WW$-fusion determinations of 
$(\gam\gam\hsm)^2/(b\anti b\hsm)^2$ are statistically independent
and can be combined to get a net error. The resulting net NLC-only 
error is not terribly good; 
at $\mhsm=80$, 100, 110, 120, 130, 140, $150\gev$ the errors for
$(\gam\gam\hsm)^2/(b\anti b\hsm)^2$ are $\pm 42\%$, $\pm 27\%$, $\pm 24\%$,
$\pm 22\%$, $\pm 23\%$, $\pm26\%$, $\pm 35\%$, respectively. 
For the lower $\mhsm$ values
the LHC does better.  If we combine the LHC and NLC measurements,
the errors for $(\gam\gam\hsm)^2/(b\anti b\hsm)^2$ at the above $\mhsm$
values are $\pm 16\%$, $\pm 14\%$, $\pm 15\%$, $\pm 16\%$, $\pm17\%$,
$\pm26\%$, $\pm35\%$, 
respectively. The NLC-only
errors are repeated later in the NLC summary table, Table~\ref{nlcerrors}.
                                   
Finally, we note that in later discussions we show that
the large errors for the $\gam\gam$
final state will dominate in computing some important quantities that
potentially allow discrimination between the SM Higgs boson and a SM-like
Higgs boson of an extended model.

\subsection{Determining the $ZZ\hsm$ coupling at the NLC}

Determination of the $(ZZ\hsm)^2$ coupling-squared is possible in two modes.
These are (using $\epem$ collision notation):
\begin{itemize}
\item
$\epem\to Z\hsm$, where $Z\to \ell^+\ell^-$ ($\ell=e,\mu$);
\item
$\epem\to\epem \hsm$ (via $ZZ$-fusion) \cite{ghs}.
\end{itemize}
Results presented here for the $ZZ$-fusion channel are preliminary.
It is convenient to separate $Z\hsm$ and $ZZ$-fusion
for the purposes of discussion
even though in the $\epem\hsm$ final state there is some interference
between the $ZZ$-fusion and $Z\hsm$ diagrams. Experimentally this
separation is easily accomplished by an appropriate cut on the $\epem$
pair mass.\footnote{Whenever $ZZ$-fusion dominates the 
$\zstar$ diagrams, such a cut requiring $M_{\epem}\slash\sim\mz$
usually improves $S/\sqrt B$ and reduces the $\sqrt{S+B}/S$ error.}

In both cases, the $\hsm$ is inclusively isolated by examining the recoil
mass spectrum computed using the incoming $\epem$ momentum
and the momenta of the outgoing leptons. In the $Z\hsm$ case,
only $\ell=e,\mu$ in the $Z$ decay
are considered ($\tau$'s and jets are excluded)
since it is essential that the recoil mass peak be as narrow as possible
in order that only a small mass window need be kept, thereby making
backgrounds very small. Clearly, excellent
momentum resolution for electrons and muons will be essential, especially
for Higgs masses in the vicinity of $\mz$.\footnote{In order to inclusively
sum over all $\hsm$ decays, it is important to avoid making any use
of $\hsm$ decay products in reconstructing the Higgs mass peak. Thus,
a 4-C fit using the energies and angles of the jets from Higgs decay 
and leptons from $Z$ decay should not be employed; 
kinematic fits (involving fewer than 4 constraints) considering the leptons
from $Z$ decay and the knowledge of $E_{\rm cm}$ 
could however still be employed. In any case, such fits
yield a jet-jet mass resolution that is no better than (worse than) 
that for the recoil mass for NLC (super-JLC) momentum resolution.}

\begin{figure}[htb]
\leavevmode
\begin{center}
\centerline{\psfig{file=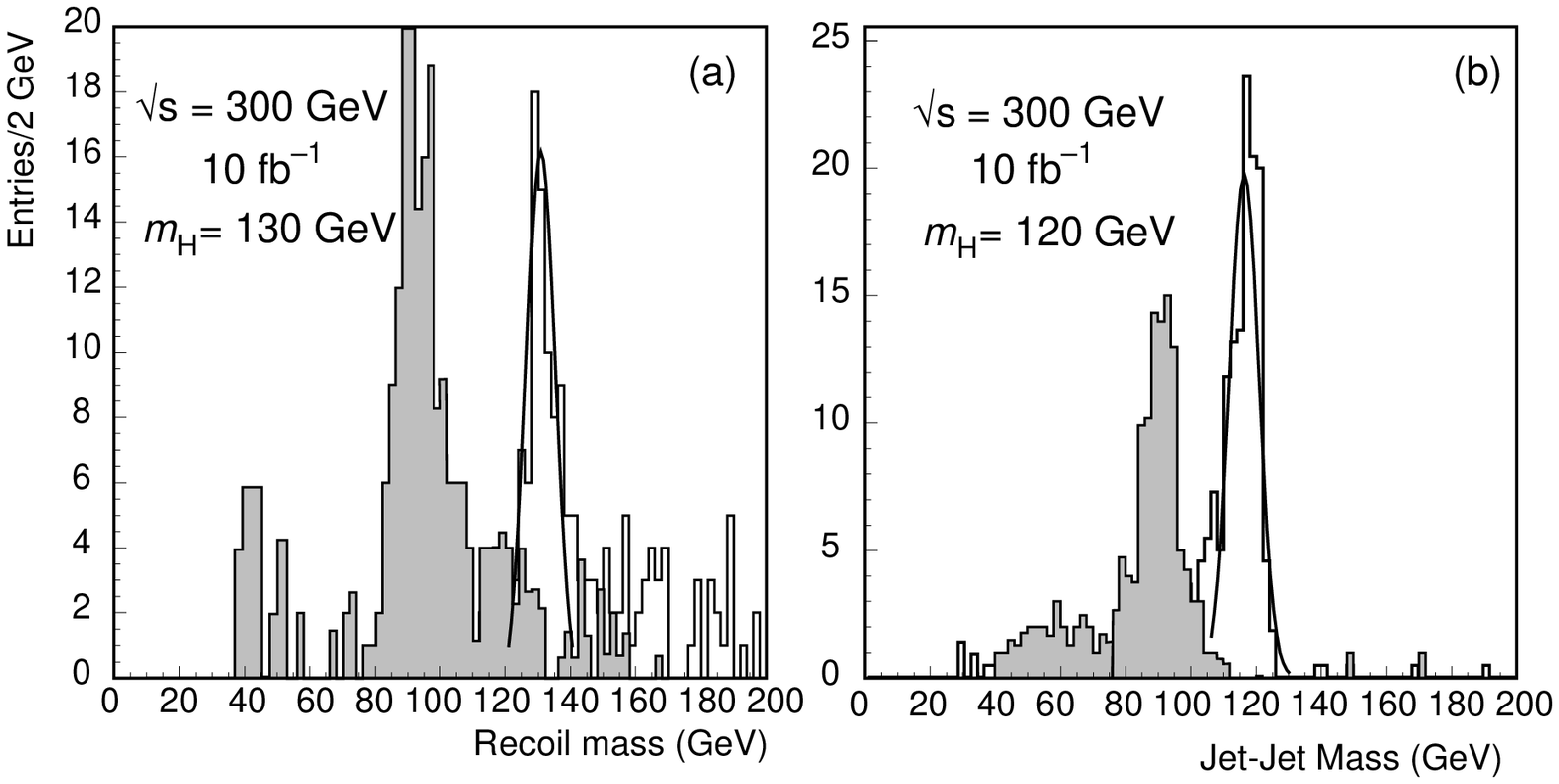,width=3.5in}}
\end{center}
\caption{Higgs mass resolution determined using a typical NLC
detector~\protect\cite{nlc} using (a) the recoil mass against a pair
of electrons or muons from $Z$ decay and (b) from the jet-jet invariant
mass of tagged $b$-quark jets after kinematically-constrained fitting.}
\label{figmassresol}
\end{figure}

Exactly how good the momentum resolution should be
in order to eliminate backgrounds is an
important question; it is currently being pursued.  The study
of Ref.~\cite{kawagoe} obtains good results in the $Z\hsm$
case only if the ``super''
performance of the JLC-I detector \cite{jlci} is assumed.
Current generic NLC detector designs will not be quite so good, but
appear to be adequate.  As an example, using four-vectors smeared
according to the performance of a typical NLC detector~\cite{nlc},
the recoil mass resolution using electrons and muons has been
found~\cite{rickvmass}
to be approximately 3.5~GeV as shown in Fig.~\ref{figmassresol}(a) where
a reasonable mass window results in about 20\% background for
$\mhsm=130\gev$ and obviously degrading for smaller masses and improving
for higher masses. (As described later, running at $\rts\sim 300\gev$,
\ie\ well below $\rts=500\gev$, is critical for such good recoil
mass resolution.)
It is interesting to note that the current performance goals of this
detector give a similar mass resolution of 3.9~GeV from the
invariant jet-jet mass of tagged $b$-quark jets following
kinematically-constrained fitting -- indicating a good match of momentum
and energy resolution.
Our error estimates below will assume momentum resolution such
that the recoil mass peak is sufficiently narrow that backgrounds
are small and can be neglected in the limit of large luminosity.
Preliminary results for the backgrounds are at the $B\sim 0.2S$
level, for which the errors computed below with $B=0$ would
be only slightly increased.
Since, as we shall see, the recoil mass peak cross
section errors sometimes dominate the errors in extracting
branching ratios,
it is quite crucial that the final detector design be adequate to achieve
a small background under the recoil mass peak.

\begin{figure}[htb]
\leavevmode
\begin{center}
\centerline{\psfig{file=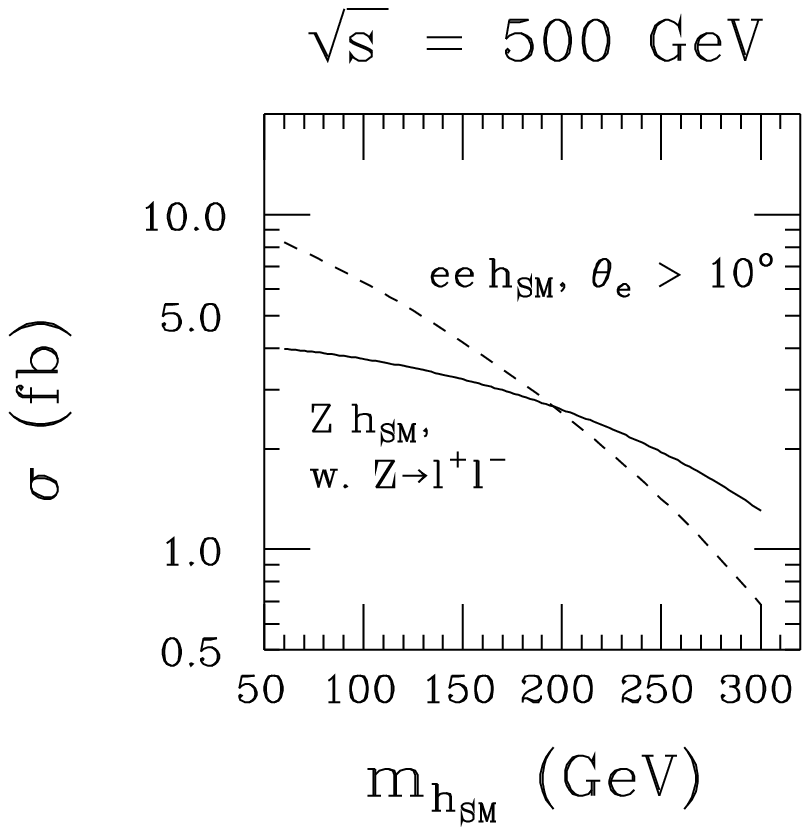,width=2.9in}}
\end{center}
\caption{$\sigma(Z\hsm)\br(Z\to \ell^-\ell^+)$ ($\ell=e,\mu$, no cuts)
and $\sigma(\epem\hsm)$ (with a cut of $\theta>10^\circ$ on the $e^+$
and $e^-$ in the final state) as a function of $\mhsm$ for 
$\protect\rts=500\gev$. From Ref.~\protect\cite{ghs}.}
\label{figzheeh}
\end{figure}

The relative value of the two production modes depends
upon many factors, but in particular it depends
on how the available instantaneous luminosity varies with
$\rts$. For an interaction region configuration/design optimized 
for maximal luminosity at $\sqrt{s_0}$, 
$\call$ falls as $[\rts/\sqrt{s_0}]^2$ \cite{ji} as one moves
to energies lower than $\sqrt{s_0}$.
This is an issue since $\sigma(Z\hsm)$ is maximal
at $\rts\sim \mz+ \mhsm+10~{\rm or}~20\gev$, whereas 
$\sigma(\epem\hsm)$ increases
monotonically with energy.  For the moment, let us assume
that the final focus is designed to maximize
$\call$ at $\rts=500\gev$.  In Fig.~\ref{figzheeh},
we plot $\sigma(Z\hsm)\br(Z\to \ell^-\ell^+)$ ($\ell=e,\mu$, no cuts)
and $\sigma(\epem\hsm)$ (with a $\theta>10^\circ$
cut\footnote{Assuming coverage down to such angles
is optimistic, but not unrealistic. In particular, it 
may be possible to employ a pixel vertexing device with a first layer at
radius of $\sim 1.5$~cm followed by next-generation tracking devices to
avoid the superconducting quads inside the detector.}
on the angles of the final state $e^+$ and $e^-$)
as a function of $\mhsm$ for $\rts=500\gev$. 
We observe a cross-over such that, for $\mhsm\lsim 200\gev$, a
higher raw event rate for the recoil spectrum is obtained using $ZZ$ fusion.

For an integrated luminosity
of $L=200\fbi$ and an overall efficiency of 30\% for
the cuts required to make the background small, the error $1/\sqrt S$
in the $\sigma(Z\hsm)\br(Z\to \ell^-\ell^+)$ ($\ell=e,\mu$) measurement
would range from 6.5\% to 8\% as $\mhsm$ ranges from 60 to $200\gev$,
growing to 11\% by $\mhsm=300\gev$.
Errors in this measurement of similar magnitude can also be achieved
for $L\sim 30-50\fbi$ if $\rts$ is adjusted to be near the value
for which $\sigma(Z\hsm)$ is maximal \cite{kawagoe}.  However, 
depending upon $\mhsm$, accumulating
this much $L$ at lower energy often takes more than the $\sim 4$
years required for $L=200\fbi$ at $\rts=500\gev$ unless the
final focus is optimized for the lower $\rts$ value.
For the $\sigma(\epem\hsm)$ measurement, assuming cut efficiency of 40\%
relative to the cross section plotted in Fig.~\ref{figzheeh}
(we have already included the $\theta>10^\circ$ calorimetry cut)
and $L=200\fbi$, we find errors that range from 4\% to 
8\% to 14\% as $\mhsm$ goes from 60 to 200 to $300\gev$. 
Combining \cite{rickjack,ghs} the $\rts=500\gev$
errors for the two processes gives an error on the $(ZZ\hsm)^2$
coupling-squared that ranges from $\sim 3\%$ to 
$\sim 6\%$ to $\sim 9\%$ for $\mhsm=60$, 200, and $300\gev$, respectively.
These errors are at least as good as those found for $L=200\fbi$ 
using the $Z\hsm$ mode alone at the optimal $\rts$. Thus,
for determining the $(ZZ\hsm)^2$ coupling-squared via the recoil mass procedure
there does not appear to be any advantage to lowering the machine energy
even if the final focus \etc\ is reconfigured so as to maintain
the same instantaneous luminosity.

\subsection{Determining $\hsm$ branching ratios 
and the $WW\hsm$ coupling at the NLC}

A determination
of $\br(\hsm\to X)$ requires measuring $\sigma(\hsm)\br(\hsm\to X)$
and $\sigma(\hsm)$ for some particular production mode,
and then computing
\begin{equation}
\br(\hsm\to X)={\sigma(\hsm)\br(\hsm\to X)\over \sigma(\hsm)}\,.
\label{brform}
\end{equation}
In $\epem$ collisions, the $\epem\to Z\hsm$
and $\epem\to\epem\hsm$ ($ZZ$-fusion) modes just discussed
are the only ones for which the absolute magnitude of $\sigma(\hsm)$ 
can be measured, inclusively summing over all final states $X$.
The $WW$-fusion $\epem\to \nu\anti\nu\hsm$ cross section must be determined
by the procedure of first measuring $\sigma\br(\hsm\to X)$
in some mode $X$ and then dividing by $\br(\hsm\to X)$ as determined
from the $ZZ$-fusion or $Z\hsm$ channels.

\subsubsection{$\br(\hsm\to b\anti b)$ and $\br(\hsm\to c\anti c)$}

By running at $\rts=500\gev$ and accumulating $L=200\fbi$,
we found earlier that by using topological tagging 
a roughly $\pm2.5\%-\pm3.5\%$ determination 
of $\sigma(Z\hsm)\br(\hsm\to b\anti b)$
is possible for $\mhsm\lsim 140\gev$ rising to $\pm 5\%-\pm 7\%$
in the $\mhsm \sim 150\gev$ region.
The error on $\sigma(Z\hsm)$ (just discussed) is in the $\pm6.5\%-\pm7\%$
in the $\mhsm\lsim 140\gev$ mass region, rising to
$\sim\pm 7.5\%$ for $\mhsm\sim 150\gev$. From Eq.~(\ref{brform})
the error in $\br(\hsm\to b\anti b)$ will then be in the $\pm 7\%-\pm 8\%$
range for $\mhsm\lsim 140\gev$, rising to
$\sim\pm 10\%$ at $\mhsm\sim 150\gev$.

The error for the $\sigma(\epem\hsm)\br(\hsm\to b\anti
b)$ measurement has not been studied in detail, but 
can be estimated as follows. We assume that an event identification
efficiency (which should include the efficiency for $b$-tagging)
of 40\% is adequate to make backgrounds small. The number of events ($S$)
is then computed by multiplying $\sigma(\epem\hsm)$ in Fig.~\ref{figzheeh}
by $0.4\br(\hsm\to b\anti b)L$ using 
$\br(\hsm\to b\anti b)$ as tabulated in Table~\ref{hsmbrs} and $L=200\fbi$.
The measurement fractional error is then estimated as $1/\sqrt{S}$.
This yields \cite{ghs} an error in
$\sigma(\epem\hsm)\br(\hsm\to b\anti b)$ ranging from $\pm 4.5\%$ to
$\pm 14\%$ as $\mhsm$ varies from $\lsim 110\gev$ to $\sim 150\gev$;
for higher $\mhsm$ values the error deteriorates rapidly.
\footnote{These errors must be confirmed by a more complete simulation
to verify the level of efficiency, including $b$-tagging,
that could be retained and
still have small backgrounds for this channel.}
Recalling the previously estimated error in the $\sigma(\epem\hsm)$ rate,
which ranges from $\pm 4\%$ to $\pm 6\%$ in the $\mhsm=110$ to $150\gev$
mass region, the resulting error \cite{ghs} on $\br(\hsm\to b \anti b)$ 
as computed from Eq.~(\ref{brform}) in the $\epem\hsm$
final state is then in the $\pm 6\%-\pm 8\%$ range
for $\mhsm\lsim 140\gev$, rising to $\sim \pm 15\%$ for $\mhsm\sim 150\gev$. 

By combining \cite{rickjack,ghs}
the $Z\hsm$ and $\epem\hsm$ determinations,
we find that $\br(\hsm\to b\anti b)$ can be measured with an accuracy
of about $\pm5\%-\pm6\%$ in the $\mhsm\lsim 140\gev$ range,
rising to $\sim\pm 9\%$ for $\mhsm\sim 150\gev$.

For $\br(\hsm\to c\anti c)$, we recall that by using
topological tagging it is estimated that the error
for $\sigma(Z\hsm)\br(\hsm\to c\anti c)$ will be of order $\pm 10\%$
in the $\mhsm\leq 130\gev$ mass region. Using $\pm 7\%$ for 
the $\sigma(Z\hsm)$ error in this mass region implies
an error for $\br(\hsm\to c\anti c)$ of order $\pm12\%$.
Above, we found that in the $b\anti b$ channel the $\epem\hsm$
production mode might yield errors that are comparable to the $Z\hsm$
mode.  A similar result is expected to apply to the 
$c\anti c$ mode \cite{ghs}, implying that
the $\br(\hsm\to c\anti c)$ error would be brought down to $\sim\pm 9\%$.
This same level of error would be achieved if we computed
$\br(\hsm\to c\anti c)=
[(c\anti c\hsm)^2/(b\anti b\hsm)^2]\br(\hsm\to b\anti b)$ and used
the $(c\anti c\hsm)^2/(b\anti b\hsm)^2$ errors given in Table~\ref{nlcerrors}.
Aside from the $\nu\anti\nu\hsm$ component in determining
the $c\anti c$ to $b\anti b$ ratio, these
two techniques are not statistically independent.
It is not clear which would have smaller systematic error.
Presumably, one would pursue both techniques to cross-check
and possibly combine the techniques taking into account
the statistical correlations.

\subsubsection{$\br(\hsm\to W\wstar)$}

The possible procedures are \cite{rickjack}:
\begin{itemize}
\item
Measure $\sigma(Z\hsm)\br(\hsm\to W\wstar)$ and $\sigma(Z\hsm)$
and compute $\br(\hsm\to W\wstar)$ by dividing.
As discussed earlier, errors in $\sigma(Z\hsm)\br(\hsm\to W\wstar)$
are roughly $\pm 22\%$, $\pm 10\%$ and $\pm 7\%$ for
$\mhsm=130$, $140$ and $150\gev$. 
The error for $\sigma(Z\hsm)$ (using recoil mass detection)
ranges (see earlier) from $\sim \pm 4\%$ to $\sim\pm 6\%$ in this mass range.
The resulting $\br(\hsm\to W\wstar)$ error would be
roughly $\pm 22\%$, $\pm 11\%$, $\pm 9\%$ for $\mhsm=130$, $140$ 
and $150\gev$, respectively. 
At $\mhsm=200,300\gev$ accuracies for $\br(\hsm\to W\wstar)$
of $\sim\pm12\%$ and $\sim\pm17\%$ are predicted
by extrapolation based on event rate and branching ratio changes.\footnote{We
have assumed that the background scales with the signal rate.  A full
simulation would be required to verify the extrapolation assumptions.}
\item
Measure $\sigma(\epem\hsm)\br(\hsm\to W\wstar)$ and $\sigma(\epem\hsm)$
(the $ZZ$-fusion processes) and again compute $\br(\hsm\to W\wstar)$
by dividing \cite{ghs}. In the $130-200\gev$ mass region,
we have already seen that the $\sigma(\epem\hsm)$ measurement will
be comparable (perhaps slightly superior)
in accuracy to the $\sigma(Z\hsm)$ measurement. A first
estimate indicates that
the accuracy of the $\sigma(\epem\hsm)\br(\hsm\to W\wstar)$
measurement will also be comparable to that for 
$\sigma(Z\hsm)\br(\hsm\to W\wstar)$.
For example, at $\mhsm=150\gev$
Fig.~\ref{figzheeh} gives $\sigma(\epem\hsm)\sim 4 \fb$
and from Table~\ref{hsmbrs} we find $\br(\hsm\to\wp\wm)\sim 0.7$. If the 
efficiency for tagging the $W\wstar$ final state and requiring
the recoil mass to be close to the known value of $\mhsm$ is, say,  $40\%$,
then we would have $S=224$ signal events with relatively small background
(due to our ability to always require recoil mass
$\sim \mhsm$\footnote{Typically, the recoil mass resolution is
better in the $Z\hsm$ ($Z\to \epem,\mupmum$) channel
than in the $\epem\hsm$ channel once the $Z$ mass
is used in a kinematically constrained fit. However, all
that is needed for the statistical estimates given here to apply is
that the recoil mass resolution in $\epem\hsm$ events
be sufficient that the background
in the peak region be small; this should be the case given that the $\epem$
momenta would be quite well-measured.} in this
production mode).  The resulting error for 
$\sigma(Z\hsm)\br(\hsm\to W\wstar)$ is $\sim\pm 7\%$.
Thus, errors on $\br(\hsm\to W\wstar)$ in the $\epem\hsm$ production channel
will be close to those in the $Z\hsm$
channel for $\mhsm$ in the $130-200\gev$ mass range.

At $\mhsm=300\gev$,
$\sigma(\epem\hsm)$ is smaller than $\sigma(Z\hsm)$ (see Fig.~\ref{figzheeh}).
After including efficiency we found in the previous subsection that
the error on $\sigma(\epem\hsm)$
will be about $\pm 14\%$ (vs. $\pm 11\%$ for $\sigma(Z\hsm)$).
Similarly, the error on $\sigma(\epem\hsm)\br(\hsm\to W\wstar)$
will be larger than for $\sigma(Z\hsm)\br(\hsm\to W\wstar)$.
At $\mhsm=300\gev$, we find (by extrapolation, subject to footnote caveats)
error on the former of about $\pm 18\%$ (vs. $\sim\pm 12\%$ for the
latter); combining with the $\pm 14\%$ error on $\sigma(\epem\hsm)$ yields
error for $\br(\hsm\to W\wstar)$ of order $\pm 23\%$ at this mass
for the $\epem\hsm$ channel.
\end{itemize}
If we combine \cite{rickjack,ghs} the above two determinations, 
the overall $\br(\hsm\to W\wstar)$ error would be reduced 
to the roughly $\pm 16\%$, $\pm 8\%$, $\pm 6 \%$ level for 
$\mhsm=130$, $140$ and $150\gev$, and even somewhat
smaller at $\mhsm=170\gev$. Above $170\gev$, the accuracy
of the determination slowly declines to about $\pm 8\%$
at $\mhsm=200\gev$ and $\pm 14\%$ at $\mhsm=300\gev$.

\subsubsection{$WW\hsm$ coupling and testing custodial SU(2)}

The goal will be to determine 
$\sigma(\nu\anti\nu\hsm)$ which is proportional to
the the $(WW\hsm)^2$ coupling-squared. 
The best procedure \cite{rickjack} depends upon $\mhsm$:
\begin{itemize}
\item 
If $\mhsm\lsim 140\gev$, then good accuracy is attained by
measuring $\sigma(\nu\anti\nu\hsm)\br(\hsm\to b\anti b)$ and then dividing
by $\br(\hsm\to b\anti b)$. For $L=200\fbi$ and $\mhsm\lsim 140\gev$,
the measurement error for the former
is $\sim\pm 2.5\%-\pm 3.5\%$ (as stated earlier),
and that for the latter $b\anti b$ branching ratio is
$\pm 5\%-\pm6\%$ (as stated above). The 
net error in $(WW\hsm)^2$ obtained in this way is of order $\pm 6\%$
for $\mhsm\lsim 140\gev$. By $\mhsm=150\gev$, the accuracy
of the $b\anti b$ mode determination of $(WW\hsm)^2$ has worsened
to about $\pm 11\%$, coming from $\sim\pm 6\%$ for $\sigma(\nu\anti\nu\hsm)
\br(\hsm\to b\anti b)$ and $\sim\pm 9\%$ for $\br(\hsm\to b\anti b)$;
see earlier subsections.
\item 
If $\mhsm\gsim 150\gev$, then good accuracy is achieved by measuring
$\sigma(\nu\anti\nu\hsm)\br(\hsm\to W\wstar)$ (in $WW$-fusion)
and dividing by $\br(\hsm\to W\wstar)$ (see earlier subsection)
to get $\sigma(\nu\anti\nu\hsm)$. Explicitly, we estimated above that
an error on $\br(\hsm\to W\wstar)$ at the $\sim \pm8\%,\pm6 \%,\pm8\%,\pm14\%$
level could eventually be achieved for $\mhsm\sim 140,150,200,300\gev$. 
Earlier, we saw that the error in $\sigma(\nu\anti\nu\hsm)\br(\hsm\to
W\wstar)$ is estimated to be $\pm10\%,\pm 8\%,\pm$ at $\mhsm=140,150\gev$.
Extrapolating to $200,300\gev$,\footnote{We re-emphasize
the fact that simulations at $200$ and $300\gev$ are needed
to check our extrapolations.}
we estimate errors of $\pm10\%,\pm20\%$, respectively. 
Combining, we find that the error on the $(WW\hsm)^2$ coupling-squared
from the $W\wstar$ final state determination would be about 
$\pm13\%$, $\pm 10\%$, $\pm13\%$, $\pm24\%$
at $\mhsm\sim 140$, 150, 200, $300\gev$, respectively. 
The error at $\mhsm=170\gev$
would be slightly smaller than that at $\mhsm=150\gev$.
The $\mhsm=140\gev$ result
is poorer than that obtained in the $b\anti b$ mode, but by $\mhsm=150\gev$
the $W\wstar$ mode determination has become comparable,
and for higher masses is distinctly superior.
\end{itemize}
If we combine the $b\anti b$ and $W\wstar$ mode determinations,
we get an error for $(WW\hsm)^2$ of order $\pm 5\%$ for $\mhsm\lsim 140\gev$,
worsening to about $\pm 8\%$ for $\mhsm\gsim 150\gev$.
For $170\gev$ and above the error is simply that found in the $W\wstar$ mode,
\eg\ $\pm 13\%,\pm24\%$ at $\mhsm=200,300\gev$, respectively.

It is, of course, of great interest to test the custodial SU(2) symmetry
prediction for the coupling-squared ratio $(WW\hsm)^2/(ZZ\hsm)^2$.
In an earlier subsection we estimated the error on $(ZZ\hsm)^2$
for $\mhsm\lsim 60-200\gev$ to be $\sim \pm 4\%-\pm 6\%$,
rising to $\sim\pm 9\%$ at $\mhsm=300\gev$.  
Combining with the above results for the $(WW\hsm)^2$ errors,
we estimate errors for $(WW\hsm)^2/(ZZ\hsm)^2$
of order $\pm 7\%$ for $\mhsm\lsim 140\gev$, $\pm 10\%$ for $\mhsm\sim
150\gev$, rising slowly to $\sim\pm14\%$ 
for $\mhsm= 200\gev$, reaching $\sim\pm 25\%$ at $\mhsm=300\gev$.

\subsubsection{$\br(\hsm\to \gam\gam)$}

We focus on $\mhsm\lsim 130\gev$.  
Only two ways to get a handle on $\br(\hsm\to\gam\gam)$ have been
demonstrated to be viable.
\begin{itemize}
\item
The first involves measuring
$\sigma(pp\to W\hsm)\br(\hsm\to \gam\gam)$ 
and $\sigma(pp\to t\anti t\hsm)\br(\hsm\to\gam\gam)$
at the LHC. As outlined earlier, each can be determined to about $\pm15\%$
for $\mhsm$ in the range $90-130\gev$. Although not
explicitly simulated in the ATLAS and CMS studies,
we assume this same error applies at $80\gev$.
These measurements can be employed in two ways.  
\begin{itemize}
\item 
In the first approach
one also measures $\sigma(pp\to t\anti t\hsm)\br(\hsm\to b \anti b)$ 
at the LHC and then computes $\br(\hsm\to\gam\gam)$ as
\begin{eqnarray}
\br(\hsm\to\gam\gam) = \br(\hsm\to b\anti b)\times && \nonumber \\
{[\sigma(pp\to t\anti t\hsm)\br(\hsm\to\gam\gam)]
\over
[\sigma(pp\to t\anti t\hsm)\br(\hsm\to b\anti b)]} &&
\end{eqnarray}
using $\br(\hsm\to b\anti b)$ determined at the NLC as described earlier.
Since the error for $\br(\hsm\to b\anti b)$
will be of order $\pm 4\%-\pm5\%$ (for $L=200\fbi$ at the NLC),
the error in the determination of $\br(\hsm\to\gam\gam)$ is dominated
by that for the $\gam\gam/b\anti b$ ratio (see Tables~\ref{m1errors}
and \ref{m2errors}), and will range from about
$\pm 18\%$ to $\pm 26\%$ over the $80-130\gev$ mass range.
\item
In the second approach, one uses only $\sigma(pp\to W\hsm)\br(\hsm\to\gam\gam)$
from the LHC, and then divides
by the computed $\sigma(pp\to W\hsm)$ cross section.  
In the $\mhsm\lsim 130\gev$ mass region, the cross
section is best computed using the $(WW\hsm)^2$ coupling-squared
determination from the NLC which, as noted earlier,
has an error of order $\pm 6\%$ for this mass region.
Including systematics, the error in
$\sigma(pp\to W\hsm)$ is then likely to be of order $\pm 10\%$.
Combining with the $\sim\pm 15\%$ error for $\sigma(pp\to
W\hsm)\br(\hsm\to\gam\gam)$ yields an error of $\sim\pm 18\%$ 
in the determination of $\br(\hsm\to\gamgam)$ in the $\mhsm=80-130\gev$
region.

\end{itemize}
To the extent that determinations from these 
two ways of getting at $\br(\hsm\to \gam\gam)$ are statistically
independent, they can be combined to yield
statistical accuracy of $\lsim \pm 16\%$ in the $\mhsm\lsim 130\gev$ range.
A rough guess based on simulations performed at lower masses
is that at $\mhsm=140\gev$ this error would deteriorate to about $\pm 25\%$.
We also assume very large error for $\mhsm\geq 150\gev$.
\item
The second technique is that explored in Ref.~\cite{gm},
using the $\sigma\br(\hsm\to\gam\gam)$ measurements at the NLC
discussed earlier. These lead to two
possible techniques for getting $\br(\hsm\to\gam\gam)$.
\begin{itemize}
\item Measure $\sigma(\epem\to Z\hsm)\br(\hsm\to \gam\gam)$ and compute
$\br(\hsm\to\gam\gam)$ as
\begin{equation}
{[\sigma( Z\hsm)\br(\hsm\to
\gam\gam)]\over\sigma( Z\hsm)}\,;
\label{way1}
\end{equation}
\item Measure $\sigma(\epem\to \nu\anti\nu \hsm)\br(\hsm\to \gam\gam)$ 
and $\sigma(\epem\to \nu\anti\nu \hsm)\br(\hsm\to b\anti b)$ 
(both being $WW$-fusion processes) and compute $\br(\hsm\to\gam\gam)$ as
\begin{equation}
{[\sigma(\nu\anti\nu\hsm)\br(\hsm\to \gam\gam)]\br(\hsm\to
b\anti b)\over [\sigma(\nu\anti\nu\hsm)\br(\hsm\to b\anti b)]}\,.
\label{way2}
\end{equation}
\end{itemize}
The $\epem\hsm$ final state from $ZZ$-fusion is a third
alternative, but does not yield errors competitive with
the above two techniques \cite{pmartin} because of a smaller signal
relative to background.
\end{itemize}

At the NLC, the errors in the $\br(\hsm\to\gam\gam)$ determinations
are completely dominated by the $\sigma\br(\hsm\to\gam\gam)$ errors,
which we have discussed earlier; see Fig.~\ref{figgamgamerrors}.
Assuming running at $\rts=500\gev$,
we found that the smallest $\sigma\br$ error was achieved in the
$WW$-fusion mode. However, a useful level of error
was also achieved in the $Z\hsm$ mode when running at this energy. 
The errors expected for $\br(\hsm\to\gam\gam)$ by combining
the determinations of Eqs.~(\ref{way1}) and (\ref{way2}) are 
essentially the same as the combined $\sigma\br(\hsm\to\gam\gam)$
error plotted in the 3rd window of Fig.~\ref{figgamgamerrors}.
For a calorimeter at the
optimistic end of current plans for the NLC detector,
the net error is predicted to range from
$\sim\pm 22\%$ at $\mhsm=120\gev$
to $\sim\pm 35\%$ ($\sim\pm 53\%$) at $\mhsm=150\gev$ ($70\gev$).

Of course, the NLC and LHC determinations can be combined
to give sometimes substantially 
smaller error than achieved at either machine alone.
The errors for $\br(\hsm\to\gam\gam)$
obtained by combining LHC and NLC data will be tabulated later
in Table~\ref{nlcerrors}.

Although one of the big motivations for measuring $\br(\hsm\to\gam\gam)$
at the NLC is its crucial role in determining $\gamhsm$ (to be outlined
later), whereas $\gamhsm$ can be directly measured at the FMC by
scanning (see next subsection), a measurement of $\br(\hsm\to\gam\gam)$ 
would ultimately also be of interest at the FMC, especially if
there is no NLC.\footnote{In particular,
since a $\gam\gam$ collider is not possible at the FMC \cite{bbgh},
if there is no NLC then the very interesting
partial width $\Gamma(\hsm\to\gam\gam)$ can only be obtained in the form
$\gamhsm\br(\hsm\to\gam\gam)$.}
The possibility
of measuring the branching ratio using FMC data at $\rts=\mhsm$ was examined
\cite{pmartin}; for a SM-like Higgs boson, $S/B$ turns out to be much
too small for this to succeed. Thus, at the FMC,
$\br(\hsm\to \gam\gam)$ would have to be determined 
following the same non-$s$-channel procedures as for the NLC.

\subsection{Determining $\gamhsm$ and $\hsm$ partial widths}

The most fundamental properties of the Higgs boson
are its mass, its total width and its partial widths.
Discussion of the mass determination will be left till the next subsection.
The total Higgs width, while certainly important in its own right,
becomes doubly so since it is required in order to compute many
important partial widths.
The partial widths, being directly proportional to the underlying
couplings, provide the most direct means of verifying that the observed
Higgs boson is or is not the $\hsm$. Branching ratios, being
the ratio of a partial width to the total width can not be unambiguously
interpreted. In contrast,
a partial width is directly related to the corresponding
coupling-squared which, in turn, is directly determined in the SM or any
extension thereof without reference to mass scales for possibly
unexpected (\eg\ SUSY) decays.
Any deviations of partial widths from SM predictions 
can be directly compared to predictions of alternative models
such as the MSSM, the NMSSM, or the general \thdm.  The more accurately
the total width and the various branching ratios can be measured, 
the greater the sensitivity
to such deviations and the greater our ability to recognize and constrain
the alternative model.

For $\mhsm\lsim 2\mw$, 
$\gamhsm$ is too small to be reconstructed in the final state;
indirect determination of $\gamhsm$ is necessary.
We note that the $\mhsm\lsim 2\mw$ mass range
is that which would be relevant for the SM-like
Higgs boson of the MSSM. For larger $\mhsm$, direct final state
reconstruction of $\gamhsm$ starts to become possible; the mass above which
reasonable error on $\gamhsm$ is obtained depends upon detector
and machine characteristics. The possibilities are reviewed below.

\subsubsection{Determining $\gamhsm$}

There are only two basic possibilities for determining $\gamhsm$
in the $\mhsm\lsim 2\mw$ mass range in which $\gamhsm$ is
too small to be reconstructed in the final state.
\begin{itemize}
\item
The first is to employ FMC $\mupmum$ collisions at $\rts\sim \mhsm$ and
directly measure $\gamhsm$ by scanning.  In this case, 
the FMC determination of $\gamhsm$
can be used to compute the partial width for any channel with a 
branching ratio measured at the NLC:
\begin{equation}
\Gamma(\hsm\to X)=\gamhsm\br(\hsm\to X)\,.
\label{partialw}
\end{equation}
\item
If there is no muon collider, then $\gamhsm$ must be determined indirectly
using a multiple step process; the best process depends upon
the Higgs mass. $\gamhsm$ is ultimately computed as:
\begin{equation}
\gamhsm={\Gamma(\hsm\to X)\over\br(\hsm\to X)}\,,
\label{partialwi}
\end{equation}
where $X=\gam\gam$ ($W\wstar$) gives the best error for $\mhsm\lsim 130\gev$
($\gsim140\gev$).
In this case, $\gamhsm$ can be used to compute partial widths
via Eq.~(\ref{partialw}) only for channels other than those
used in the determination of $\gamhsm$ via Eq.~(\ref{partialwi}).
\end{itemize}
In what follows we outline the errors anticipated in the ultimate
determination of $\gamhsm$ in the $\mhsm\lsim 2\mw$
mass region, and then discuss implications for
the errors in partial widths, both with and without combining NLC
and FMC data. We also discuss the determination of $\gamhsm$ by
final state mass peak reconstruction in the mass range $\mhsm\gsim 2\mw$.

\begin{figure}[htb]
\leavevmode
\begin{center}
\centerline{\psfig{file=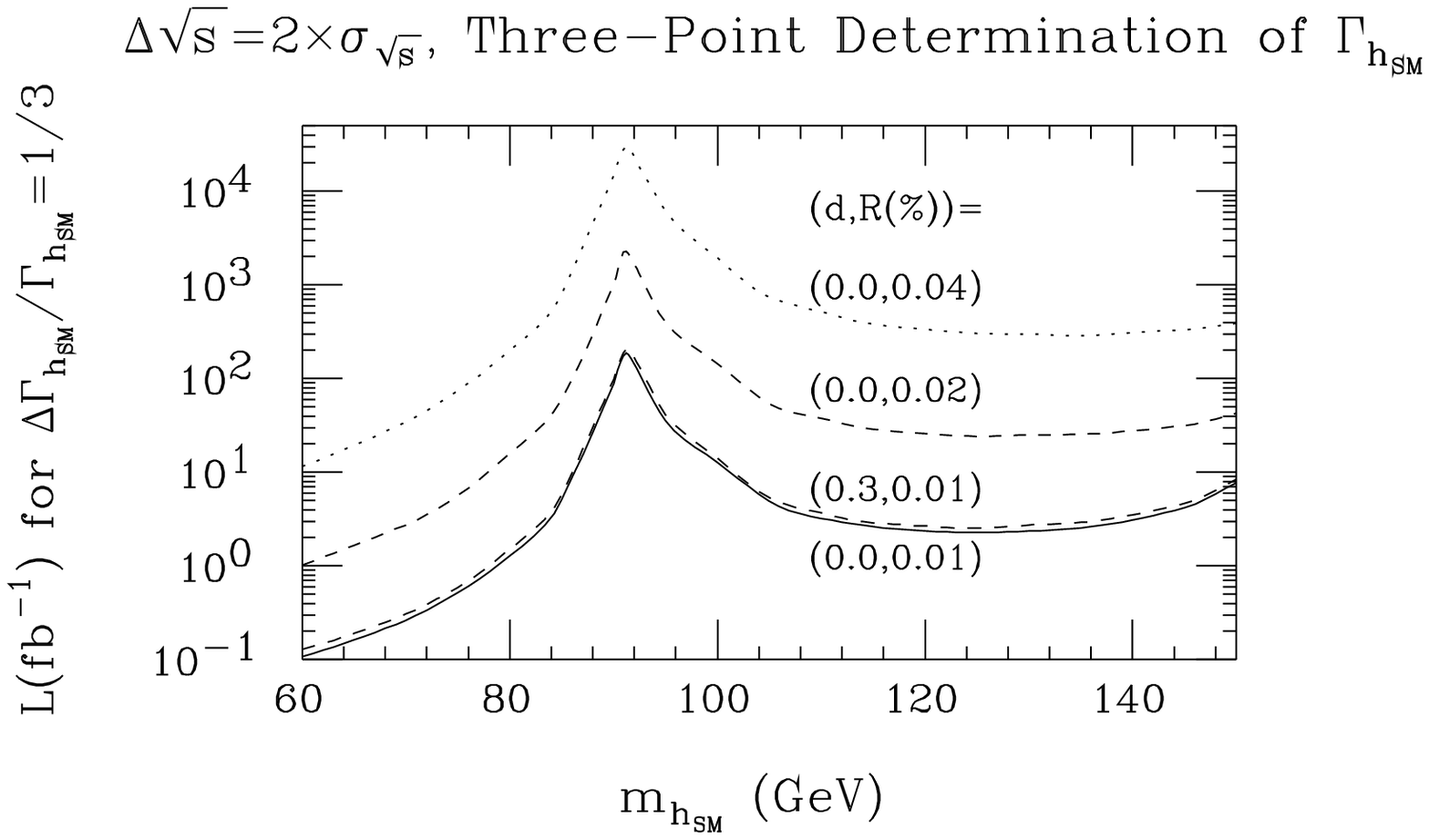,width=3.5in}}
\end{center}
\caption{Luminosity required for a $\Delta\gamhsm/\gamhsm=1/3$
measurement in the $b\anti b$ final state using the 3-point
technique described in \protect\cite{bbgh}.  Results for resolutions
of $R=0.01\%$, $0.02\%$ and $0.04\%$ are shown for $d=0$, where 
$d = | \protect\sqrt{s_0} - \mhsm | / \sigma_{\protect\rts}$. Here,
$\protect\sqrt{s_0}$ is the location of the central energy
setting in the 3-point scan and
$\sigma_{\protect\rts}$ is the resolution in $\protect\rts$
for a given value of $R$.
The result for $d=0.3$ and $R=0.01\%$ illustrates how insensitive
the total luminosity required is to the accuracy of the central setting.}
\label{figscan}
\end{figure}

\bigskip
\begin{center}
\underline{FMC-scan determination of $\gamhsm$}
\end{center}
\smallskip
Only the $\mupmum$ collider
can have the extremely precise energy resolution and energy setting
capable of measuring $\gamhsm$ by scanning \cite{bbgh}.
The amount of integrated luminosity required for a $\pm 33\%$ determination
of $\gamhsm$ using a 3-point scan with 0.01\% beam energy resolution
is shown in Fig.~\ref{figscan}. The most difficult case is if $\mhsm\sim \mz$,
implying a large $Z$ background to $\hsm$ production in the $s$-channel.
The accuracy of the $\gamhsm$ determination scales as $1/\sqrt L$.

We assume that since the mass of the Higgs boson will be 
relatively precisely known from the LHC (see next subsection)
the FMC would be designed to have optimal luminosity at
$\rts\sim\mhsm$, so that accumulation of $L=200\fbi$ for
scanning the Higgs peak would be possible. It is important
to note that in the 3-point scan procedure
of Ref.~\cite{bbgh} most (5/6) of the luminosity is devoted to
the wings of the Higgs peak; only 1/6 of the total $L$ is accumulated
at $\rts=\mhsm$ (exactly).  Very roughly the total number
of Higgs events from the wing measurements is equivalent to $\sim 0.3L$
on the peak, but $S/B$ is smaller on the wings. Overall, if luminosity $L$
is devoted to the scan procedure, 
the errors that can be achieved for rate measurements in specific
channels are roughly equivalent to what would be achieved if $0.25L$
was devoted to $\rts=\mhsm$ running.
Since it is not useful to sacrifice accuracy in the $\gamhsm$
measurement in order to devote more luminosity to the peak, 
in Table~\ref{fmcsigbrerrors}
we quoted measurement errors for specific channels 
obtained for $L=50\fbi$ at $\rts=\mhsm$. These same channel errors will be
used in subsequent calculations.

Fig.~\ref{figscan} implies that integrated luminosity of $L=200\fbi$
would yield a $\pm 2.6\%$, 
$\pm 32\%$, $\pm 3.6\%$, $\pm 6.5\%$ determination of $\gamhsm$
at $\mhsm=80\gev$, $\mz$, $120\gev$, $150\gev$, respectively.
A complete listing of errors appears (later) in Table~\ref{fmcerrors}.
In the $\mhsm\sim \mz$ worst case, the $s$-channel FMC accuracy will turn
out to be worse than can be attained at the NLC.  
However, for most masses, the $s$-channel 
FMC accuracy would be much superior and would provide
an extremely valuable input to precision tests of the Higgs sector.

\bigskip
\begin{center}
\underline{Indirect determination of $\gamhsm$}
\end{center}
\smallskip
If there is no $\mupmum$ collider, then $\gamhsm$ must be determined
indirectly.  The best procedure for doing
so depends upon the Higgs mass. If $\mhsm\lsim 130\gev$,
then one must make use of $\gam\gam$ Higgs decays. If $\mhsm\gsim 140\gev$,
$W\wstar$ Higgs decays will be most useful. In both cases, we ultimately
employ Eq.~(\ref{partialwi}) to obtain $\gamhsm$.

\begin{figure}[htb]
\leavevmode
\begin{center}
\centerline{\psfig{file=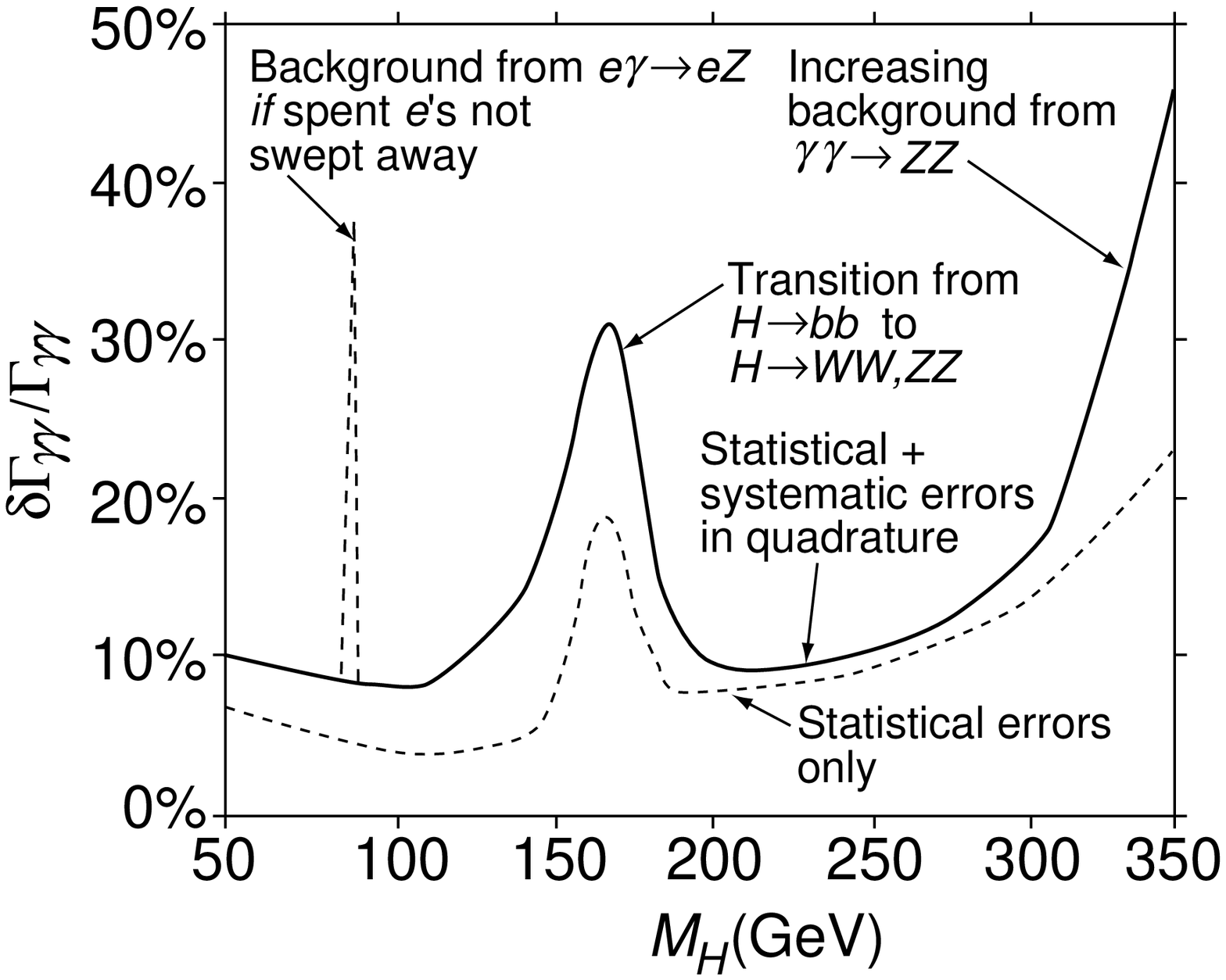,width=3.5in}}
\end{center}
\caption{Accuracy (including systematic as well as statistical errors)
with which $\Gamma(\hsm\to \gam\gam)\br(\hsm\to b\anti b~{\rm or} WW,ZZ)$
can be measured at the NLC $\gam\gam$ collider with integrated
luminosity of $L=50\fbi$ \protect\cite{bauer}.}
\label{figbauer}
\end{figure}

Since the $\Gamma(\hsm\to\gam\gam)$ partial width plays a crucial
role in the $\mhsm\lsim 130\gev$ procedure, it is convenient
to discuss it first.
The study of $\gam\gam\to\hsm \to b\anti b$ 
at the NLC is performed
by tuning the beam energy so that the $\gam\gam$ luminosity peak at 
$\sim 0.8\rts_{\epem}$ coincides with $\mhsm$ \cite{ghgamgam,borden}.
The statistical accuracy that could be achieved for
$\Gamma(\hsm\to\gamgam)\br(\hsm\to b\anti b)$ was estimated in
Ref.~\cite{borden}. Systematic errors have now been evaluated and
the effects of gluon radiation and $ZZ$ backgrounds have been included
\cite{bauer}.
Suppressing the dangerous $c\anti c g$ backgrounds reduces the signal by only a
factor of two.  
The net error on $\Gamma(\hsm\to\gamgam)\br(\hsm\to b\anti b)$ for $L=50\fbi$
is illustrated in Fig.~\ref{figbauer}.
For $L=200\fbi$, the error would be only half as large as shown,
but since the luminosity employed
in this measurement would be lost to normal running to get the
branching ratios \etc, we consider only the $L=50\fbi$ errors.
Thus, the error in the $\mhsm\lsim 120\gev$ mass region
will be in the 8\%-10\% range, rising to 15\% by $\mhsm=140\gev$
and peaking at 30\% at $\mhsm=150\gev$,
as illustrated in Fig.~\ref{figbauer}. 
To get the accuracy in the $\Gamma(\hsm\to \gam\gam)$ partial width itself,
we recall that $\br(\hsm\to b\anti b)$ is measured with accuracy of
$\pm 5\%-\pm6\%$ for $\mhsm\lsim 140\gev$, rising to $\pm9\%$
at $\mhsm\sim 150\gev$. The result is $\Gamma(\hsm\to\gam\gam)$
error of order $\pm 12\%$ for $\mhsm\lsim 120\gev$, rising to
$\sim\pm 17\%$ at $\mhsm\sim 140\gev$ and $\sim\pm 31\%$ at $\mhsm\sim
150\gev$.

We now give the procedures for determining $\gamhsm$.
\begin{itemize}
\item
For $\mhsm\lsim 130\gev$ (\ie\ in the MSSM $\mhl$ range), 
the only known procedure for determining $\gamhsm$ is that outlined
in Ref.~\cite{dpfreport}. NLC data is required.
\begin{itemize}
\item
As described above, measure
$\Gamma(\hsm\to \gam\gam)\br(\hsm\to b\anti b)$ and then compute
$\Gamma(\hsm\to\gam\gam)$ by dividing by the value of 
$\br(\hsm\to b\anti b)$.
\item
Compute $\gamhsm=\Gamma(\hsm\to\gam\gam)/\br(\hsm\to\gam\gam)$,
using the $\br(\hsm\to\gam\gam)$ determination(s) described earlier.
\end{itemize}
The accuracies of the various measurements involved are a crucial issue.
The results obtained in earlier sections appear in Table~\ref{nlcerrors}.
Using the determination of $\br(\hsm\to\gam\gam)$ based
on combining NLC and LHC data, we find an error on $\gamhsm$
of $\sim\pm 18-19\%$ for $\mhsm=80-120\gev$ and $\sim\pm 20\%$
for $\mhsm=130\gev$. At $\mhsm=140,150\gev$,
errors on $\br(\hsm\to\gam\gam)$ and $\Gamma(\hsm\to\gam\gam)$
increase and the $\gamhsm$ error would be $\sim\pm 25\%,\sim\pm46\%$,
respectively.

\item
For $\mhsm\gsim 130\gev$, a second possible procedure based on
$\hsm\to W\wstar$ decays emerges.
Use $(WW\hsm)^2$ to compute $\Gamma(\hsm\to W\wstar)$ and then compute
$\gamhsm=\Gamma(\hsm\to W\wstar)/\br(\hsm\to W\wstar)$.
\footnote{Of course, keeping only the $W\wstar$ mode,
this latter procedure can be viewed as a computation
of $\gamhsm\propto \sigma(\nu\anti\nu\hsm)/[\br(\hsm\to W\wstar)]^2$.}
The required errors were obtained in earlier
sections and are tabulated in Table~\ref{nlcerrors}.
We find an error for $\gamhsm$ of about $\pm17\%$ at $\mhsm=130\gev$,
falling to $\pm10\%-\pm11\%$ for $\mhsm$ in the $150-170\gev$ range.
This latter is certainly much better than the $\sim\pm 46\%$ achieved
in the $\gam\gam$ channel at $\mhsm\sim 150\gev$ using NLC $\gam\gam$
collider data and the (NLC+LHC) determination of $\br(\hsm\to\gam\gam)$.
For $\mhsm\sim 130\gev$,
the $\pm 17\%$ achieved in the present $W\wstar$ technique is still superior
to the $\pm 20\%$ for the $\gam\gam$ technique. Combining
the determinations made via the two techniques at $\mhsm=130\gev$,
we would get an error on $\gamhsm$ of order $\pm 13\%$.
For $\mhsm\lsim 120\gev$,
the $\gam\gam$ technique determination of $\gamhsm$
is substantially superior to what can be achieved via the $W\wstar$
technique, primarily because $\br(\hsm\to W\wstar)$ is very poorly determined.

\end{itemize}
In Table~\ref{nlcerrors}, we tabulate the errors for $\gamhsm$ 
obtained by using both the $\gam\gam$ and the $W\wstar$
techniques, and including the (NLC+LHC) determination of $\br(\hsm\to\gam\gam)$
in the former.

As apparent from Tables~\ref{fmcerrors} and \ref{nlcerrors},
for $\mhsm\lsim 130\gev$ (and $\mhsm\not\sim\mz$) the FMC-scan
determination of $\gamhsm$ is very much superior to the NLC determination.
The superiority is still significant at $\mhsm=140\gev$ while  errors
are similar at $\mhsm=150\gev$. At $\mhsm=150\gev$, combining
the FMC-scan and NLC determinations of $\gamhsm$
would increase accuracy for $\gamhsm$, yielding a combined
error of $\sim\pm 5.4\%$ (vs. $\pm 6.5\%$ for the FMC-scan alone).
This would be beneficial for computing
partial widths (other than that for the $W\wstar$ channel used
in the NLC determination of $\gamhsm$ at this mass).
For $\mhsm\gsim 160\gev$, FMC $s$-channel detection
of the $\hsm$ becomes difficult, and only the NLC allows a reasonable
determination of $\gamhsm$.

\bigskip
\begin{center}
\underline{Final-state mass peak determination of $\gamhsm$: NLC}
\end{center}
\smallskip

\def\gamr{\Gamma_{\rm R}}
\def\gameff{\Gamma_{\rm eff}}
Of course, once $\mhsm\gsim 2\mw$, $\gamhsm$ is large enough
that measurement directly from the shape of the mass peak becomes
conceivable. The precise sensitivity depends upon detector characteristics
and other details. We \cite{rickjack} will illustrate results for $Z\hsm$
production in five cases.  
In the first four cases, we demand that $Z\to\epem,\mupmum$ and
reconstruct the Higgs peak via the recoil mass.  The momenta
of the muons are measured by the tracking component of the detector.
The momenta of the electrons are measured by both the tracker and
the electromagnetic calorimeter --- since these are not statistically
independent of one another, we use the measurement having the smaller
error. The $\epem$ and $\mupmum$ final states are treated separately,
and at the end their errors are statistically combined. Four
different combinations of tracking and calorimetry are considered.
In the fifth case, we allow the $Z$ to decay to either $\epem,\mupmum$
or $q\anti q$, and reconstruct the Higgs resonance peak using the $b\anti b$
or $\wp\wm$ Higgs decay products. The five cases are specified
in detail as follows:
\begin{enumerate}
\item 
We assume super-JLC tracking \cite{jlci}, implying
$\Delta p/p=5\times 10^{-5}p(\gev)\oplus 0.001$,
and slightly better than `standard' NLC detector \cite{nlc} calorimetry of
$\Delta E/E=0.12/\sqrt{E(\gev)}\oplus 0.005$. In this case, the best electron
momentum measurement is almost always from the tracking, so
that the natural event-by-event resolution ($\gamr$)
in the $\epem$ and $\mupmum$
channels is the same. One finds that $\gamr$ can be as small as $0.3\gev$ or so
when $\rts=\mz+\mhsm+\sim 20\gev$ and the $e$'s/$\mu$'s are not 
terribly energetic but that $\gamr$ deteriorates considerably if the machine is
run at $\rts=500\gev$ because of the much larger energies of the leptons,
implying larger tracking errors.
We assume a systematic error in our knowledge of $\gamr$ in the $\epem$
and $\mupmum$ channels of 10\%.
\item
The assumptions for this case are exactly the same as for the first case,
except that we allow for a much larger systematic error of 50\%,
as could be relevant when $\gamr$ is so small.
\item
We assume the `standard' NLC detector tracking \cite{nlc}, implying
$\Delta p/p=5\times 10^{-4}p(\gev)\oplus 0.0015/\sqrt{p(\gev)}$, 
and electromagnetic calorimetry unchanged at
$\Delta E/E=0.12/\sqrt{E(\gev)}\oplus 0.005$. In this case, the best electron
momentum measurement is always from the calorimetry, especially when
$\mhsm$ is small and one runs at the higher $\rts=500\gev$. Thus,
the natural resolution $\gamr$ in the $\epem$ and $\mupmum$
channels can be quite different. Systematic error of $10\%$ for $\gamr$
is assumed.
\item
We assume the same `standard' 
NLC tracking as in the 3rd case, but adopt a CMS \cite{CMS}
type electromagnetic calorimeter, specified by
$\Delta E/E=0.02/\sqrt{E(\gev)}\oplus0.005\oplus 0.2/E(\gev)$.
The electron resolution improves still further and the $\epem$ channel
yields much smaller resolution and errors than the $\mupmum$ channel.
Systematic error of $10\%$ for $\gamr$ is assumed.
\item
The resolution $\gamr$ for the Higgs mass peak in the $b\anti b$
and $\wp\wm$ final states (we weight according to branching ratio)
has been studied systematically as a function of $\mhsm$. The result
for the NLC detector specified in Ref.~\cite{nlc} can be parameterized as
$\gamr=4.86-0.019\mhsm+0.964\cdot 10^{-4}\mhsm^2-0.103\cdot 10^{-6}\mhsm^3$.
Typically $\gamr$ is of order $4\gev$, as illustrated in
Fig.~\ref{figmassresol}.
Systematic error of $10\%$ for $\gamr$ is assumed.
\end{enumerate}
Sensitivity to Higgs widths becomes possible when $\gamhsm$ is not
too much smaller than $\gamr$; some benchmarks are (see Fig.~\ref{hwidths})
$\gamhsm\sim 17\mev,32\mev,400\mev,1\gev,4\gev,10\gev$ for 
$\mhsm\sim 150,155,170,190,245,300\gev$, respectively.
Results for recoil mass $\gamr$'s are potentially sensitive to
beamstrahlung, bremsstrahlung and beam energy smearing. We shall
assume that these effects are small. The JLC studies of Ref.~\cite{kawagoe}
show that they are clearly so if one has small 0.4\% full width
beam energy spread and runs at $\rts \sim\mz+\mhsm+20\gev$.

In order to compute the error in the $\gamhsm$ measurement
given a value for the event-by-event resolution $\gamr$,
one proceeds as follows.
The convolution of the Higgs gaussian and the resolution
gaussian yields a gaussian of effective width
$\gameff=\sqrt{[\gamhsm]^2+[\gamr]^2}$. Assuming
small background, the statistical accuracy with
which $\gameff$ can be measured is $\Delta\gameff^{\rm stat}=\gameff/\sqrt{2N}$
where $N$ is the number of events in the Higgs mass peak.  
The systematic error in $\gameff$
coming from the systematic uncertainty $\Delta\gamr^{\rm sys}$ in $\gamr$ is
$\Delta\gameff^{\rm sys}=\Delta\gamr^{\rm sys}\gamr/\gameff$.
Adding in quadrature, we have a total $\Delta\gameff=\sqrt{[\Delta\gameff^{\rm
stat}]^2+[\Delta\gameff^{\rm sys}]^2}$. The relationship between
this and the $\Delta\gamhsm$ error in $\gamhsm$ is:
$\Delta\gamhsm=[\gameff/\gamhsm]\Delta\gameff$.
(For very small $\gamhsm$, this error becomes ill-defined and it is
$[\gamhsm]^2$ that is more appropriately studied; however, for
masses such that $\gamhsm$ is, indeed, resolvable, the result
obtained by the above procedure is valid.)

\begin{figure}[h]
\leavevmode
\begin{center}
\centerline{\psfig{file=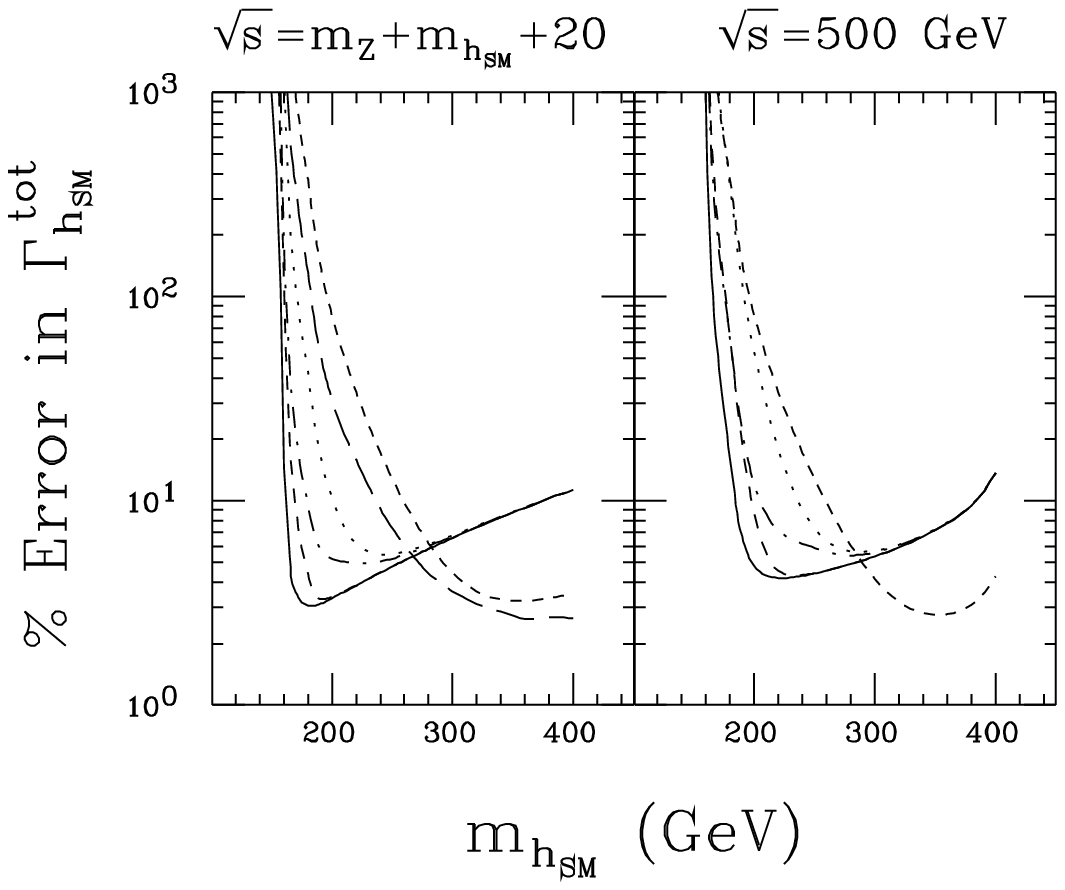,width=3.5in}}
\end{center}
\caption{Accuracy (including systematic as well as statistical errors)
with which $\gamhsm$ can be {\it directly} 
measured for $\protect\rts=\mz+\mhsm+20\gev$
and $\protect\rts=500\gev$ with luminosity times efficiency of 
$L=120\fbi$ using the $Z\hsm$ production mode at the NLC
\protect\cite{rickjack}. Results are given for the five cases
described in the text: 1=solid; 2=regular dashes; 3=dots; 4=dot dashes;
5=short dashes. Bremsstrahlung, beamstrahlung, and beam energy smearing are 
assumed unimportant compared to the contributions of tracking
and calorimetry to $\gamr$. Also shown in the first window
(very long dashes) are estimated errors for 
$\gamhsm$ using the $ZZ^{(*)}\to 4\ell$ mode with $L=600\fbi$
for ATLAS+CMS at the LHC.}
\label{figdgam}                                             
\end{figure}

The implications of the event-by-event mass resolutions 
in the five cases will now be described \cite{rickjack}.  
We assume integrated luminosity of $L=200\fbi$ at the energies
$\rts=\mz+\mhsm+20\gev$ and $\rts=500\gev$.  An overall detection
and acceptance efficiency of 60\% is employed.
$\br(Z\to\epem)=\br(Z\to\mupmum)=0.0336$ is employed in cases 1-4
and $\br(Z\to\epem+\mupmum+q\anti q)=0.7672$ is employed in case 5.
The resulting percentage errors
in $\gamhsm$ as a function of Higgs mass are plotted in Fig.~\ref{figdgam}.

We see from the figure that a reasonable accuracy
of \eg\ $\pm 20\%$ for $\gamhsm$ is achieved in cases 1,2,3,4,5
at $\mhsm\sim163,165,187,170,235\gev$ assuming $\rts=\mz+\mhsm+20\gev$
and at $\mhsm\sim 178,189,218,192,235\gev$ assuming $\rts=500\gev$,
respectively. These are the Higgs masses at which $\gamhsm$ becomes of
order $\gamr$, as one might naively anticipate.
We see immediately the importance of optimizing luminosity
for $\rts=\mz+\mhsm+20\gev$ and also having either excellent tracking
or excellent calorimetry if the Higgs mass happens to be in the $\sim
160\gev$ to $\sim190\gev$ range. For $\mhsm=190-220\gev$, running 
at $\rts=500\gev$
would allow a 20\% measurement of $\gamhsm$ if we have excellent tracking
or excellent calorimetry. However, if we only have the `standard' tracking
and calorimetry of case 3, then a 20\% measurement in the $\mhsm=190-220\gev$ 
range would require optimizing luminosity at the lower $\rts=\mz+\mhsm+20\gev$
energy. The case 5 results show that
the increased statistics from being able to include $q\anti q$ as well
as $\epem,\mupmum$ decays of the $Z$ when using the Higgs decay final
state to reconstruct the mass peak becomes important for $\mhsm\gsim 270\gev$.

If the errors for the direct measurement of $\gamhsm$ are compared
with those for the indirect determination assuming
$\rts=500\gev$, we see that super-JLC tracking
resolution of cases 1 or 2 would make the direct measurement errors competitive
with those from the indirect determination for $\mhsm\gsim 180\gev$.
The total error on $\gamhsm$ would be significantly
improved by combining the direct and indirect determinations 
for $\mhsm$ in the $180-190\gev$ range: combined error would be of order 
$\pm 10\%,\pm 6\%,\pm 4.6\%$ at $\mhsm=180,190,200\gev$, respectively.
Below this mass range, the indirect determination is much the better,
while above this mass range the direct determination has by far the smaller
error.  The mass range of the cross over would move to lower masses
for running at $\rts=\mz+\mhsm+20\gev$. In either case, we would obtain
a very important improvement over indirect
determination errors in a mass region where
$\gamhsm$ cannot be precisely measured via $s$-channel scanning at the FMC.
The above should be contrasted with the situation for the 'standard'
tracking/calorimetry of case 3, where we
are left with a $\mhsm$ region in which neither direct nor indirect errors 
are good, the direct measurement errors only 
becoming competitive with indirect errors for $\mhsm\gsim 250\gev$.

\bigskip
\begin{center}
\underline{Final-state mass peak determination of $\gamhsm$: LHC}
\end{center}
\smallskip

Measurement of $\gamhsm$ in the $gg\to \hsm\to ZZ^{(*)}\to 4\ell$ mode
at the LHC will also be possible.  For our estimates we have taken
a $4\ell$ resolution of $\gamr=1.25\%\mhsm$, which approximates the ATLAS
resolutions quoted in Table~29 of Ref.~\cite{atlas48}
in the $\mhsm\leq 180\gev$ mass region.
The $L=100\fbi$ event rates from Table~29 ($\mhsm\leq 180\gev$) and Table~38
($\mhsm\geq200\gev$) appearing in Ref.~\cite{atlas48} have been rescaled
to $L=600\fbi$ (assuming CMS and ATLAS will have similar resolution)
and a systematic error of $\pm 10\%$ for $\gamr$ is incorporated in
quadrature. The background rate given in Tables~29 and (especially) 38 
is always small compared to the signal rate and can be neglected
in computing the error in $\gamhsm$. Following the same error estimation
procedures as outlined in the NLC subsection, we arrive at the
results plotted as the long dashes in the first window of Fig.~\ref{figdgam}.
The expected error for $\gamhsm$ has a similar mass dependence to
that obtained via hadronic final state reconstruction at the NLC, but
is uniformly smaller for the assumed integrated luminosity. The excellent
$4\ell$ mass resolution expected for ATLAS and CMS is crucial for
this favorable result.  If both LHC and NLC results are available,
then it will be useful to combine results to improve the error. Even so,
error for $\gamhsm$ below $20\%$ using reconstruction of the $\hsm$ 
resonance peak
in decay final states only becomes possible once $\mhsm\gsim 210\gev$.

\subsubsection{Partial widths using $\gamhsm$}

In this section, we focus on results obtained using NLC data,
FMC data, or a combination thereof. (It is important to recall
our convention that the notation NLC means $\rts=500\gev$ running
in $\epem$ or $\mupmum$ collisions, while FMC refers explicitly
to $s$-channel Higgs production in $\mupmum$ collisions.)
Due to lack of time, LHC data has generally not been incorporated.
The only exception is that the error on $\br(\hsm\to\gam\gam)$
is estimated after including the (NLC+LHC) determination. This is
particularly crucial in obtaining a reasonable error for
the indirect determination of $\gamhsm$ when $\mhsm\lsim 130\gev$.

We have seen that determinations of the $(ZZ\hsm)^2$,
$(WW\hsm)^2$ and $(\gam\gam\hsm)^2$ couplings-squared are possible
in $\rts=500\gev$ NLC running without employing $\gamhsm$.
However, determination of $(b\anti b\hsm)^2$ is only possible
by determining $\gamhsm$ and then employing Eq.~(\ref{partialw}).
This procedure can also be used for $(c\anti c\hsm)^2$,
but it turns out that it is statistically better to compute
$(c\anti c\hsm)^2$ using $(b\anti b\hsm)^2$ and the experimental
determination of their ratio (described earlier).
Finally, by using the FMC-scan determination of $\gamhsm$ and
Eq.~(\ref{partialw}), we can obtain determinations of 
$(WW\hsm)^2$ and $(\gam\gam\hsm)^2$ from $\br(\hsm\to W\wstar)$ and
$\br(\hsm\to \gam\gam)$ (as measured in $\rts=500\gev$
running), respectively, which are independent of the NLC determinations
of these same quantities made directly without use of $\gamhsm$.

\bigskip
\centerline{\underline{$(b\anti b\hsm)^2$ and $(c\anti c\hsm)^2$: 
NLC only or NLC+FMC data}}
\smallskip

Given a determination of $\gamhsm$, we can employ 
Eq.~(\ref{partialw}) and the determination
of $\br(\hsm \to b\anti b)$ (which has
reasonable accuracy for $\mhsm\lsim150\gev$) to determine
$\Gamma(\hsm\to b\anti b)$ (equivalent
to determining the $(b\anti b\hsm)^2$ squared coupling). The 
expected $(b\anti b\hsm)^2$
errors using the indirect $\gamhsm$-determination errors are
listed in Table~\ref{nlcerrors}.  
They are not especially good,
primarily because of the large $\gamhsm$ errors. The $(c\anti c\hsm)^2$
coupling-squared can be computed 
either from $(b\anti b\hsm)^2$ and the $(c\anti c\hsm)^2/(b\anti b\hsm)^2$
measurement or from $\br(\hsm\to c\anti c)$
and Eq.~(\ref{partialw}).  Either way, the errors
are ultimately dominated by those for $\gamhsm$. Thus, using NLC data
only, the $(c\anti c\hsm)^2$ errors will be essentially the same
as those for $(b\anti b\hsm)^2$.

The $(b\anti b\hsm)^2$ errors
are greatly improved for $\mhsm\lsim 140\gev$ by using the
FMC-scan determination of $\gamhsm$ in conjunction with the
$\rts=500\gev$, $L=200\fbi$ $\br(\hsm\to b\anti b)$ errors \cite{bbghnew}.
Combining the FMC-scan determination
of $\gamhsm$ with the indirect NLC determination of $\gamhsm$ to 
minimize the $\gamhsm$ error
and then computing $\Gamma(\hsm\to b\anti b)$
yields the errors $(b\anti b\hsm)^2|_{\rm NLC+FMC}$ 
tabulated in Table~\ref{nlcfmcerrors}. The corresponding $(c\anti c\hsm)^2$
errors as computed from $(b\anti b\hsm)^2$ and the
$(c\anti c\hsm)^2/(b\anti b\hsm)^2$ ratio using the errors for
the latter tabulated in Table~\ref{nlcerrors} are also listed. 
These are slightly
superior to those obtained if $(c\anti c\hsm)^2$ is computed
via Eq.~(\ref{partialw}) using the combined NLC+FMC $\gamhsm$ determination.

\smallskip
\centerline{\underline{$(\mupmum\hsm)^2$: NLC+FMC data}}
\smallskip

The very small errors for the FMC $s$-channel measurements of 
$\sigma(\mupmum\to\hsm)\br(\hsm\to b\anti b,W\wstar,W\zstar)$ \cite{bbgh}
are summarized in Table~\ref{fmcsigbrerrors}.\footnote{Recall
that the FMC $s$-channel errors quoted are for $L=50\fbi$, the
amount of luminosity exactly on the $\rts=\mhsm$ Higgs peak
that is roughly equivalent to the on-peak and off-peak luminosity
accumulated in performing the scan determination of $\gamhsm$.}
As noted in the associated discussion, a measurement of
$\sigma(\mupmum\to\hsm)\br(\hsm\to X)$ is readily converted to an equally
accurate determination of $\Gamma(\hsm\to\mupmum)\br(\hsm\to X)$.
Given these measurements, there are four independent ways of combining
NLC data with the $s$-channel FMC
data to determine $\Gamma(\hsm\to\mupmum)$ \cite{bbghnew}.
\begin{description}
\item{ 1)} compute $\Gamma(\hsm\to\mupmum)=[\Gamma(\hsm\to\mupmum)\br(\hsm\to
b\anti b)]_{\rm FMC}/\br(\hsm\to b\anti b)_{\rm NLC}$;
\item{ 2)} compute $\Gamma(\hsm\to\mupmum)=[\Gamma(\hsm\to\mupmum)\br(\hsm\to
W\wstar)]_{\rm FMC}/\br(\hsm\to W\wstar)_{\rm NLC}$;
\item{ 3)} compute $\Gamma(\hsm\to\mupmum)=[\Gamma(\hsm\to\mupmum)\br(\hsm\to
Z\zstar)]_{\rm FMC}\gamhsm/\Gamma(\hsm\to Z\zstar)_{\rm NLC}$, where
the combined direct FMC plus indirect NLC determination of $\gamhsm$ can
be used since the NLC $(Z\zstar\hsm)^2$ determination was not used in
the indirect NLC determination of $\gamhsm$;
\item{ 4)} compute $\Gamma(\hsm\to\mupmum)=[\Gamma(\hsm\to\mupmum)\br(\hsm\to
W\wstar)\gamhsm]_{\rm FMC}/\Gamma(\hsm\to W\wstar)_{\rm NLC}$, where
we can only employ $\gamhsm$ as determined at the FMC since $(W\wstar\hsm)^2$
is used in the NLC indirect determination of $\gamhsm$.
\end{description}
The resulting (very small) errors for $(\mupmum\hsm)^2$ 
obtained by combining determinations from all four techniques are 
labelled $(\mupmum\hsm)^2|_{\rm NLC+FMC}$ and tabulated
in Table~\ref{nlcfmcerrors}.

\smallskip
\centerline{\underline{$(WW\hsm)^2$ and $(\gam\gam\hsm)^2$: NLC+FMC data}}
\smallskip

In Table~\ref{nlcerrors} we summarized the errors for the $(WW\hsm)^2$
coupling squared coming from determining the $\nu\anti\nu\hsm$
cross section from $\rts=500\gev$ running at the NLC.
We \cite{bbghnew} can
obtain a second independent determination of $(WW\hsm)^2$
by taking $\br(\hsm\to W\wstar)$ (as determined in $Z\hsm$ and $\epem\hsm$
production at the NLC) and multiplying by $\gamhsm$ as determined 
by $s$-channel scanning at the FMC ---
the NLC $\gamhsm$ determination employs the $W\wstar$ branching ratio in the
relevant mass region and cannot be used as part of a statistically independent
determination. These errors are summarized in Table~\ref{nlcfmcerrors}
using the notation $(WW\hsm)^2|_{\rm FMC}$. If we combine the
two different determinations, then we get the errors denoted
$(WW\hsm)^2|_{\rm NLC+FMC}$. (Results are not quoted for $\mhsm\leq
130\gev$, for which $\br(\hsm\to W\wstar)$ is too poorly
measured for this procedure to yield any improvement
over the errors of Table~\ref{nlcerrors}.)

Also given in Table~\ref{nlcerrors} were the errors for $(\gam\gam\hsm)^2$
coming from combining NLC $\gam\gam$ collider data with $b\anti b$
and $W\wstar$ branching ratios as measured at the NLC. In close
analogy to the $W\wstar$ procedure given above, we \cite{bbghnew} can
obtain a second independent determination of $(\gam\gam\hsm)^2$
by taking $\br(\hsm\to \gam\gam)$ (as determined using LHC and
$\nu\anti\nu\hsm$ NLC data) and multiplying by $\gamhsm$ as determined 
by $s$-channel scanning at the FMC ---
the NLC $\gamhsm$ determination employs the $\gam\gam$ branching ratio in the
relevant mass region and cannot be used as part of a statistically independent
determination. The resulting errors are summarized in Table~\ref{nlcfmcerrors}
using the notation $(\gam\gam\hsm)^2|_{\rm FMC}$. If we combine the
two different determinations, then we get the errors denoted
$(\gam\gam\hsm)^2|_{\rm NLC+FMC}$.

One last point concerning $\Gamma(\hsm\to\gam\gam)$ is worth noting.
At the FMC, a $\gam\gam$ collider is not possible \cite{bbgh},
Only the $(\gam\gam\hsm)^2_{\rm FMC}$ determination of this potentially
very revealing coupling would be available.

\begin{table}[h]
\caption[fake]{Summary of approximate errors for branching ratios,
coupling-squared ratios, and couplings-squared as
determined using $L=200\fbi$ of data accumulated in $\protect\rts=500\gev$
running at the NLC.
For $\br(\hsm\to\gam\gam)$, but {\it not} $(\gam\gam\hsm)^2/(b\anti b\hsm)^2$,
we have combined the NLC $\rts=500\gev$ results
with results obtained using LHC data; the net accuracy so obtained
for $\br(\hsm\to\gam\gam)$ is also reflected in the errors quoted
for the determination of $\gamhsm$ following the indirect procedure.
The errors for $\Gamma(\hsm\to\gam\gam)$ quoted are for $L=50\fbi$
accumulated in $\gam\gam$ collisions while running at 
$\rts_{\epem}\sim \mhsm/0.8$,
and are those employed in the indirect $\gamhsm$ determination.
A $-$ indicates large error and a $?$ indicates either that a reliable
simulation or estimate is not yet available or that the indicated
number is a very rough estimate.}
\footnotesize
\begin{center}
\begin{tabular}{|c|c|c|c|c|}
\hline
 Quantity & \multicolumn{4}{c|}{Errors} \\
\hline
\hline
{$\bf\mhsm$}{\bf (GeV)} & { \bf80} & { \bf100} & { \bf 110} & {\bf 120} \\
\hline
 $(c\anti c\hsm)^2/(b\anti b\hsm)^2$ & \multicolumn{4} {c|}{$\sim\pm7\%$} \\
\hline
 $(WW\hsm)^2/(b\anti b\hsm)^2$ & $-$ & $-$ & $-$  & $\pm 23\%$ \\
\hline
 $(\gam\gam \hsm)^2/(b\anti b\hsm)^2$ & $\pm 42\%$ & $\pm 27\%$ & $\pm 24\%$ &
 $\pm 22\%$ \\
\hline
 $(ZZ\hsm)^2$ & \multicolumn{4} {c|}{$\pm 3\%-\pm 4\%$} \\
\hline
 $\br(\hsm\to b\anti b)$ & \multicolumn{4}{c|}{$\pm5\%$} \\
\hline
 $\br(\hsm\to c\anti c)$ & \multicolumn{4}{c|}{$\sim\pm9\%$} \\
\hline
 $\br(\hsm\to W\wstar)$ & \multicolumn{4}{c|}{$-$} \\
\hline
 $(WW\hsm)^2$ & \multicolumn{4}{c|}{$\pm5\%$} \\
\hline
 $(ZZ\hsm)^2/(WW\hsm)^2$ & \multicolumn{4}{c|}{$\pm 6\%-\pm 7\%$}\\
\hline
 $\br(\hsm\to\gam\gam)$ & $\pm 15\%$ & $\pm 14\%$ & $\pm 13\%$ & $\pm 13\%$ \\
\hline
 $(\gam\gam\hsm)^2$ & \multicolumn{4}{c|}{$\sim \pm 12\%$}\\
\hline
 $\gamhsm$ (indirect) & $\pm 19\%$ & $\pm 18\%$ & $\pm 18\%$ & 
 $\pm 18\%$ \\
\hline
 $(b\anti b\hsm)^2$ & $\pm 20\%$ & $\pm 19\%$ & $\pm 18\%$ & 
 $\pm 18\%$ \\
\hline
\hline
{$\bf\mhsm$}{\bf (GeV)} & {\bf 130} & {\bf 140} & {\bf 150} & {\bf 170} \\
\hline
 $(c\anti c\hsm)^2/(b\anti b\hsm)^2$ & $\pm7\%$ & 
\multicolumn{3} {c|}{$?$} \\
\hline
 $(WW\hsm)^2/(b\anti b\hsm)^2$ & $\pm 16\%$ & $\pm 8\%$ & $\pm 7\%$ &
 $\pm 16\%$ \\
\hline
 $(\gam\gam \hsm)^2/(b\anti b\hsm)^2$ & $\pm 23\%$ & $\pm 26\%$ & $\pm 35\%$ &
 $-$ \\
\hline
 $(ZZ\hsm)^2$ & \multicolumn{4} {c|}{$\pm 4\%$} \\
\hline
 $\br(\hsm\to b\anti b)$ & \multicolumn{2}{c|}{$\pm6\%$} & $\pm 9\%$ & $\sim
 20\%?$ \\
\hline
 $\br(\hsm\to c\anti c)$ & $\sim\pm 9\%$ & \multicolumn{3}{c|}{$?$} \\
\hline
 $\br(\hsm\to W\wstar)$ & $\pm 16\%$ & $\pm 8\%$ & $\pm 6\%$ & $\pm 5\%$ \\
\hline
 $(WW\hsm)^2$ & $\pm 5\%$ & $\pm 5\%$ & $\pm 8\%$ & $\pm 10\%$ \\
\hline
 $(ZZ\hsm)^2/(WW\hsm)^2$ & $\pm 7\%$ & $\pm 7\%$ & $\pm 9\%$ & $\pm 11\%$ \\
\hline
 $\br(\hsm\to\gam\gam)$ & $\pm 13\%$ & $\pm 18\%?$ & $\pm 35\%$ & $-$ \\
\hline
 $(\gam\gam\hsm)^2$ & $\pm 15\%$ & $\pm 17\%$ & $\pm 31\%$ & $-$ \\
\hline
 $\gamhsm$ (indirect) & $\pm 13\%$ & $\pm 9\%$ & $\pm 10\%$ & 
 $\pm 11\%$ \\
\hline
 $(b\anti b\hsm)^2$ & $\pm 14\%$ & $\pm 11\%$ & $\pm 13\%$ & 
 $\pm 23\%$ \\
\hline
\hline
{$\bf\mhsm$}{\bf (GeV)} & {\bf 180} & {\bf 190} & {\bf 200} & {\bf 300} \\
\hline
 $(ZZ\hsm)^2$ & \multicolumn{2} {c|}{$\pm 4\%-\pm5\%$}& $\pm 6\%$ & $\pm 9\%$ \\
\hline
 $(WW\hsm)^2$ & $\pm 11\%$ & $\pm 12\%$ & $\pm 13\%$ & $\pm 24\%$ \\
\hline
 $(ZZ\hsm)^2/(WW\hsm)^2$ & $\pm 12\%$ & $\pm 13\%$ & $\pm 14\%$ & $\pm 25\%$ \\
\hline
 $\br(\hsm\to WW)$ & $\pm 6\%$ & $\pm 7\%$ & $\pm 8\%$ & $\pm 14\%?$ \\
\hline
 $(\gam\gam\hsm)^2$ & $\pm 13\%$ & $\pm 12\%$ & $\pm 12\%$ &  $\pm 22\%$ \\
\hline
 $\gamhsm$ (indirect) & $\pm 13\%$ & $\pm 14\%$ & $\pm 15\%$ & 
 $\pm 28\%$ \\
\hline
\end{tabular}
\end{center}
\label{nlcerrors}
\end{table}

\begin{table}[h]
\caption[fake]{Summary of approximate errors for 
coupling-squared ratios and $\gamhsm$ in the case of 
$s$-channel Higgs production at the FMC, assuming $L=200\fbi$
total scan luminosity (which for rate
measurements in specific channels is roughly equivalent to $L=50\fbi$ 
at the $\rts=\mhsm$ peak). Beam resolution of $R=0.01\%$ is assumed.
A $-$ indicates large error and a $?$ indicates either that a reliable
simulation or estimate is not yet available or that the indicated
number is a very rough estimate.}
\footnotesize
\begin{center}
\begin{tabular}{|c|c|c|c|c|}
\hline
 Quantity & \multicolumn{4}{c|}{Errors} \\
\hline
\hline
{$\bf\mhsm$}{\bf (GeV)} & {\bf 80} & {\bf $\mz$} & {\bf 100} & {\bf 110} \\
\hline
$(W\wstar\hsm)^2/(b\anti b\hsm)^2$ & $-$ & $-$ & $\pm 3.5\%$ & $\pm 1.6\%$ \\
\hline
$(Z\zstar\hsm)^2/(b\anti b\hsm)^2$ & $-$ & $-$ & $-$ & $\pm 34\%$ \\
\hline
$(Z\zstar\hsm)^2/(W\wstar\hsm)^2$ & $-$ & $-$ & $-$ & $\pm 34\%$ \\
\hline
 $\gamhsm$ & $\pm 2.6\%$ & $\pm 32\%$ & $\pm 8.3\%$ & 
  $\pm 4.2\%$ \\
\hline
\hline
{$\bf\mhsm$}{\bf (GeV)} & {\bf 120} & {\bf 130} & {\bf 140} & {\bf 150} \\
\hline
$(W\wstar\hsm)^2/(b\anti b\hsm)^2$ & 
 $\pm 1\%$ & $\pm 0.7\%$ & $\pm 0.7\%$ & $\pm 1\%$ \\
\hline
$(Z\zstar\hsm)^2/(b\anti b\hsm)^2$ & 
 $\pm 6\%$ & $\pm 3\%$ & $\pm 2\%$ & $\pm 2\%$ \\
\hline
$(Z\zstar\hsm)^2/(W\wstar\hsm)^2$ & 
 $\pm 6\%$ & $\pm 3\%$ & $\pm 2\%$ & $\pm 2\%$ \\
\hline
 $\gamhsm$ & $\pm 3.6\%$ & $\pm 3.6\%$ & $\pm 4.1\%$ &
  $\pm 6.5\%$ \\
\hline
\end{tabular}
\end{center}
\label{fmcerrors}
\end{table}

\begin{table}[h]
\caption[fake]{Summary of approximate errors for branching ratios,
coupling-squared ratios and couplings-squared 
obtained by combining the results of Tables~\ref{nlcerrors}
and \ref{fmcerrors}. See text for further discussion.
A $-$ indicates large error and a $?$ indicates either that a reliable
simulation or estimate is not yet available or that the indicated
number is a very rough estimate.}
\footnotesize
\begin{center}
\begin{tabular}{|c|c|c|c|c|}
\hline
 Quantity & \multicolumn{4}{c|}{Errors} \\
\hline
\hline
{$\bf\mhsm$}{\bf (GeV)} & {\bf 80} & {\bf 100} & {\bf 110} & {\bf 120} \\
\hline
 $(b\anti b\hsm)^2|_{\rm NLC+FMC} $ & $\pm6\%$ & $\pm 9\%$ & $\pm 7\%$ &
  $\pm6\%$ \\
\hline
 $(c\anti c\hsm)^2|_{\rm NLC+FMC} $ & $\pm9\%$ & $\pm 11\%$ & $\pm 10\%$ &
  $\pm9\%$ \\
\hline
 $(\mupmum\hsm)^2|_{\rm NLC+FMC}$ & 
$\pm 5\%$ & $\pm 5\%$ & $\pm 4\%$ & $\pm 4\%$ \\
\hline
 $(\gam\gam\hsm)^2|_{\rm FMC}$ & $\pm 15\%$ & $\pm 16\%$ & $\pm 14\%$ &
 $\pm 13\%$ \\
\hline
 $(\gam\gam\hsm)^2|_{\rm NLC+FMC}$ & $\pm 9\%$ & $\pm 10\%$ & $\pm 9\%$ &
 $\pm 9\%$ \\
\hline
\hline
{$\bf\mhsm$}{\bf (GeV)} & {\bf 130} & {\bf 140} & {\bf 150} & {\bf 170} \\
\hline
 $(b\anti b\hsm)^2|_{\rm NLC+FMC}$ & $\pm7\%$ & $\pm7\%$ & $\pm10\%$ &
 $\pm23\%$ \\
\hline
 $(c\anti c\hsm)^2|_{\rm NLC+FMC} $ & $\pm10\%$ & \multicolumn{3}{c|}{$?$} \\
\hline
 $(\mupmum\hsm)^2|_{\rm NLC+FMC}$ & 
$\pm 3\%$ & $\pm 3\%$ & $\pm 4\%$ &  $\pm 10\%$ \\
\hline
 $(W\wstar\hsm)^2|_{\rm FMC}$ & $\pm 16\%$ & $\pm 9\%$ & $\pm 9\%$ &
 $-$ \\
\hline
 $(W\wstar\hsm)^2|_{\rm NLC+FMC}$ & $\pm 5\%$ & $\pm 4\%$ & $\pm 6\%$ &
 $\pm 10\%$ \\
\hline
 $(\gam\gam\hsm)^2|_{\rm FMC}$ & $\pm 14\%$ & $\pm 18\%$ & $\pm 36\%$ & $-$ \\
\hline
 $(\gam\gam\hsm)^2|_{\rm NLC+FMC}$ & 
    $\pm 10\%$ & $\pm 13\%$ & $\pm 23\%$ & $-$ \\
\hline
\end{tabular}
\end{center}
\label{nlcfmcerrors}
\end{table}

\subsubsection{Summary Tables}

We present in Tables~\ref{nlcerrors}, \ref{fmcerrors},
and \ref{nlcfmcerrors} 
a final summary of the errors that can be achieved for fundamental $\hsm$
properties (other than the mass) in three different situations:
\begin{itemize}
\item
$L=200\fbi$ devoted to $\rts=500\gev$ running at the NLC
supplemented with $L=50\fbi$ of $\gam\gam$
collider data collected while running at 
$\rts_{\epem}\sim \mhsm/0.8$ and the (LHC+NLC) determination
of $\br(\hsm\to\gam\gam)$;
\item
A total $L=200\fbi$ of luminosity devoted to scanning the Higgs peak
to determine $\gamhsm$ --- as explained earlier, specific channel rate
errors are equivalent to those that would be obtained by devoting
$L=50\fbi$ to the Higgs peak at $\rts=\mhsm$;
\item
combining the above two sets of data.
\end{itemize}
The results we have obtained depend strongly on detector
parameters and analysis techniques and in some cases (those marked
by a ?) were obtained by extrapolation rather than full simulation.
Nonetheless, these results should serve as an illustration of what
might ultimately be achievable on the basis of NLC 
$\rts=500\gev$ running and/or FMC $s$-channel data.
Results for FMC $s$-channel errors assume very excellent $0.01\%$ beam energy
resolution and the ability to measure the beam energy with
precision on the order of 1 part in $10^6$.
Due to lack of time, except for the determination
of $\br(\hsm\to\gam\gam)$ and implications for $\gamhsm$,
we have not explored the undoubted benefits
that would result from combining NLC/FMC data with LHC data.
Such a study is in progress. 

Of course, it should not be forgotten that the
$\rts=500\gev$ data could also be obtained by running an FMC with a final
ring optimized for this energy. (Confirmation 
that the FMC can achieve the same precisions as the NLC when
run at $\rts=500\gev$ must await a full machine and detector design;
it could be that the FMC backgrounds and detector design will
differ significantly from those employed in the $\rts=500\gev$ studies reported
here.) However,
it should be apparent from comparing Tables~\ref{nlcerrors}, \ref{fmcerrors}
and \ref{nlcfmcerrors} that if there is a SM-like
Higgs boson in the $\mhsm\lsim 2\mw$ mass region (as
expected in supersymmetric models) then 
it is very advantageous to have $L=200\fbi$
of data from both $\rts=500\gev$ running and from an FMC $s$-channel scan
of the Higgs resonance. Thus, the importance of
obtaining a full complement of Higgs boson data on a reasonable time scale
argues for having either an NLC plus a FMC or two FMC's. A single FMC
with two final rings --- one optimized for $\rts=\mhsm$
and one for $\rts=500\gev$ --- would suffice, but take twice
as long (8 years at $L_{\rm year}=50\fbi$) to accumulate the necessary data.

\subsection{Measuring $\mhsm$ at TeV33, LHC and NLC}

In our discussion, we will focus on the $\mhsm\leq 2\mw$ mass region,
but give some results for higher masses.
In the $\mhsm\leq 2\mw$ region,
measurement of the Higgs boson mass at the LHC and/or NLC
will be of great practical importance for the FMC since it
will enable a scan of the Higgs resonance 
peak with minimal luminosity wasted on
locating the center of the peak. Ultimately the accuracy
of the Higgs mass measurement will impact precision tests of
loop corrections, both in the SM and in extended models such as the MSSM.
For example, in the minimal supersymmetric standard model,
the prediction for
the mass of the light SM-like $\hl$ to one loop is \cite{dpfreport}:
\begin{eqnarray}
\mhl^2&=&{1\over
2}\Bigl[\mha^2+\mz^2
 -\biggl\{(\mha^2+\mz^2)^2 \nonumber \\
& & -4\mha^2\mz^2\cos^22\beta\biggr\}^{1/2}~\Bigr]
  + \Delta\mhl^2\,,
\label{mhlform}
\end{eqnarray}
where
$\Delta\mhl^2=3g^2\mt^4\ln\left(\mstop^2/\mt^2\right)/[8\pi^2\mw^2]$. Here,
$\mstop$ is the top-squark mass and
we have simplified by neglecting top-squark mixing and non-degeneracy.
From Eq.~(\ref{mhlform}), one can compute $d\mhl/d\mha$, $d\mhl/d\tanb$,
$d\mhl/d\mt$, and $d\mhl/d\mstop$ for a given choice of input parameters.
These derivatives determine the sensitivity of these parameters to the error
in $\mhl$.  For example, for $\mha=200\gev$, $\mstop=260\gev$, $\tanb=14$
and $\mt=175\gev$, for which $\mhl=100\gev$, we find that
a $\pm 100\mev$ measurement of
$\mhl$ (a precision that should be easily achieved, as discussed below)
would translate into constraints (for variations of one variable
at a time) on $\mha$, $\tanb$, 
$\mt$ and $\mstop$ of about $\pm 37\gev$, $\pm 0.7$, $\pm 670\mev$
and $\pm 1\gev$, respectively.  Since $\mt$ will be known to much
better accuracy than this and (for such low $\mha$)
the $\ha$ would be observed and its
mass measured with reasonable accuracy, 
the determination of $\mhl$ would be used as a joint constraint on
$\mstop$ and $\tanb$. More generally, 
squark mixing parameters should be included in the analysis.
The challenge will be to compute higher loop corrections to $\mhl$
to the $\pm 100\mev$ level.

Determination of $\mhsm$ will proceed
by examining a peaked mass distribution 
constructed using the measured momenta of particles appearing
in the final state. At TeV33 and the LHC, these will be the particles
into which the Higgs boson decays. For $Z\hsm$ production at the NLC,
there are two possibilities; 
we may employ the $Z\to \ell^+\ell^-$ decay products and
reconstruct the recoil mass peak or we may directly reconstruct
the Higgs mass from its decay products, as outlined in the discussion
associated with determining $\gamhsm$. 
The accuracy of the Higgs boson mass
determination will depend upon the technique/channel, 
the detector performance and
the signal and background statistics.  

If the background
under the peak is small, then the accuracy of the mass
measurement is given by $\Delta\mh\sim\gamr/\sqrt S$, where $\gamr$
is the natural (Gaussian) mass resolution of the re-construction and $S$
is the total number of events in the mass peak.\footnote{As always, 
our notation is that $\Delta X$ represents
the absolute magnitude of the $1\sigma$ error on the quantity
$X$; that is the $1\sigma$ limits on $X$ are $X\pm\Delta X$.}
The background at the NLC is generally sufficiently small that this
is a good approximation. At the LHC, the background level is small
(after cuts) in the $4\ell$ final state of $\hsm\to ZZ^{(*)}$ decay.
But, in the inclusive production $2\gam$ final state mode
the background is much larger than the signal 
and in the associated $W\hsm+t\anti t \hsm\to
\ell\nu 2\gam X$ modes the background and signal event rates are approximately
equal (after cuts).
If we assume that the background is constant under
the Higgs peak, that the signal 
peak is Gaussian with width $\gamr$, and that we examine the portion
of the mass peak lying between $\mh-n\gamr$ and $\mh+n\gamr$,
then one can demonstrate that the statistical error in $\mh$ is
\begin{equation}
\Delta\mh^{\rm stat}={\gamr\over \sqrt S}\left[c(n)+{n^2 B\over 3 S}\right]\,,
\label{delmhb}
\end{equation}
where $c(n)\equiv \int_{-n}^{+n} dx\,x^2\exp[-x^2/2]/
\int_{-n}^{+n} dx\,\exp[-x^2/2]$ and $S$ and $B$ are the total number
of signal and background events contained in the above-specified interval.
In our TeV33 and LHC estimates, we will employ $n=2$, for which $c(n)=0.774$.
All signal and background rates from tables given in the various TeV33
and LHC studies will be scaled (using Gaussian shape for the signal peak
and assuming a flat background) to the above value of $n$.

\bigskip
\centerline{\underline{LEP2, TeV33 and LHC}}
\smallskip

The first measurement of $\mhsm$
will probably take place at LEP2, the Tevatron, or the LHC.
At LEP2, the accuracy will be limited by statistics.
For example, at $\rts=192\gev$ and with $L=150\pbi$ for each
of the four experiments and summing over all channels, 
the number of signal and background
events will be roughly $S,B=250,100$ at $\mhsm=80\gev$ and
$S,B=180,150$ at $\mhsm=91\gev$ in a $n=2$ interval
\cite{janotcom}. A conservative expectation for the resolution in all channels
is $\gamr\sim 3\gev$ \cite{janotcom}. Using Eq.~(\ref{delmhb}),
these event numbers lead to $\Delta\mhsm\sim 250,400\mev$
at $\mhsm=80,91\gev$, respectively.

At the Tevatron, the primary discovery mode is $W\hsm$ with $\hsm\to b\anti b$.
We give $\delmhsm$ estimates for TeV33.
A detailed study of the accuracy with which $\mhsm$ can be
determined at Tev33 has not been performed, but we have estimated
the error from the mass plots and statistics of Ref.~\cite{kky}. 
Examining Fig.~1 of Ref.~\cite{kky} and comparing to the mass bins quoted
in the Table~I caption of Ref.~\cite{kky} one concludes that 
$\gamr\sim 10.0,12.5,13.8,16.3\gev$ at $\mhsm=60,80,100,120\gev$, respectively,
and that the accepted $b\anti b$ mass range corresponds to $n\sim 1.2$.
Rescaling the $L=10\fbi$ final $S$ and $B$
values of Table~IV \cite{kky} to $n=2$ and to an ultimate integrated
luminosity of $L=60\fbi$ (3 years for two detectors)
implies statistical errors of $\Delta\mhsm^{\rm stat}=
0.61,0.96,1.5,2.7\gev$ at $\mhsm=60,80,100,120\gev$, respectively. 
Allowing for systematic effects
at the level of $\Delta\mh^{\rm syst}=0.01\mh$, added in quadrature,
already increases these errors to $\Delta\mhsm^{\rm tot}=0.85,1.3,1.8,2.9\gev$,
respectively. It is clearly crucial that systematic effects be well controlled.

At the LHC, the  excellent $\gam\gam$ mass resolution planned
by both the ATLAS and CMS detectors implies that the best mass measurement 
in the $\mhsm\lsim 150\gev$ range will come from detection modes in
which $\hsm\to \gam\gam$; the production modes for which detection
in the $\gam\gam$ final state is possible are $gg\to\hsm$ inclusive
and $W\hsm,t\anti t\hsm$ associated production.
ATLAS $\mgamgam$ resolutions from Table~21 of Ref.~\cite{atlas48}
are $\gamr=1.07,1.16,1.25,1.30,1.34,1.43,1.52\gev$ at
$\mhsm=60,90,100,110,120,130,150\gev$, respectively.
The $\mgamgam$ resolution currently
claimed by CMS is of order $\gamr=\sim 0.7\%\mhsm$ at high luminosity.

ATLAS inclusive signal and background rates for $n=1.4$, $L=100\fbi$ 
appear in Table~21 of \cite{atlas48}. We have rescaled these to $n=2$
and $L=300\fbi$.
CMS inclusive signal rates have been estimated
for $L=100\fbi$ and $n=2$ by counting events
in the peaks of Fig.~12.3 of Ref.~\cite{CMS}:
the $L=100\fbi$, $n=2$ estimates are $S=1275,1700,1840,650$ 
at $\mhsm=90,110,130,150\gev$, respectively.  
The corresponding background rates have been computed using
the $S/\sqrt B$ values from Fig.~12.5 of \cite{CMS}
(after appropriate rescalings to account for the fact that the 
plotted $S/\sqrt B$ values are those for $n\sim 1.2$, \ie\ for
keeping about 75\% of the signal peak): the
$n=1.2$ values are $S/\sqrt B=6.5,10,13,8$ at $\mhsm=90,110,130,150\gev$,
respectively. The resulting $S$ and $B$ for $L=100\fbi$ are multiplied
by a factor of 3 to get $L=300\fbi$ rates.
The combined $W\hsm,t\anti t\hsm$ event rates in
the $\hsm\to\gam\gam$ final state for ATLAS at $L=100\fbi$ were taken
from Table~11.8 of Ref.~\cite{ATLAS}, namely $S=B=15$ for
$\mhsm=80,100,120\gev$. We assume these rates correspond to a bin of
size $n=2$.  CMS signal and background 
rates for the associated production modes were
obtained from the $L=165\fbi$, $n=2$ Table~12.3 of Ref.~\cite{CMS}.
The associated production $S$ and $B$
rates for both ATLAS and CMS are rescaled to $L=300\fbi$.  The 
statistical error in $\mhsm$
is then computed from Eq.~(\ref{delmhb}) for ATLAS/inclusive
ATLAS/associated, CMS/inclusive and CMS/associated, separately.
The net error $\delmhsm$ for each detector
is then computed by combining the associated and inclusive
results and then adding in a systematic error (in quadrature)
given by $\delmhsm^{\rm syst}=0.001\mhsm$ (the ATLAS estimate).
Finally, the net error is computed by combining the ATLAS and CMS
net errors.  

The result is that $\delmhsm\sim 90-110$~MeV 
for $\mhsm\lsim 130\gev$ with $\delmhsm\sim 150$~MeV for $\mhsm=150\gev$.
For example, at $\mhsm=100\gev$, we obtained the following $\delmhsm$ values:
\begin{equation}
\delmhsm=\left\{\begin{array}{lr}
\mbox{ATLAS/inclusive} & 204 \mev \\
\mbox{ATLAS/associated} & 270 \mev \\
\mbox{ATLAS/stat+syst} & 191 \mev \\
\mbox{CMS/inclusive} & 65 \mev \\
\mbox{CMS/associated} & 85 \mev \\
\mbox{CMS/stat+syst} & 111 \mev \\
\mbox{Total} & 96 \mev \\ \end{array}\right.
\label{delmhvalues}
\end{equation}
As one cross check on this computation, we \cite{gp} took the $S$ and $B$
numbers for $L=300\fbi$ at $\mhsm=100\gev$, for inclusive and associated
production separately, from ATLAS and then generated 100 experiments
throwing $S$ and $B$ according to Gaussian/Poisson statistics.  The background
subtraction was then made to get the signal peak, and the rms of the peak
position for the 100 experiments was computed.  $\Delta\mhsm^{\rm stat}$
was found to be 230 MeV for inclusive production and 246 MeV for associated
production.  Combining these with a 100 MeV systematic error gives
$\Delta\mhsm^{\rm tot}\sim 200\mev$. All these results are 
very similar to the above-quoted ATLAS numbers.  CMS statistical
errors are smaller by virtue of the better resolution; in fact, the assumed
0.1\% systematic uncertainty dominates the CMS statistical plus systematic
error.

For $\mhsm\gsim 130\gev$, $\mhsm$ can also be determined using the inclusive
$\hsm\to ZZ^{(*)}\to 4\ell$ final state.  Our inputs from Ref.~\cite{atlas48}
are the same as in the discussion of the $4\ell$-mode determination of 
$\gamhsm$. For $\mhsm\leq 180\gev$, we employ the $L=100\fbi$ signal
and background rates of Table~29 
and the corresponding value of $n=2$ (for which $c(n)=0.774$)
in Eq.~(\ref{delmhb}). For $\mhsm\geq 200\gev$, we employ the $L=100\fbi$
signal and background rates of Table~38 which effectively correspond
to $n=1.65$~\footnote{The table caption states 
that the accepted mass interval includes
90\% of the events, which for a Gaussian shape would imply $n\sim 1.65$.}
for which $c(n)=0.626$. All rates are scaled to $L=600\fbi$.
Further, we include in quadrature
a 1 per mil systematic uncertainty in the overall mass scale.
The resulting error for $\mhsm$ is in
the range $\Delta\mhsm\sim 60-120\mev$ for
$140\leq\mhsm\leq 400\gev$, except at $\mhsm\sim 170\gev$ where
$\Delta\mhsm\sim 270\mev$. For $\mhsm\geq 200\gev$, it is possible
that smaller error could be obtained for less stringent cuts
(implying larger signal rates, but also larger background)
than those employed in Table~38. We have not pursued this possibility.

The improvement in $\Delta\mhsm$ obtained by
combining the $\gam\gam$ and $4\ell$ mode determinations of $\mhsm$
is small since only the $\gam\gam$ ($4\ell$) mode gives small errors for
$\mhsm\lsim 130\gev$ ($\gsim 140\gev$).

\begin{figure}[h]
\leavevmode
\begin{center}
\centerline{\psfig{file=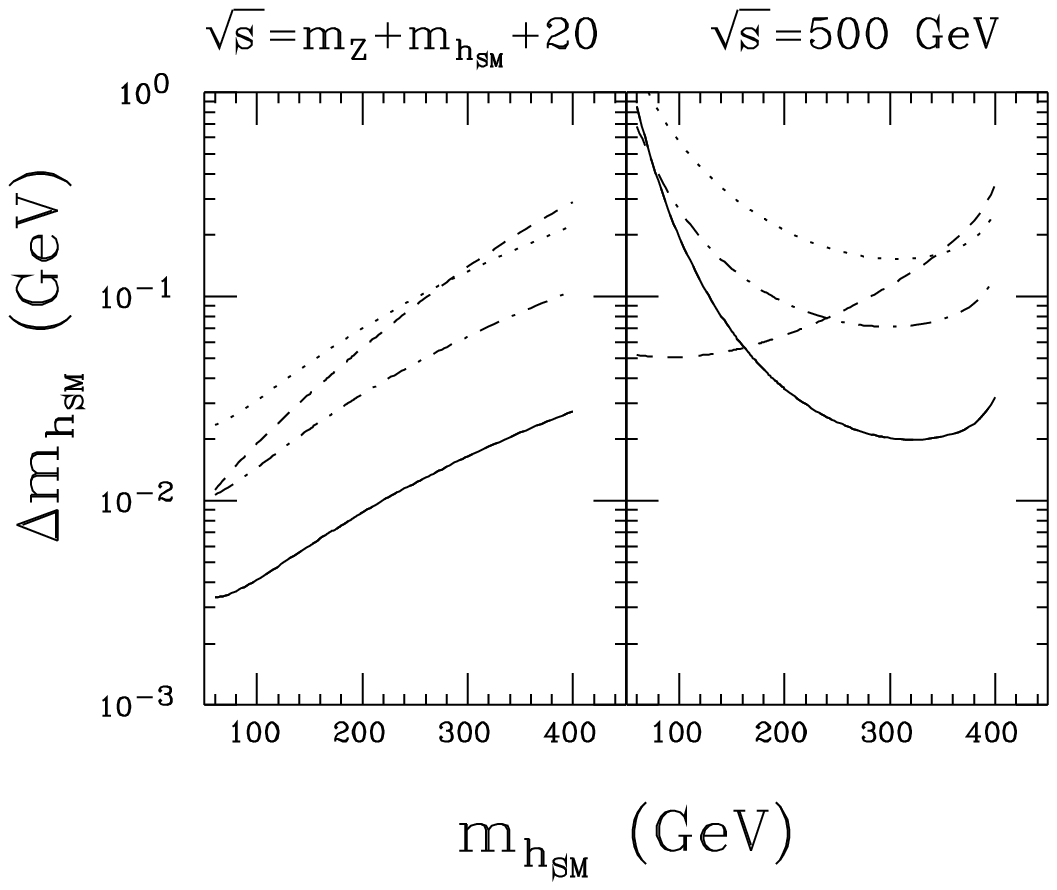,width=2.9in}}
\end{center}
\caption{The error $\Delta \mhsm$ 
for measurements at $\protect\rts=\mz+\mhsm+20\gev$
and $\protect\rts=500\gev$ with luminosity times efficiency of 
$L=120\fbi$ using the $Z\hsm$ production mode at the NLC
\protect\cite{rickjack}. Results are given for the five cases
described in association with directly measuring $\gamhsm$ --- 
see Fig.~\ref{figdgam}; 1=2=solid; 3=dots; 4=dot dashes;
5=short dashes.
Bremsstrahlung, beamstrahlung, and beam energy smearing are 
assumed unimportant compared to the contributions of tracking
and calorimetry to $\gamr$.}
\label{figdeltam}
\end{figure}

\bigskip
\centerline{\underline{NLC}}
\smallskip

At the NLC, we \cite{rickjack} consider the same five cases discussed earlier
with regard to directly determining $\gamhsm$ from the Higgs mass
peak in the $Z\hsm$ production mode. The resulting errors
for $\mhsm$ are plotted in Fig.~\ref{figdeltam}. (Cases 1 and 2
are indistinguishable, the systematic error in $\gamr$ not having
significant influence on $\Delta\mhsm$.) The results for
$\Delta\mhsm$ in case 5 (in which the Higgs peak
is reconstructed from the $\hsm\to b\anti b,\wp\wm$ final states
assuming hadronic calorimetry as defined in Ref.~\cite{nlc})
are probably too optimistic when $\mhsm$ is near $\mz$,
given that we have not included backgrounds in the estimates. Backgrounds
should be small in cases 1-4 since we demand quite precise reconstruction
of $Z\to\epem,\mupmum$ in the $Z\hsm$ final state, 
implying that the only background would be from $ZZ$
production where one of the $Z$'s decays leptonically. (For a sample
plot showing the small expected background level, 
see Fig.~2 of Ref.~\cite{kawagoe}.) Fig.~\ref{figdeltam} shows
that distinctly greater accuracy at the NLC is possible than
by using the $\gam\gam$ mode at the LHC, provided NLC systematic errors
are not substantial.
In all cases, for $\mhsm\lsim 300\gev$
running at $\rts=500\gev$ yields much larger $\Delta\mhsm$ than
running at $\rts\sim\mz+\mhsm+20\gev$.

\begin{figure}[h]
\leavevmode
\begin{center}
\centerline{\psfig{file=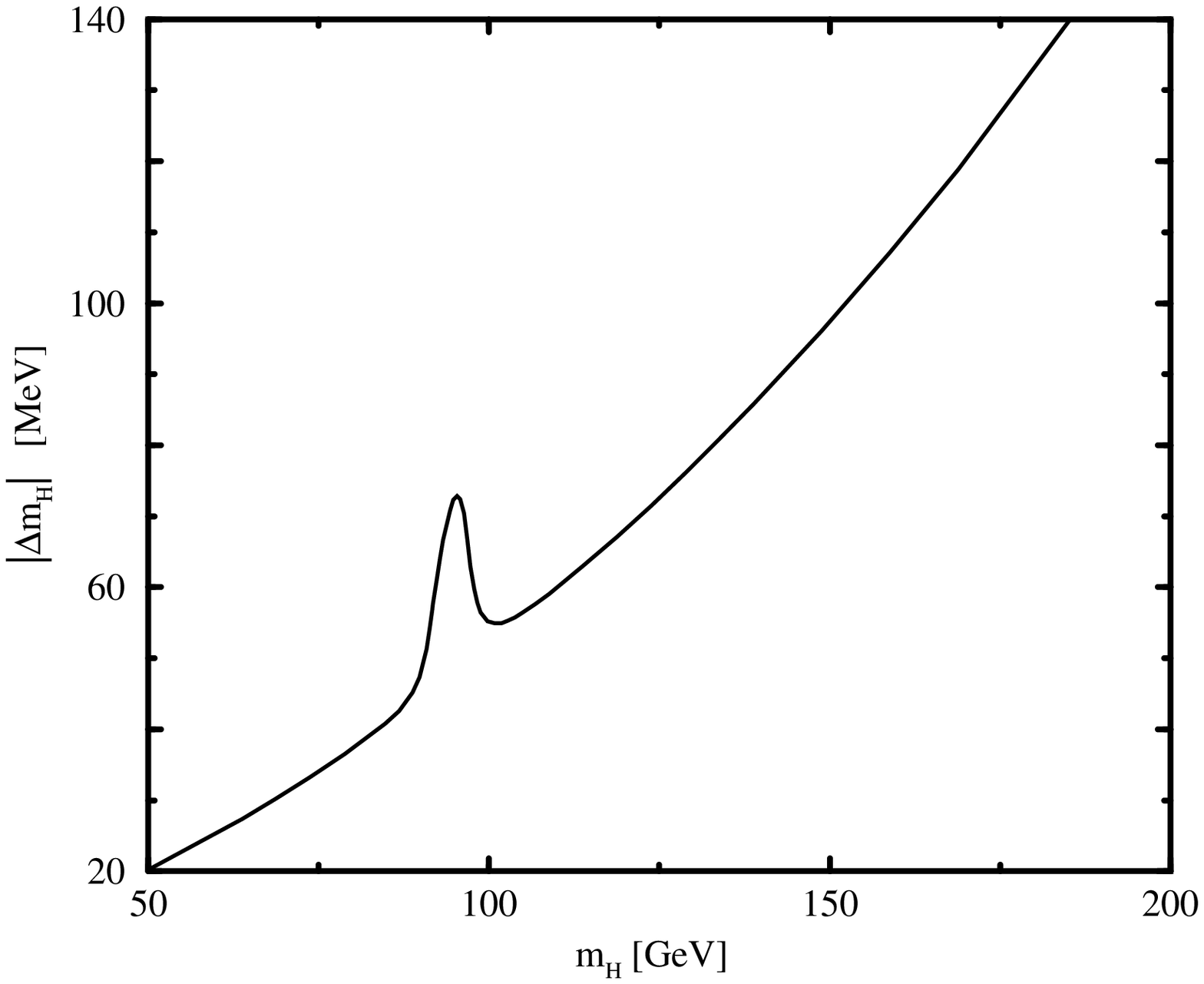,width=2.9in}}
\end{center}
\caption{The precision $\Delta \mhsm$ attainable from a $50\fbi$
measurement of the $Z b\anti b$ cross section at 
$\protect\rts=\mz+\mhsm+0.5\gev$
as a function of $\mhsm$, including $b$-tagging and cuts.
Bremsstrahlung, beamstrahlung, and beam energy smearing are neglected.
A precise measurement of the cross section well above threshold
is presumed available. Results from Ref.~\protect\cite{bbghzh}.}
\label{error}
\end{figure}

Another technique that is available at the NLC is to employ
a threshold measurement of the $Z\hsm$ cross section \cite{bbghzh}.
The procedure makes use of the fact that both $\mhsm$
and the $\rts=500\gev$ cross section for $\epem\to Z\hsm$ 
(with $\hsm\to b\anti b$) will be well-measured after a number of years
of NLC running. One then re-configures the collider 
for maximal luminosity just above the threshold energy $\rts= \mz+\mhsm$,
and expends $L=50\fbi$ at $\rts=\mz+\mhsm+0.5\gev$, \ie\ on the
steeply rising portion of the threshold curve for the $Z\hsm$
cross section.  The ratio of the cross section at $\rts=\mz+\mhsm+0.5\gev$
to that at $\rts=500\gev$ is insensitive to systematic effects and
yields a rather precise $\mhsm$ determination.
The expected precision for the Higgs mass
after including appropriate cuts to reduce backgrounds,
but before including the effects of bremsstrahlung, beamstrahlung
and beam energy smearing, is given in Fig.~\ref{error} for an
integrated luminosity of 50~fb$^{-1}$. (We deem it unlikely
that more than $L=50\fbi$ would be devoted to this special purpose
energy.) The precision degrades
as $\mhsm$ increases because the signal cross section is smaller.
The background from the $Z$-peak reduces the precision
for $\mhsm\sim \mz$. Bremsstrahlung, beamstrahlung and
beam energy smearing yield a reduction in sensitivity of 15\%
at a muon collider and 35\% at an $e^+e^-$ collider. 
Comparing to the errors that one would have for $L=50\fbi$ at
$\rts\sim\mz+\mhsm+20\gev$ from Higgs peak reconstruction in
the $Z\hsm$ mode (which are a factor of 2 larger than the $L=200\fbi$
errors plotted in Fig.~\ref{figdeltam}), we see that the threshold
measurement errors would be quite competitive for $\mhsm\slash\sim \mz$
unless the detector has either 
excellent CMS-style calorimetry (case 4) or super-JLC type tracking (case 1)
for the recoil mass reconstruction.

\begin{table}[p]
\caption[fake]{Summary of approximate errors, $\Delta\mhsm$, for $\mhsm\leq
300\gev$. LEP2 errors are for $L=600\pbi$. Tev33 errors are for $L=60\fbi$. 
LHC errors are for $L=600\fbi$ for ATLAS+CMS.
NLC errors are given for a luminosity times efficiency 
of $L\eps=200\fbi\times 0.6$
at both $\rts=500\gev$ and $\rtszhsm\equiv\mz+\mhsm+20\gev$.
For recoil mass reconstruction we consider tracking/calorimetry cases 1 and 3
(\ie\ super-JLC \cite{jlci} and `standard' NLC \cite{nlc}, respectively);
for Higgs peak reconstruction in the $\hsm\to b\anti b,\wp\wm$ final
states, case 5, we assume `standard' NLC \cite{nlc} hadronic calorimetry.
Beamstrahlung, bremsstrahlung and beam energy smearing effects
upon the recoil mass reconstruction are neglected.
NLC threshold results are for $L=50\fbi$ at $\rts=\mz+\mhsm+0.5\gev$, \ie\
just above threshold, and are quoted before including beamstrahlung,
bremsstrahlung and beam energy smearing --- at the NLC (FMC)
these effects increase the error by about 35\% (15\%).
FMC scan errors are for $L=200\fbi$ devoted to the scan
with beam energy resolution of 0.01\%. TeV33 and NLC errors are
statistical only. Systematic FMC error is neglected assuming
extremely accurate beam energy determination.}
\small
\begin{center}
\begin{tabular}{|c|c|c|c|c|}
\hline
 Machine/Technique & \multicolumn{4}{c|}{$\Delta\mhsm$ (MeV)} \\
\hline
\hline
{$\bf\mhsm$}{\bf (GeV)} & {\bf 80} & {\bf $\mz$} & {\bf 100} & {\bf 110} \\
\hline
LEP2 & 250 & 400 & $-$ & $-$ \\
\hline
TeV33 & 960 & ? & 1500 & 2000 \\
\hline
LHC/$\gam\gam$ (stat+syst) & 90 & 90 & 95 & 100 \\
\hline
NLC/case-3 $\rts=500$ & 813 & 674 & 572 & 494 \\
\hline
NLC/case-1 $\rts=500$ & 370 & 264 & 196 & 151 \\
\hline
NLC/hadronic $\rts=500$ & 51 & ? & 51 & 51 \\
\hline
NLC/case-3 $\rts=\rtszhsm$ & 27 & 29 & 31 & 34 \\
\hline
NLC/case-1 $\rts=\rtszhsm$ & 3.6 & 3.8 & 4.1 & 4.4 \\
\hline
NLC/hadronic $\rts=\rtszhsm$ & 15 & 17 & 19 & 22 \\
\hline
NLC/threshold & 40 & 70 & 55 & 58 \\
\hline
FMC/scan & 0.025 & 0.35 & 0.1 & 0.08 \\
\hline
\hline
{$\bf\mhsm$}{\bf (GeV)} & {\bf 120} & {\bf 130} & {\bf 140} & {\bf 150} \\
\hline
TeV33 & 2700 & $-$ & $-$ & $-$ \\
\hline
LHC/$\gam\gam$ (stat+syst) & 105 & 110 & 130 & 150 \\
\hline
LHC/$4\ell$ (stat+syst) & $-$ & 164 & 111 & 90 \\
\hline
NLC/case-3 $\rts=500$ & 432 & 383 & 343 & 311 \\
\hline
NLC/case-1 $\rts=500$ & 120 & 97 & 80 & 68 \\
\hline
NLC/hadronic $\rts=500$ & 52 & 52 & 53 & 55 \\
\hline
NLC/case-3 $\rts=\rtszhsm$ & 37 & 40 & 44 & 48 \\
\hline
NLC/case-1 $\rts=\rtszhsm$ & 4.8 & 5.2 & 5.6 & 6.1 \\
\hline
NLC/hadronic $\rts=\rtszhsm$ & 24 & 27 & 30 & 34 \\
\hline
NLC/threshold & 65 & 75 & 85 & 100 \\
\hline
FMC/scan & 0.06 & 0.12 & 0.20 & 0.49 \\
\hline
\hline
{$\bf\mhsm$}{\bf (GeV)} & {\bf 170} & {\bf 190} & {\bf 200} & {\bf 300} \\
\hline
LHC/$4\ell$ (stat+syst) & 274 & 67 & 56 & 90 \\
\hline
NLC/case-3 $\rts=500$ & 261 & 225 & 211 & 153 \\
\hline
NLC/case-1 $\rts=500$ & 50 & 39 & 35 & 20 \\
\hline
NLC/hadronic $\rts=500$ & 58 & 62 & 65 & 113 \\
\hline
NLC/case-3 $\rts=\rtszhsm$ & 56 & 65 & 70 & 133 \\
\hline
NLC/case-1 $\rts=\rtszhsm$ & 7.1 & 8.2 & 8.8 & 17 \\
\hline
NLC/hadronic $\rts=\rtszhsm$ & 41 & 51 & 56 & 140 \\
\hline
NLC/threshold & 120 & 150 & 170 & ? \\
\hline
\end{tabular}
\end{center}
\label{dmhsm}
\end{table}

\bigskip
\centerline{\underline{FMC}}
\smallskip

The ultimate in $\mhsm$ accuracy is that which can be achieved
at a muon collider by scanning the Higgs mass peak in the $s$-channel.
The scan was described earlier. For $L=200\fbi$ devoted to the scan
and a beam energy resolution of $0.01\%$,
one finds \cite{bbgh} $\Delta\mhsm=0.007,0.025,0.35,0.10,0.060,0.20,0.49\mev$
for $\mhsm=60,80,90,100,120,140,150\gev$, respectively.

\bigskip
\centerline{\underline{Summary}}
\smallskip

A summary of the accuracies possible for $\mhsm$ at the various machines
using the techniques described is given for $\mhsm\leq 300\gev$
in Table~\ref{dmhsm}.

\subsection{Verifying the spin, parity and CP of the $\hsm$}

Much of the following material is summarized in more detail
and with more referencing in \cite{dpfreport}.  We present
here only a very rough summary. We often focus on 
strategies and results for a relatively light SM-like Higgs boson.

If the $\hsm$ is seen in the $\gam\gam$ decay mode (as possible
at the LHC and at the NLC or FMC with sufficient luminosity
in mass regions M1, M2 and M3) or produced at the LHC via
gluon fusion (as presumably could be verified for
all mass regions) or produced
in $\gam\gam$ collisions at the NLC, then Yang's theorem
implies that it must be a scalar and not a vector, and, of course,
it must have a CP$=+$ component (C and P can no longer be regarded
as separately conserved once the Higgs is allowed to have fermionic
couplings). If the Higgs is observed with substantial rates in production
and/or decay channels that require it
to have $ZZ$ and/or $WW$ couplings, then it is very likely
to have a significant CP-even component given that the $ZZ/WW$ coupling
of a purely CP-odd Higgs boson arises only at one-loop.
Thus, if there is a Higgs boson with anything like SM-like couplings
it will be evident early-on that it
has spin-zero and a large CP$=+$ component.
Verifying that it is purely CP-even as predicted
for the $\hsm$ will be much more challenging. 

As we have discussed in earlier sections, 
observation of a Higgs boson in the $Z\h$ and/or $\epem\h$ mode
at LEP2 or the NLC via the missing-mass technique yields
a direct determination of the squared coupling $(ZZ\h)^2$.
Other techniques allow determination of $(WW\h)^2$.
At LEP2 only $Z\h$ production is useful; for
a SM-like Higgs boson its reach will be confined to $\mhsm\lsim 95\gev$ and
the accuracy of the $(ZZ\hsm)^2$ determination is quite limited ($\sim\pm 26\%$
at $\mhsm\sim\mz$).  Errors in the case of $L=200\fbi$ at the NLC
for a SM-like Higgs boson were quoted in Table~\ref{nlcerrors}
--- for $\mhsm\lsim 2\mw$, $(ZZ\hsm)^2$
can be measured to $\pm3\%-\pm4\%$ and $(WW\hsm)^2$ to $\pm 5\%-\pm8\%$.  
If the measurement yields the SM value to this accuracy, 
then the observed Higgs must be essentially
purely CP-even unless there are Higgs representations
higher than doublets.  This follows from the sum rule 
\begin{equation}
\sum_i (ZZ\h_i)^2=\sum_i (WW\h_i)^2=1
\label{srsat}
\end{equation}
(where the $(VV\h_i)^2$ -- $V=W,Z$ -- are defined relative to the SM-values)
that holds when all Higgs bosons are in singlet or doublet representations.
However, even if a single $\h$ appears to
saturate the coupling strength sum-rule, the possibility remains
that the Higgs sector is exotic and that saturation
of the sum rule by a single $\h$ is purely accidental.
Further, even if the $ZZ\h$ coupling is not full strength the $\h$
could still be purely CP-even.  To saturate the sum rule of Eq.~(\ref{srsat}), 
one need only have other Higgs bosons with appropriate
CP-even components; such Higgs bosons are present in the many attractive
models (including the minimal supersymmetric model) 
that contain additional doublet and/or 
some number of singlet Higgs representations beyond the single doublet
Higgs field of the SM.

When the $Z\h$ rate is significant, as particularly true
at the NLC, it will be possible to
cross check that there is a large CP-even component by examining
the angular distribution in $\theta$, the polar angle
of the $Z$ relative to the $\epem$ beam-axis
in the $Z\h$ (\ie\ $\epem$) center of mass.
(For summaries, see Refs.~\cite{kksz,dpfreport}.) 
However, the $Z\h$ rate is adequate
to measure the $\theta$ distribution only if the $\h$ has significant
$ZZ\h$ coupling, which in most models is only possible if the $\h$
has a significant CP-even component (since
only the CP-even component has a tree-level $ZZ\h$ 
coupling).  Further, if the CP-even component dominates the $ZZ\h$ coupling,
it will also dominate the angular distribution which will then 
not be sensitive to any CP-odd component of the $\h$ that might be present.
Thus, we arrive at the unfortunate conclusion
that whenever the rate is adequate for the angular distribution measurement,
the angular distribution will appear to be that for a purely CP-even
Higgs, namely $d\sigma/d\cos\theta\propto 8\mz^2/s+\beta^2\sin^2\theta$,
even if it contains a very substantial CP-odd component. 
(This insensitivity is numerically explicit in, for example,
the results of Ref.~\cite{sonixu}.) Thus,
observation of the above $\theta$ distribution only implies
that the $\h$ has spin-0 and that it is not {\it primarily} CP-odd.

At machines other than the NLC, measurement of the $\theta$
distribution for $Z\h$ events will be substantially more difficult.
Rates for $Z\h$ production will be at most just adequate for detecting
the $\h$ at LEP2,
TeV33 and the LHC.  Further, at TeV33 (in the $\h\to b\anti b$ channel)
and at the LHC (in the $\h\to \gam\gam$ channel) background rates
are substantial (generally larger than the signal). Further, 
$W\h$ production at TeV33 and the LHC cannot be employed because of inability
to reconstruct the $W\h$ center of mass (as required to determine $\theta$)
in the $W\to \ell\nu$ decay mode.

The $\tauptaum$ decays of the $\h$ provide a more
democratic probe of its CP-even vs. CP-odd components \cite{kksz,ggcp}
than does the $\theta$ angular distribution.
Further, the $\taup$ and $\taum$ decays are
self analyzing. The distribution in the azimuthal angle ($\phi$) 
between certain effective `spin' directions that can be defined
for these decays depends upon the CP mixture for the $\h$ eigenstate.
However, LEP2 is unlikely to produce
the large number of events required for decent statistical precision
for this measurement. For $\mh=90\gev$ and $\rts=192\gev$,
$\sigma(Z\h)\sim 0.5\pb$, implying some 500 total events for $L=1000\pbi$.
With $\br(\h\to \tauptaum)\sim 0.1$, we are left with only 50 events
before taking into account efficiencies and
the need for a fully reconstructable $Z$ decay.  
Expectations at the NLC \cite{kksz,ggcp} or FMC \cite{ggcp}
are much better. Particularly valuable
would be a combination of $Z\h$ with $\h\to\tauptaum$
measurements at $\rts=500\gev$ at the NLC and $\mupmum\to\h\to\tauptaum$
measurements in the $s$-channel mode at the FMC. Relatively
good verification of the CP-even nature of a light SM-like $\h$ is possible.
At higher Higgs masses (and higher machine energies) the self-analyzing
nature of the $t\anti t$ final states of Higgs decay can be
exploited in analogous fashion at the two machines.

One should not give up on a direct CP determination at the LHC.
There is one technique that shows real promise.
The key is the ability to observe the Higgs in the $t\anti t\h$
production channel with $\h\to \gam\gam$ or $\h\to b\anti b$.  
We saw earlier that separation of the $t\anti t\h$ from the $W\h$
channel at the LHC can be performed with good efficiency and purity.
The procedure for then determining the CP nature of the $\h$ was
developed in Ref.~\cite{ghcp}.  The $\gam\gam$ decay mode shows the greatest
promise because of a much smaller background. It is possible
to define certain projection operators that do not require knowledge
of the $t\anti t\h$ center of mass and yet are are sensitive to the angular
distributions of the $t$ and $\anti t$ relative to the $\h$.
Assuming $\mh=100\gev$ and $L=600\fbi$ for ATLAS+CMS combined, 
these projection operators distinguish between a SM-like (purely CP-even) 
Higgs boson and a purely CP-odd Higgs boson at roughly
the $6\sigma$ to  $7\sigma$ statistical level. For $\mh=100\gev$,
discrimination between a SM-like Higgs boson and a Higgs which is an equal 
mixture of CP-even and CP-odd is possible 
at the $2\sigma$ to $3\sigma$ level. (These statements assume
that the CP-even coupling squared plus CP-odd coupling squared
for $t\anti t\h$ is equal to the SM coupling-squared.)
Of course, rates are only adequate for relatively light Higgs bosons.
Verification of the efficiencies assumed in this analysis by full simulation
will be important. The projection operator technique (but not
the statistical significance associated with its application) is independent
of the overall event rate.

There is also a possibility that polarized beams at the LHC could be
used to look for spin asymmetries in the $gg\to\h$ production rate
that would be present if the $\h$ is a CP-mixed state \cite{gycp}.

Angular distributions in the $t\anti t\h$ final state in $\epem$
collisions at the NLC or $\mupmum$ collisions at the FMC
are even more revealing than those in the $t\anti t\h$ final state
at the LHC. The analysis procedures
appear in \cite{gghcp,ghe} and are summarized in Sec. III.A. 
By combining $Z\h$ measurements
with $t\anti t\h$ measurements verification of the $t\anti t$ and $ZZ$
couplings of a SM-like $\h$ will be possible at a remarkable level
of accuracy \cite{ghe}.  For instance, for 
$\rts=1\tev$ (we must be substantially above $t\anti t\h$
threshold), 2 1/2 years of running is expected to yield $L=500\fbi$
and in the case of $\mhsm=100\gev$ we can achieve
a determination of the CP-even $t\anti t\hsm$ coupling magnitude at the
$\sim\pm 3\%$ level, the (CP-even) 
$ZZ\hsm$ coupling magnitude at the $\sim\pm 2\%$ level, and a meaningful
limitation on the CP-odd $t\anti t\hsm$ coupling magnitude.

The most elegant determination of the CP nature of Higgs boson
is probably that possible in $\gam\gam\to\h$ production
at the $\gam\gam$ collider facility of the NLC
\cite{ggcpgamgam}. Since the CP-even and CP-odd components
of a Higgs boson couple with similar strength to $\gam\gam$ 
(via one-loop graphs),
there is no masking of the CP-odd component such as
occurs using probes involving $ZZ\h$ or $WW\h$ couplings.
The precise technique depends upon whether the Higgs is a pure or a mixed
CP eigenstate.
\begin{itemize}
\item
The most direct probe of a CP-mixed state is provided by
comparing the Higgs boson production rate
in collisions of two back-scattered-laser-beam 
photons of different helicities \cite{ggcpgamgam}.
The difference in rates for photons colliding with $++$ vs. $--$ 
helicities is non-zero only if CP violation is present.
A term in the cross section changes sign when
both photon helicities are simultaneously flipped.
Experimentally, this is achieved by 
simultaneously flipping the helicities of both of the initiating
back-scattered laser beams. One finds that the asymmetry 
is typically larger than 10\% and is 
observable if the CP-even and CP-odd components of the $\h$
are both substantial.
\item
In the case of a CP-conserving Higgs sector, 
one must have colliding photons with substantial transverse polarization.
This is achieved by transversely polarizing the incoming
back-scattered laser beams (while maintaining the ability
to rotate these polarizations relative to one another) and optimizing
the laser beam energy.  This optimization has been discussed in
Refs.~\cite{gkgamgamcp,kksz}. By computing
the difference in rates for parallel vs. perpendicular
polarizations divided by the sum, 
which ratio is $+1$ ($-1$) for a CP-even (CP-odd)
Higgs boson, it is found that $\gam\gam$ collisions
may well allow direct verification that a SM-like $\h$ is CP-even
vs. CP-odd.
\end{itemize}

A $\mupmum$ collider might provide an analogous opportunity
for directly probing the CP properties of any Higgs boson that
can be produced and detected in the $s$-channel mode \cite{atsoncp,dpfreport}.
However, it must be possible to transversely
polarize the muon beams.  Assume that
we can have 100\% transverse polarization and that 
the $\mu^+$ transverse polarization is rotated with respect
to the $\mu^-$ transverse polarization by an angle $\phi$.  The production
cross section for a $\h$ with coupling of a mixed
CP nature exhibits a substantial asymmetry of the form \cite{atsoncp}
\begin{equation}
A_1\equiv {\sigma(\pi/2)-\sigma(-\pi/2)\over \sigma(\pi/2)+\sigma(-\pi/2)}\,.
\end{equation}
For a pure CP eigenstate, the asymmetry \cite{dpfreport}
\begin{equation}
A_2\equiv {\sigma(\pi)-\sigma(0) \over \sigma(\pi)+\sigma(0)}
\end{equation}
is $+1$ or $-1$ for a CP-even or CP-odd $\h$, respectively.
Of course, background processes in the final states where
a Higgs boson can be most easily observed ({\it e.g.} $b\anti b$
for the MSSM Higgs bosons) will typically dilute these asymmetries
substantially. Whether or not they will prove useful depends even more 
upon our very uncertain ability to transversely polarize the muon
beams while maintaining high luminosity.

\section{Non-Minimal Higgs Sectors}

Five new projects were developed and pursued: 
\begin{description}
\item{A)} determining
the accuracy with which the $t\anti t$ CP-even and CP-odd Yukawa
couplings and the $ZZ$ coupling of a general neutral Higgs boson ($\h$)
could be measured by using both the $\epem\to t\anti t\h$ and $\epem\to Z\h$
production processes (or the $\mupmum$ analogues); 
\item{B)} determining the extent to which 
discovery of at least one Higgs boson of the NMSSM is guaranteed at the LHC;
\item{C)} detecting $\ha\to\gam\gam$ at the LHC;
\item{D)} determining $\tanb$ in the MSSM using measurements of $gg\to \hh,\ha$ and
$gg\to\hh b\anti b,\ha b\anti b$ production at the LHC;
\item{E)} evaluating the prospects for discovering and studying the
heavy $\hh,\ha,\hpm$  in $\hh\ha$ and $\hp\hm$ pair production at
the NLC or FMC and thereby constraining $\tanb$
and GUT-scale boundary conditions;
\item{F)} implications of LHC and NLC data upon the prospects for
discovering the $\hh,\ha$ in $s$-channel production at the FMC;
\item{G)} determining the discovery reach for doubly-charged Higgs bosons
in the process $p\anti p\to \dmm\dpp$ with $\dmm \to
\ell^-\ell^-,\dpp\to\ell^+\ell^+$ ($\ell=e,\mu,\tau$) at TeV33.
\end{description}
In what follows we motivate the importance of these projects and summarize
the results obtained.

\subsection{Determining the $t\anti t$ and $ZZ$ couplings of a neutral Higgs
boson \protect\cite{ghe}}

It is very possible (some would say probable) 
that the SM is not correct.  In this case, and if there is a weakly-coupled
Higgs sector, 
there will certainly be Higgs bosons that do not have SM-like couplings.
In particular, if one neutral Higgs is very SM-like 
(as for example is very probable in the minimal supersymmetric model),
the others must have very small $ZZ$
coupling and can have all manner of $t\anti t$ couplings.
Thus, it will be crucial to determine if an observed Higgs boson
fits into a given model context, such as the two-Higgs-doublet model,
and to determine the model parameters and associated couplings
for acceptable solutions.  By doing this for all the Higgs bosons
we would be able to completely fix the Higgs sector model and parameters.

The $t\anti t$ and $ZZ$ couplings of a neutral Higgs boson take the form:
\begin{eqnarray}
t\anti t h: -\anti t (a+ib\gamma_5) t {gm_t\over 2 m_W}\;,\;\;
ZZh: c {gm_Z\over cos(\theta_W)}g_{\mu\nu}\;,
\end{eqnarray}
where $g$ is the usual electroweak coupling constant. For the SM,
$a=1,b=0,c=1$.
However, these couplings become free parameters in a general
Higgs sector model.  For example,
in the general two-Higgs doublet model, \thdm, the couplings are
\begin{equation}
a={R_{2j}\over \sin\beta }\;,~~b=R_{3j}\cot\beta \;,~~c=R_{1j}\cos\beta 
+R_{2j}\sin\beta \;,
\end{equation}
where $j=1,2,3$ indicates one of the three Higgs mass eigenstates,
$\tanb$ is the ratio of the vacuum expectation values of the neutral members
of the two Higgs doublets (we assume a type-II \thdm), and 
$R_{ij}$ is a $3\times 3$ orthogonal matrix which specifies the transformation
between the \thdm\ Higgs fields and the Higgs boson
mass eigenstates. The result is that
\begin{equation}
a=-{s_1c_2\over \sin\beta }\;,~~~b= s_1s_2\cot\beta \;,~~~
c = c_1\cos\beta -s_1c_2 \sin\beta \;,
\label{abcsol}
\end{equation}
where $s_i=\sin\alpha_i$ and $c_i=\cos\alpha_i$ and $\alpha_{1,2}$
are free parameters in the range $0\leq \alpha_{1,2}<2\pi$.
The $\h$ has CP-violating couplings if either $ab\neq0$ or $bc\neq 0$.

The optimal technique \cite{gghcp} for extracting the couplings from
the $t\anti t\h$ process is reviewed in \cite{ghe}. It 
makes full use of the distribution $d\sigma/d\phi$
of the $t$, $\anti t$ and $\h$ in the final state as a function
of the final state kinematical variables, $\phi$ (rather than
just the total cross section). One of the $t$'s is required to decay
semi-leptonically and the other hadronically in order to reconstruct
all the useful variables. The $Z\h$ cross section is kinematically trivial
(neglecting the 1-loop $ZZ$ coupling to the CP-odd part of the $\h$
in comparison to the tree-level $ZZ$ coupling to the CP-even part of $\h$);
only the total rate for $Z\h$ with $Z\to \epem$ or $\mupmum$ 
(with the Higgs observed as a peak in the recoil mass spectrum) is employed. 
In order to demonstrate the power
of combining the $t\anti t\h$ and $Z\h$ processes, we have
considered a NLC or FMC with $\rts=1\tev$ (energy substantially
above the $t\anti t\h$ threshold is needed) and a light Higgs boson
with mass $\mh=100\gev$. We assume integrated luminosity of $L=500\fbi$
(about 2 1/2 years of running at the presumed design luminosity of
$L=200\fbi$ per year at $\rts=1\tev$).  Appropriate efficiency factors
(which include relevant branching ratios) are employed. For details
see \cite{ghe}. Our procedure is to input a given \thdm\ and determine
the accuracy with which the input parameters can be extracted from the data.
Our quantitative measure of accuracy is the $\chi^2$ 
associated with choices for $a$, $b$, and $c$ that differ
from the values of the input model.\footnote{Since $d\sigma/d\phi(t\anti t\h)$
and $\sigmazh$ are only sensitive to $a^2$, $c^2$, $b^2$, $ac$ and $bc$,
nothing changes if we simultaneously flip the signs of $a,b,c$. Thus,
there will inevitably be an overall sign ambiguity.}
The total $\chi^2$ is computed
by combining the $t\anti t\h$ and $Z\h$ processes:
\begin{equation}
\chi^2=\chi^2(t\anti t\h)+\chi^2(Z\h)\,;
\label{chisqtot}
\end{equation}
$\chi^2(t\anti t\h)$ is computed using the full correlated error matrix.

\begin{figure}[htb]
\leavevmode
\begin{center}
\centerline{\psfig{file=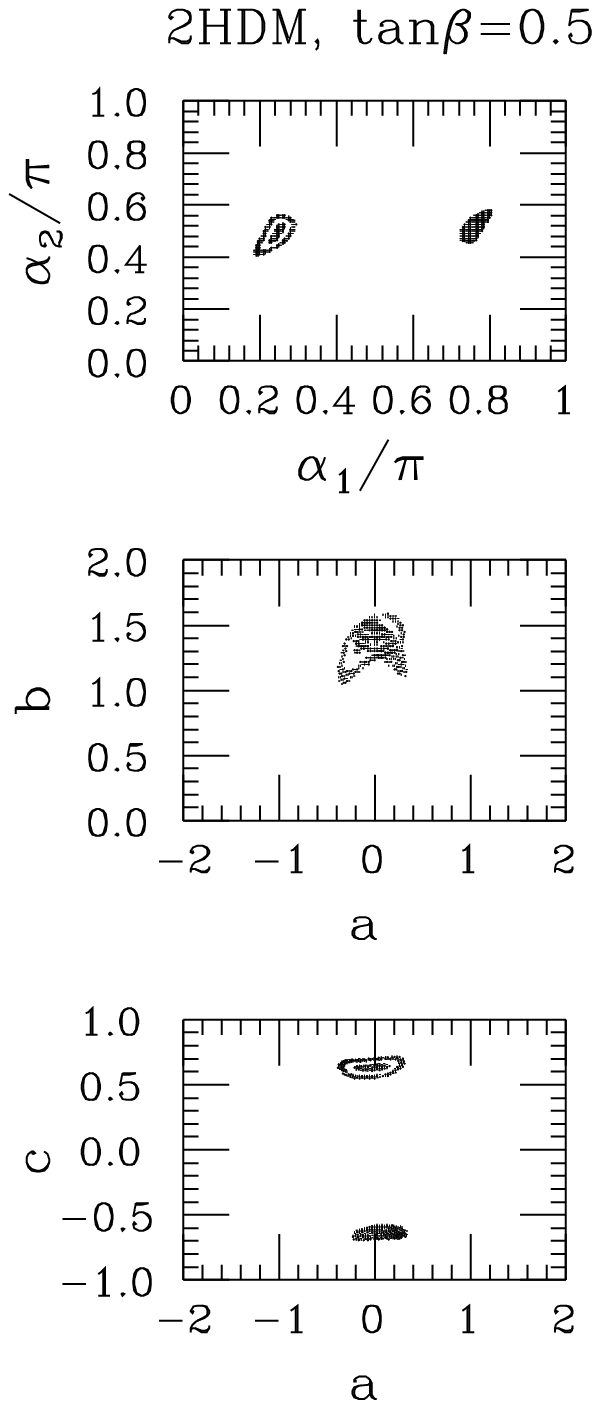,width=2.50in}}
\end{center}
\caption{$\chi^2\leq1$, $1<\chi^2\leq 4$ and $4<\chi^2\leq 9$ 
regions in the $(\alpha_1/\pi,\alpha_2/\pi)$, $(a,b)$ and $(a,c)$
planes, assuming as input a \thdm\ model with $\tanb=0.5$,
$\alpha_1=\pi/4$ and $\alpha_2=\pi/2$.}
\label{figttbhzh}
\end{figure}

We discuss one example in detail.
We take a \thdm\ model with $\tanb=0.5$, $\alpha_1=\pi/4$
and $\alpha_2=\pi/2$ as our input model.\footnote{We take $\tanb$
to be small so that the $t\anti t\h$ rate is substantial. If $\tanb$
is large, the $t\anti t\h$ process will have too small an event rate
to be terribly useful. If $\tanb$ is large enough, the $b\anti b \h$
final state can be employed in analogous fashion; it would be best
to run at smaller $\rts$ in such a case.} For the alternative
models, we considered $\tanb=0.5$, $\tanb=1.0$ and $\tanb=1.5$,
and computed $\chi^2$ as a function of $\alpha_1$ and $\alpha_2$
assuming the \thdm\ forms of $a,b,c$ as given in 
Eq.~(\ref{abcsol}).\footnote{We considered only $0\leq \alpha_{1,2}<\pi$
so as to avoid the above-noted overall sign ambiguity.}
We first note that, in the case of the particular input model specified above,
only $\tanb=0.5$ (the input value), and not $\tanb=1$ or
$1.5$, yields any $a,b,c$ value set
(as $\alpha_{1,2}$ are varied) that leads to $\chisq\leq 9$. 
Thus, an approximate determination of $\tanb$ would be possible.
In Fig.~\ref{figttbhzh}, 
we take $\tanb=0.5$ and plot different $\chi^2$ regions
in the $(\alpha_1,\alpha_2)$, $(a,b)$ and $(a,c)$ planes.
In each window of the figure,
a filled central region, an empty band, and a filled
band may all be visible. The central region is the $\chisq\leq 1$
region, the empty band is the $1<\chisq\leq 4$ region,
and the outer filled band is the $4<\chisq\leq 9$ region.
If no filled central region is visible, the central region being empty,
then this means that $\chisq\leq 1$ was not possible.  If only a completely
filled region appears, then $\chisq\leq 4$ was not possible.
From the $\chisq$ regions of Fig~\ref{figttbhzh}
we arrive at the following additional results.
\begin{itemize}
\item
The $\chisq\leq1$ region for $\tanb=0.5$ corresponds closely
to the input values of $\alpha_1=\pi/4$ and $\alpha_2=\pi/2$.
An alternative region with $\alpha_1\to \pi-\alpha_1$
develops for $4<\chisq\leq 9$. 
\item
The values of $a,b,c$ are well-determined
if we demand $\chisq\leq1$; $\chisq\leq 4$
allows only slightly greater flexibility.
However, $4<\chisq\leq 9$ allows a 
a solution with the flipped sign of $ac$ and slightly distorted $b$ values.
[In the $(a,b)$ plane window, 
the three different $\chisq$ regions associated with
the correct sign of $ac$ are somewhat obscured by
the strange extra blob associated with $4<\chisq\leq 9$ and the wrong
sign of $ac$.]
\end{itemize}

Of course, if the $\h$ being studied is the SM $\hsm$ or simply SM-like,
the above $t\anti t\h$/$Z\h$ techniques can be employed to verify
the SM $a,b,c$ couplings, see Ref.~\cite{ghe}.  Another interesting
extreme is a purely CP-odd Higgs boson, $\a$. 
(For example, the $\ha$ of the MSSM.)
The $\a$ might be light enough and $\tanb$
small enough that the $t\anti t\a$ production rate would be large.
It was demonstrated in Ref.~\cite{gghcp} that the $b$ coupling
of a CP-odd $\a$ with $a=c=0,b=1$ could be measured with substantial
accuracy and significant limits placed on the $a,c$ couplings
using $t\anti t\a$ data alone.\footnote{Numerical details 
for the $\ha$ of the MSSM would differ
slightly due to the fact that the other Higgs bosons would also
be light, whereas in Ref.~\cite{gghcp} it was assumed that only the $\a$
was light.}

We note that systematic uncertainties in the 
experimental determination of the overall normalization
of the $t\anti t\h$ and $Z\h$ total cross sections could
have substantial impact on our ability to determine couplings if
the systematic errors are not small compared to the statistical errors.
Also, at larger Higgs masses, statistics will deteriorate; higher $\ltot$
will be required to avoid significant ambiguity in the coupling determinations.
However, even when ambiguities emerge, they are usually
sufficiently limited that the type of analysis outlined above will 
make a critical contribution to gaining a clear understanding
of the exact nature of all the Higgs bosons.
Certainly, the procedures discussed will provide a powerful means
for distinguishing between substantially different models. 

\subsection{Is discovery of a NMSSM Higgs boson guaranteed with LEP2 plus LHC?
\protect\cite{ghm}}

It is well-established \cite{dpfreport} that at least one of
the Higgs bosons of the MSSM can be discovered either at LEP2 or at the LHC
throughout all of the standard $(\mha,\tanb)$ parameter space.
Ref.~\cite{ghm} reconsiders this issue in the context of the NMSSM,
in which there is greater freedom by virtue of there being three
instead of two CP-even Higgs bosons and correspondingly greater
freedom in their couplings. It is found that there are regions
of parameter space for which none of the NMSSM Higgs bosons can be detected
at either LEP2 or the LHC.  This result is to be contrasted with
the NLC or FMC no-lose theorem \cite{kot}, according to which at least one
of the CP-even Higgs bosons (denoted generically by $\h$) of the NMSSM
will be observable in $\zstar\to Z\h$ production.

The detection modes considered for the NMSSM are the same as those
employed in establishing the LEP2 plus LHC no-lose theorem
for the MSSM:
1) $\zstar\to Z\h$ at LEP2; 2) $\zstar\to \h\a$ at LEP2;
3) $gg\to \h\to\gam\gam$ at LHC; 4) $gg\to\h\to Z\zstar~{\rm or}~ZZ\to 4\ell$
at LHC; 5) $t\to\hp b$ at LHC;
6) $gg\to b\anti b \h,b\anti b\a \to b\anti b \tauptaum$ at LHC;
7)  $gg\to\h,\a\to\tauptaum$ at LHC.
Additional Higgs decay modes that could be considered at the LHC include:
a) $\a\to Z\h$; b) $\h\to\a\a$;
c) $\h_j\to\h_i\h_i$; d) $\a,\h\to t\anti t$.  Because
of the more complicated Higgs self interactions,
b) and c) cannot be reliably computed in the NMSSM without
additional assumptions. The Higgs mass values for
which mode a) is kinematically allowed can be quite different
than those relevant to the MSSM and thus there are
uncertainties in translating ATLAS and CMS results for the MSSM
into the present more general context. Finally, mode d) is currently
of very uncertain status and might turn out to be either more
effective or less effective than current estimates.
Thus, to be conservative, any choice
of NMSSM parameters for which the modes a)-d) might be relevant is excluded.
Even over this restricted region of parameter space, 
NMSSM parameter choices can be found such that there are
no observable Higgs signatures at either LEP2 or the LHC.

The free parameters of the model can be chosen
to be $\tanb$, $\mhi$, $\lam$, $\alpha_{1,2,3}$, and $\ma$.
Here, $\mhi$ is the mass of the lightest CP-even Higgs mass eigenstate.
$\lam$ appears in the superpotential in the term $W\ni \lam\hat H_1\hat H_2\hat
N$. A crucial ingredient in constraining the model is that $\lam\lsim 0.7$
is required if $\lam$ is to remain perturbative during evolution
from scale $\mz$ to the Planck scale. This limitation on $\lam$ implies
a $\tanb$-dependent upper limit on $\mhi$ in the range $\lsim 140\gev$.
The angles $\alpha_{1,2,3}$
are those parameterizing the orthogonal matrix which diagonalizes the CP-even
Higgs mass-squared matrix. $\ma$ is the mass of the lighter of
the two CP-odd mass eigenstates --- the second CP-odd state can be
assumed to be very massive for the purposes of establishing
the existence of parameter choices for which no Higgs boson can be found.
All couplings and cross sections are determined once the above parameters
are specified. Details regarding the procedure for scanning the NMSSM parameter
space and assessing observability of the various Higgs
bosons are given in Ref.~\cite{ghm}.  A choice
of parameters such that none of the Higgs bosons
$\h_{1,2,3}$, $\a$ or $\hpm$ are observable at LEP2 or the LHC is declared to
be a ``point of unobservability'' or a ``bad point''.

The results obtained are the following.
If $\tanb\lsim 1.5$
then all parameter points that are included in the search are observable
for $\mhi$ values up to the maximum allowed ($\mhi^{\rm max}\sim 137\gev$
for $\lam_{\rm max}=0.7$, after including radiative corrections).
For such low $\tanb$, the LHC $\gam\gam$ and $4\ell$ modes allow
detection if LEP2 does not. For high $\tanb\gsim 10$, the
parameter regions where points of unobservability are found
are also of very limited extent, disappearing as the $b\anti b\h_{1,2,3}$
and/or $b\anti b\a$ LHC modes allow detection where LEP2 does not.
However, significant portions of
searched parameter space contain points of unobservability
for moderate $\tanb$ values. That such $\tanb$ values
should be the most `dangerous'
can be anticipated from the MSSM results.  It is well-known
(see, for example, Ref.~\cite{dpfreport}) that for
stop masses of order $1\tev$ and no stop-mixing there is a wedge
of MSSM parameter space at moderate $\tanb$ and with $\hh$ and $\ha$
masses above about $200\gev$ for which the only observable
MSSM Higgs boson is the light SM-like $\hl$, and that the $\hl$ can
only be seen in the $\gam\gam$ mode(s) at the LHC. (Observation at LEP2
is impossible in this wedge of parameter space
since $\mhl+\mz,\mhl+\mha > 192\gev$.)  
By choosing $\mhi$ and $\ma$ in the NMSSM
so that $\mhi+\mz$ and $\mhi+\ma$ are close to or above the $\rts$ of LEP2,
then, by analogy, at moderate $\tanb$ 
we would need to rely on the $\h_{1,2,3}\to \gam\gam$ modes.
However, in the NMSSM, parameter choices are possible for which all the
$WW\h_{1,2,3}$ couplings are reduced relative to SM strength.
This reduction will
suppress the $\gam\gam$ couplings coming from the $W$-boson
loop. All the $\h_i\to\gam\gam$ widths
can be sufficiently smaller than the somewhat enhanced $b\anti b$ widths
so that the $\gam\gam$ branching ratios are {\it all} no longer of
useful size.

\begin{figure}[htb]
\leavevmode
\begin{center}
\centerline{\psfig{file=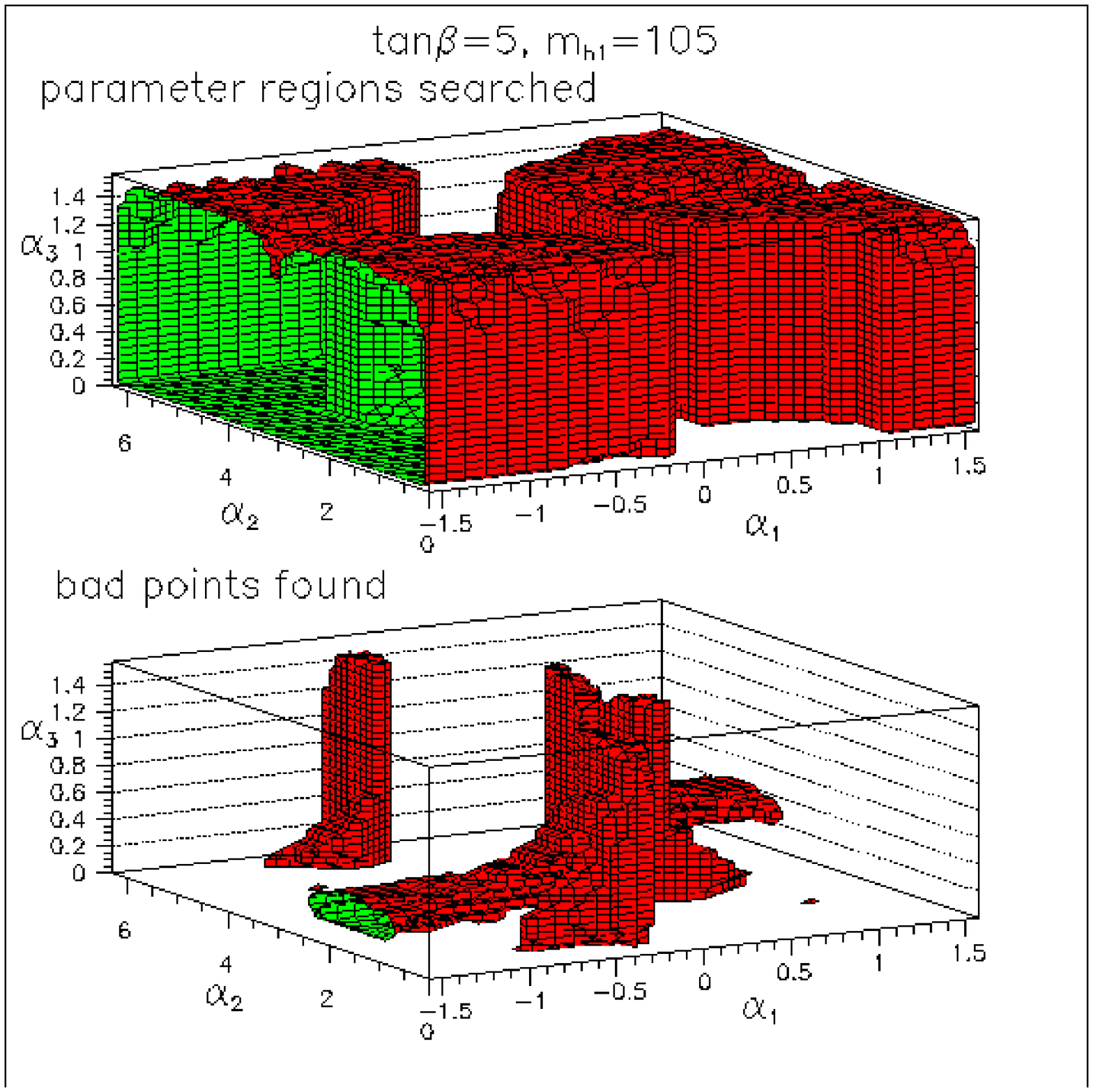,width=2.9in}}
\end{center}
\caption{For $\tanb=5$ and $\mhi=105\gev$, we display in three dimensional
$(\alpha_1,\alpha_2,\alpha_3)$ parameter space the parameter regions
searched (which lie within the surfaces shown), and the
regions therein for which the remaining model parameters can
be chosen so that no Higgs boson is observable
(interior to the surfaces shown).}
\label{tanb5}
\end{figure}

To illustrate, we shall discuss results for $\tanb=3$, $\tanb=5$ and
$\tanb=10$ (for which $\mhi^{\rm max}\sim 124\gev$, $118\gev$ and $114\gev$,
respectively) and $\mhi=105\gev$.
\begin{itemize}
\item
In Fig.~\ref{tanb5}, we display for $\tanb=5$ both the
portions of $(\alpha_1,\alpha_2,\alpha_3)$
parameter space that satisfy our search restrictions,
and the regions (termed ``regions of unobservability'')
within the searched parameter space such that,
for {\it some} choice of the remaining parameters ($\lam$ and $\ma$),
no Higgs boson will be detected
using any of the techniques discussed earlier.\footnote{For a 
given $\alpha_{1,2,3}$ value
such that there is a choice of $\lam$ and $\ma$ for which no Higgs
boson is observable, there are generally
other choices of $\lam$ and $\ma$ for which at least one
Higgs boson {\it is} observable.} Relatively large regions of unobservability
within the searched parameter space are present.
\item
At $\tanb=3$, a similar picture emerges. The search region that satisfies our
criteria is nearly the same; the regions of unobservability lie mostly within
those found for $\tanb=5$, and are about 50\% smaller.
\item
For $\tanb=10$, the regions of unobservability comprise only a
very small portion of those found for $\tanb=5$.
This reduction is due to the increased $b\anti b$ couplings
of the $\h_i$ and $\a$, which imply increased $b\anti b\h_i,b\anti b\a$
production cross sections. As these cross sections become large, detection
of at least one of the $\h_i$ and/or the $\a$ in the
$b\anti b\taup\taum$ final state becomes
increasingly difficult to avoid. For values of 
$\tanb\gsim 10$,\footnote{The precise value 
of the critical lower bound on $\tanb$
depends sensitively on $\mhi$.}
we find that one or
more of the $\h_i,\a$ should be observable regardless of location in
$(\alpha_1,\alpha_2,\alpha_3,\lam,\ma)$ parameter space (within
the somewhat restricted search region that we explore).
\end{itemize}

It is useful to present details on what goes wrong at a typical
point of unobservability. For $\tanb=5$ and $\mhi=105\gev$,
no Higgs boson can be observed for $\ma=103\gev$ if
$\alpha_1=-0.479$, $\alpha_2=0.911$, $\alpha_3=0.165$, and $\lam=0.294$
(for which $\mhii=124\gev$, $\mhiii=206\gev$, and $\mhp=201\gev$).
For this point, the Higgs boson couplings
(relative to the SM values) are:
\begin{eqnarray*}
(VV\h_1)^2= 0.79 & (VV\h_2)^2=0.21 & (VV\h_3)^2=0.006 \\
(b\anti b \h_1)^2=5.3 & (b\anti b \h_2)^2=2.5 & (b\anti b\h_3)^2=18 \\
(t\anti t\h_1)^2=0.69 & (t\anti t \h_2)^2=0.29 & (t\anti t\h_3)^2=0.062
\end{eqnarray*}
where $V=W$ or $Z$.
Note that $\h_3$ has very small couplings to $VV$.
The manner in which this point
escapes discovery is now apparent. First,
the minimum values required for the $(b\anti b\h_i)^2$ values for $\h_i$
observability in the $\tauptaum$ mode are:  53 ($i=1$); 32 ($i=2$); 35 ($i=3$).
The actual values all lie below these required values.
Observation of the $\a$ at $\ma=103\gev$ would require $\tanb=8$.
Regarding the other discovery modes,
$\h_1$ and $\h_2$ are both in the mass range for which the $\gam\gam$
mode is potentially viable and the $\h_3$
is potentially detectable in the $ZZ\to 4\ell$ channel.
However, the suppressed $t\anti t\h_{1,2,3}$
couplings imply smallish $gg$ production rates for $\h_{1,2,3}$.
Relative to a SM Higgs of the same mass we have:
\begin{equation}
{(gg\h_i)^2\over (gg\hsm)^2}=0.58~(i=1);~~~0.43~(i=2);~~~0.15~(i=3)\,.
\end{equation}
(Note that these strengths are not simply the $(t\anti t\h_i)^2$
magnitudes; the enhanced $b$-quark loop contributions 
interfere with the $t$-quark loop contributions at amplitude level.)
Further, the enhanced Higgs decay rate to $b\bar b$ and the reduced
$W$-loop contributions to the $\gam\gam$ coupling suppress the
$\gam\gam$
branching ratios of $\h_1$ and $\h_2$ relative to SM expectations.  We find:
\begin{equation}
{\br(\h_i\to \gam\gam) \over \br(\hsm\to\gam\gam)}
=0.18~(i=1)\,;~~~0.097~(i=2)\,;
\end{equation}
\ie\ suppression sufficient to make $\h_1$ and $\h_2$ invisible
in the $\gam\gam$ mode. The suppressed $ZZ\h_3$ coupling
and the enhanced $\h_3\to b\bar b$ decays
are sufficient to suppress $\br(\h_3\to ZZ)$
much below SM expectations:
\begin{equation}
{\br(\h_3\to ZZ)\over \br(\hsm\to ZZ)}=0.11\,,
\end{equation}
\ie\ such that the $4\ell$ signal has a significance of only $1.5\sigma$,
even though a SM Higgs of this mass would yield a $\sim 37\sigma$ signal.
In short, there is enough flexibility due to the addition of the singlet
Higgs field (which has no couplings to SM fermions and vector bosons!)
for {\it all} the Higgs bosons to escape
detection for certain choices of model parameters,  provided $\tanb$
is moderate in size. Moderate $\tanb$ implies
that $\h\to\gam\gam$ decays for light Higgs are
suppressed, while at the same time
$b\anti b \h$ production is not adequately enhanced
for detection of the $\h\to \tauptaum$ mode.

The regions of NMSSM parameter space where no Higgs boson
can be detected will expand if
full $L=600\fbi$ ($L=1000\pbi$) luminosity is not available at the LHC
(LEP2) or efficiencies are smaller than anticipated.
Conversely, these ``regions of unobservability''
could decrease substantially (perhaps disappear)
with improved efficiency (\eg\ due to an expanded calorimeter)
in the $\tau\tau$ final state or higher luminosity.
Supersymmetric decays of the Higgs bosons are neglected in the above.
If these decays are important,
the regions of unobservability found without using the SUSY final states will
increase in size.  However, Higgs masses in the
regions of unobservability
are typically modest in size ($100-200\gev$), and
as SUSY mass limits increase with LEP2 running this additional
concern will become less relevant.  Of course, if SUSY
decays are significant, detection of the Higgs
bosons in the SUSY modes might be possible, in which case
the regions of unobservability might decrease in size. Assessment of this
issue is dependent upon a specific model for soft SUSY breaking.

Although it is not possible to establish a no-lose
theorem for the NMSSM Higgs bosons by combining data from LEP2 and the LHC
(in contrast to the no-lose theorems applicable to
the NLC Higgs search with $\rts\gsim 300\gev$), the regions of complete
Higgs boson unobservability appear to constitute a small fraction of the
total model parameter space.
It would be interesting to see whether or not these
regions of unobservability correspond to unnatural choices
for the Planck scale supersymmetry-breaking parameters.

\subsection{Detecting $\ha\to\gam\gam$ at the LHC \protect\cite{kaogamgam}}

In this report, a realistic study was performed of observability 
for the CP-odd Higgs boson ($\ha$) 
in the minimal supersymmetric standard model (MSSM) 
via its photon decay mode ($\ha \to \gamma\gamma$) 
with the CMS detector performance.
It is demonstrated that it will be possible
to discover the CP-odd $\ha$ and
reconstruct its mass ($\mha$) with high precision 
for 170 GeV $< \mha < 2m_t$ at the LHC
if the decays of the $\ha$ into SUSY particles
are forbidden and $\tan\beta$ is close to one.
Thus, the $\ha\to\gam\gam$ mode complements the $\mupmum$ decay modes
($\hl,\hh,\ha \to \mu\mu$) that are promising 
\cite{CMS,mumu} for observing and precisely reconstructing masses  
for the neutral Higgs bosons at large $\tanb$.

The total cross section for the process 
$pp \to \ha \to \gamma\gamma+X$ 
is given by $\sigma(pp \to \ha +X)\br(\ha\to\gam\gam)$;
$\sigma(pp\to\ha+X)$ is evaluated using
the parton distribution functions of CTEQ2L 
with $\Lambda_4 = 0.190$ GeV and $Q^2 = \mha^2$.


Gluon fusion ($gg \to \ha$), via the top quark and the bottom quark 
triangle loop diagrams, is the major source for the CP-odd Higgs pseudoscalar  
if $\tan\beta$ is less than about 4. At higher $\tanb$, $gg\to \ha b\anti b$
dominates since
the $(b\anti b\ha)^2$ coupling-squared is proportional to $\tan^2\beta$.
In the computations, both production mechanisms are included.
QCD radiative corrections, which, for example,
increase $\sigma(gg \to \ha)$ 
by about 50\% to 80\% for $\tan\beta \sim 1$,
are not included for either the signal or the backgrounds.
 

The $b\anti{b}$ mode dominates $\ha$ decays 
for $\tanb\gsim4$ and $\mha \le 2m_t$.
For $\mz+\mhl < \mha \le 2m_t$ and $\tan\beta < 4$, 
$\br(\ha \to Zh)$ is comparable to $\br(\ha \to
b\bar{b})$. If present, SUSY decay modes would deplete both.
In the analysis, parameters were chosen so that
$\ha\to$SUSY decays are kinematically forbidden.
Then, for $\tan\beta$ close to 1 and 170 GeV $< \mha < 2m_t$ 
$\br(\ha \to \gamma\gamma)\sim 5 \cdot 10^{-4} - 2 \cdot 10^{-3}$.


Events were simulated at the particle level
using PYTHIA 5.7 and  JETSET 7.4  generators \cite{PYTHIA}
with the CTEQ2L parton distribution functions.
The PYTHIA/JETSET outputs were processed
with the CMSJET program \cite{CMSJET}, which
is designed for fast simulations of ``realistic'' CMS detector response.
Resolution effects were taken into account
by using the parameterizations obtained
from the detailed GEANT \cite{GEANT} simulations.
CMSJET includes also some analysis programs, in particular, 
a set of jet reconstruction algorithms.

The irreducible backgrounds considered were:
(i) $q \bar{q} \to\gamma \gamma $ and (ii) $ g g \to\gamma \gamma $ (Box).
In addition, reducible backgrounds with at least 1 $\gamma$ 
in the final state were included:
(i) $q \bar{q} \to g \gamma$, (ii)  $q g \to q \gamma$,  
and (iii) $g g \to g \gamma$ (Box). 

The ECAL resolution was assumed to be 
$\sigma(E)/E = 5\% /\sqrt E \oplus 0.5\%$ (CMS high luminosity regime). 
Each photon was required to have transverse momentum ($p_T$) larger than 40 GeV 
and $|\eta| < 2.5$.   Both photons were required to be
isolated, {\it i.e.}, 
(1) no charged particle with $p_T > $ 2 GeV 
 in the cone $R = 0.3$; and 
(2) the total transverse energy $\sum E_T^{cell}$ 
must be less than 5 GeV in the cone ring 0.1 $< R < 0.3$.
In this preliminary analysis, 
no rejection power against $\pi^0$'s with high $p_T$ was assumed; this means
all $\pi^0$'s surviving the cuts ($p_T$, isolation, etc.) 
were considered as $\gamma$'s.\footnote{
This is quite conservative.
The background from the $\pi^0$ is overestimated, especially  
in the low mass $\mgamgam$ region.} 
For each $\mha$ and $\tan \beta$, 
the mass window around the peak (within the range 2-6 GeV) 
and the $p_T$ cut (50-100 GeV) 
were chosen to provide the best value of  $S/\sqrt{B}$. 
For example, the best values of the 
mass window and $p_T$ cut for  $\mha =$ 200 GeV ($350\gev$) 
are 2 GeV ($4\gev$) and 60 GeV ($100\gev$), respectively.

\begin{figure}[htb]
\leavevmode
\begin{center}
\centerline{\psfig{file=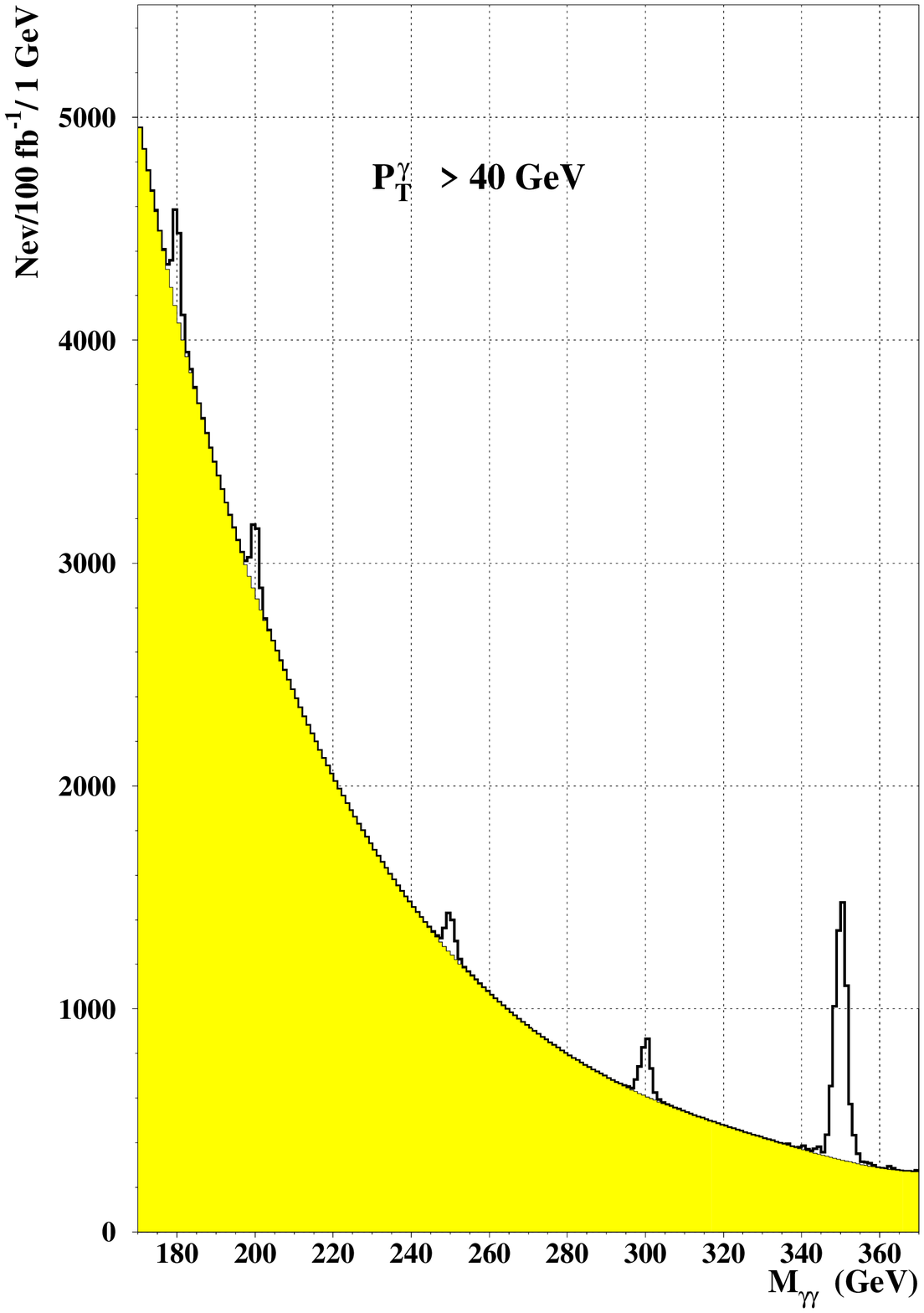,width=2.9in}}
\end{center}
\caption{
Number of events vs. $\mgamgam$ 
for the signal and the background at $\protect\sqrt{s} = 14$ TeV 
with $L = 100$ fb$^{-1}$ and $\tan\beta = 1$.
CMS performance is assumed and SUSY parameters are
$m_{\tilde{q}} = \mu =$ 1000 GeV.
}
\label{figkaogamgam1}
\end{figure}

\begin{figure}[htb]
\leavevmode
\begin{center}
\centerline{\psfig{file=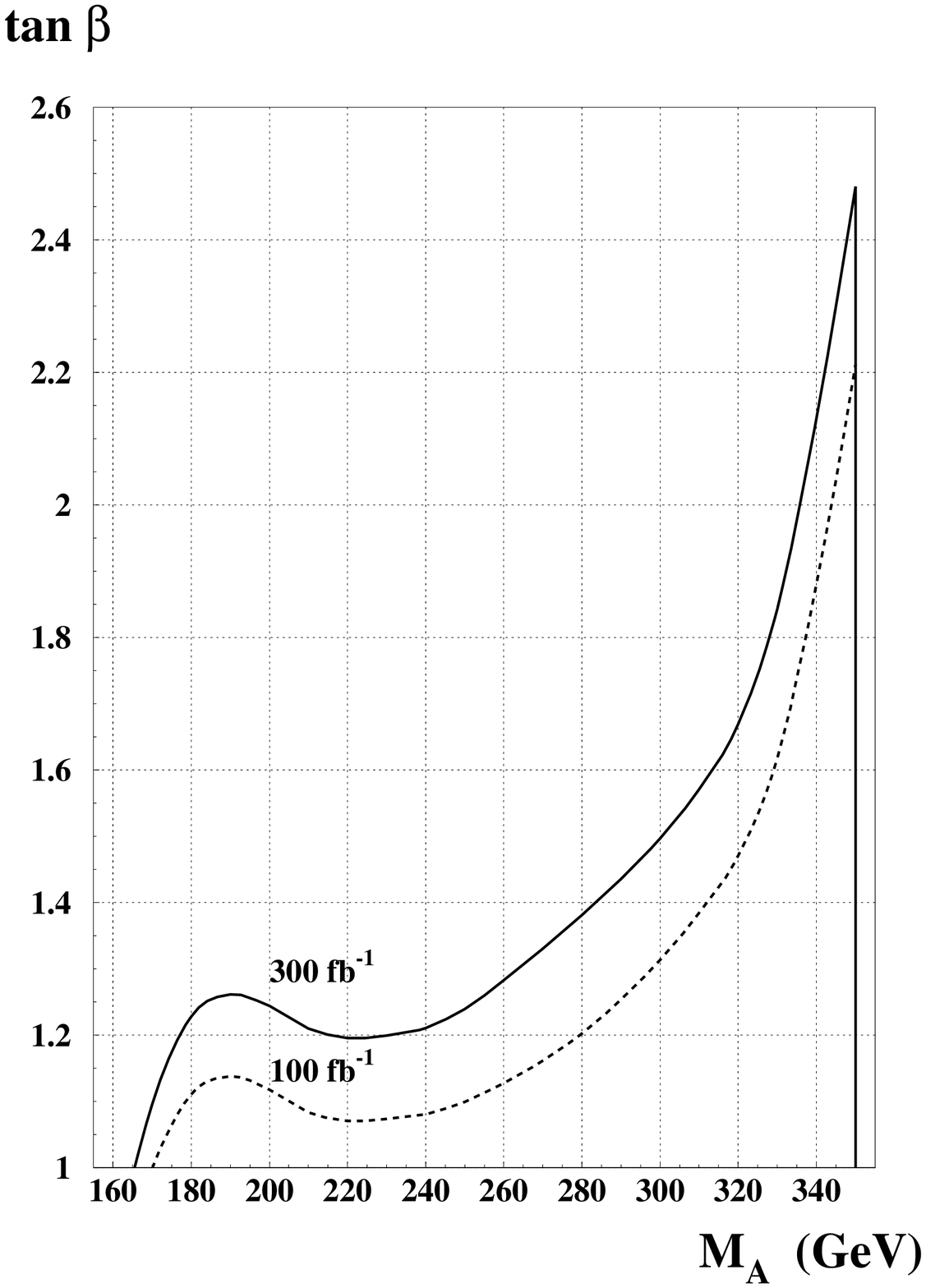,width=2.9in}}
\end{center}
\caption{
The $5\sigma$ $\ha\to\gam\gam$ discovery contours in the $(\mha,\tanb)$ plane
at $\protect\sqrt{s} = 14$ TeV with $L=100\fbi$ and $L=300\fbi$.
CMS performance is assumed and SUSY parameters are
$m_{\tilde{q}} = \mu =$ 1000 GeV.
The discovery regions lie below the contours shown.
}
\label{figkaogamgam2}
\end{figure}

A typical $\mgamgam$ distribution is shown in Fig.~\ref{figkaogamgam1}.
The $L=100\fbi$ and $300\fbi$ $5\sigma$ discovery contours are
shown in Fig.~\ref{figkaogamgam2}. Apparently,
this channel should provide a good opportunity to observe and
precisely reconstruct 
the CP-odd Higgs boson mass ($\mha$) for 170 GeV $< \mha < 2m_t$ 
if the $\ha\to$SUSY decays
are forbidden and $\tan\beta$ is close to one.
The impact of SUSY decays on this discovery channel 
might be significant \cite{Z2Z2} 
and is under investigation with realistic simulations.

\subsection{Determining $\tanb$ in the MSSM from $\ha$ and $\hh$ production
at the LHC \protect\cite{gkp}}

As noted in a previous subsection (see Ref.~\cite{dpfreport} for a
thorough review), detection of single
$\ha$ and/or $\hh$ production at the LHC will be possible in several
regions of $(\mha,\tanb)$ parameter space. In particular, $gg\to \hh$
and $gg\to \ha$ inclusive production can be isolated in the 
$\hh,\ha\to\tauptaum$ decay mode if $\tanb$ is modest in size ($\lsim 3$) and 
$\mha,\mhh$ are below $2\mt$.  It is mainly the $\ha$
which provides a viable signal in this region of parameter space.
For masses above $2\mt$, there is
also some hope for detection in the $gg\to \hh,\ha\to t\anti t$ decay mode,
provided the $t\anti t$ background normalization and shape can be determined
at about the 5-10\% level.  Since $\mha\sim\mhh$ at higher mass, 
it is the combined $gg\to \hh+gg\to \ha$ signal that will be observed.
At high $\tanb$, the $\hh$ and $\ha$
have enhanced $b\anti b$ coupling resulting in large rates for
the $gg\to\hh b\anti b$ and $gg\to \ha b\anti b$ processes; detection
of these production modes
in the $\hh,\ha\to \tauptaum$, $\mupmum$ and, perhaps, $b\anti b$ decay
channels will be possible. (At high $\tanb$, these are the only important
decay modes since they are the only ones associated with enhanced coupling,
$\propto \tanb$ at the amplitude level.) In the $(\mha,\tanb)$ parameter
space, the lowest value of $\tanb$ for which $gg\to \hh b\anti b,\ha b\anti
b\to\tauptaum b\anti b$ production
can be observed ranges from $\tanb\gsim 3$ at $\mha\sim 200\gev$
to $\tanb\gsim 15$ at $\mha\sim 500\gev$. Still higher $\tanb$ values would
be required at higher $\mha$ simply due to the fact that the cross
section decreases (at fixed $\tanb$) as $\mha$ increases.  A similar
range of viability may be possible in the $b\anti b b\anti b$ final state,
if $b$ tagging can be performed at the optimistic end of current
efficiency and purity expectations \cite{dgv}.
Higher $\tanb$ values are required for viability of the signal
in the $\mupmum b\anti b$ final state (simply because of the much
smaller rates deriving from the much smaller $\mupmum$ branching
ratios for the $\hh$ and $\ha$).
At high $\tanb$,
the degeneracy between the $\ha$ and $\hh$ is such that their independent
signals would not be separable, except, possibly, in the $\mupmum$ mode.

\begin{figure}[htb]
\leavevmode
\begin{center}
\centerline{\psfig{file=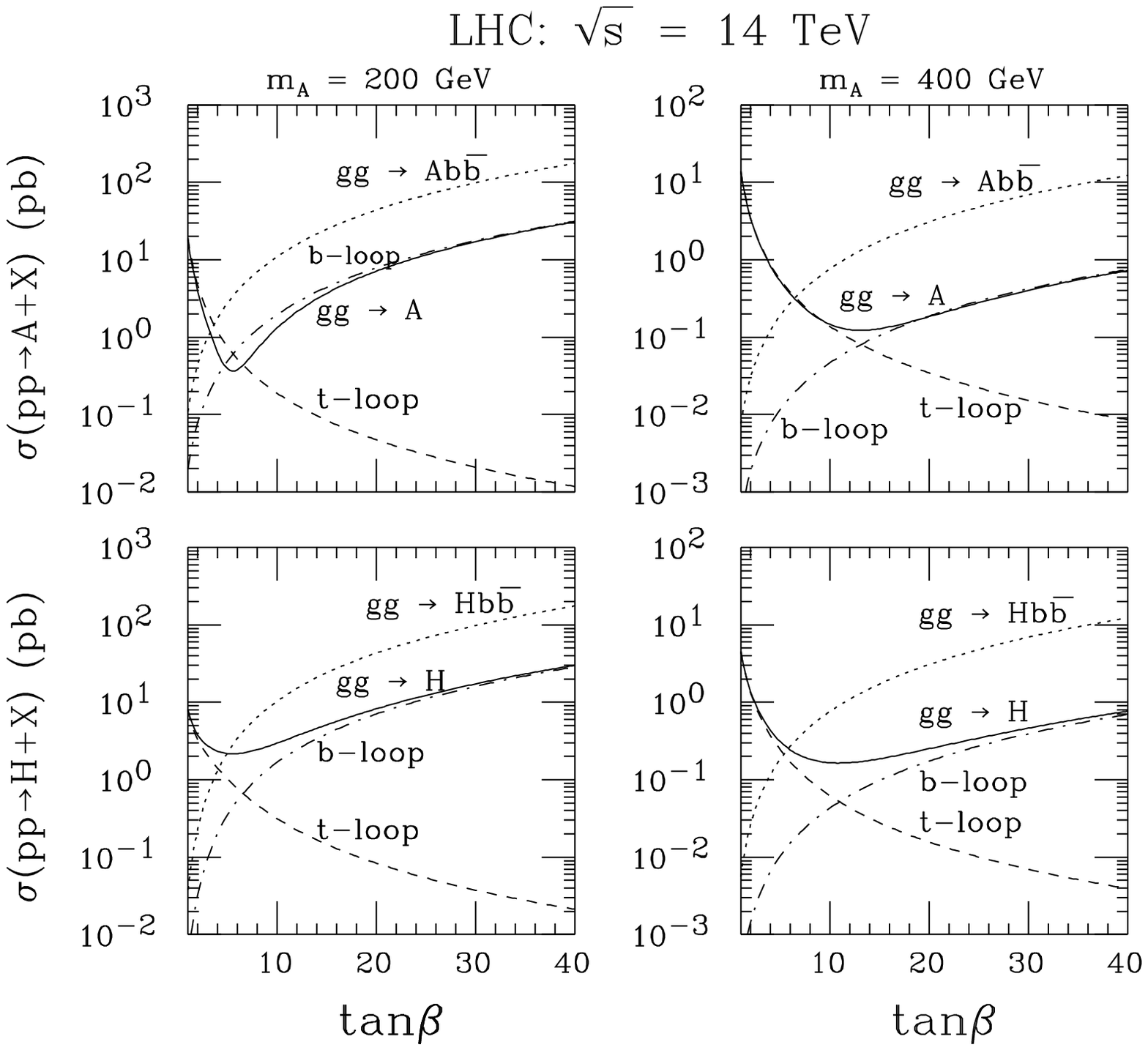,width=3.5in}}
\end{center}
\caption{The dependence of $gg\to\hh,\ha$ and $gg\to \hh b\anti b,\ha b\anti b$
cross sections at the LHC on $\tanb$ for $\mha=200$ and $400\gev$.
Also shown are the $gg\to \hh,\ha$ cross sections obtained
by retaining only $b$- or $t$-loop contributions to the one-loop coupling.}
\label{figgghhha}
\end{figure}

The $\tanb$ dependence of $\hh$ and $\ha$ rates implicit in the above
discussion is quantified in Fig.~\ref{figgghhha}, which displays
the $gg\to \hh$, $gg\to \ha$, $gg\to b\anti b\hh$ and $gg\to b\anti b\ha$
cross sections (and separate $t$ and $b$ loop contributions to the first
two).\footnote{QCD corrections to these cross sections are not included.
They have only been calculated for $gg\to\hh$ and $gg\to \ha$, for which cases
they increase the cross section by $\sim 50\%-100\%$.}
At low $\tanb$,
we see that the $gg\to \hh$ and, especially, $gg\to\ha$ cross sections
fall rapidly as the $t$-loop contribution falls with increasing $\tanb$.
At high $\tanb$, the rapid rise  of the
$gg\to \hh b\anti b$ and $gg\to \ha b\anti b$ cross sections is apparent.

The strong $\tanb$ dependence of the $gg\to \hh,\ha$
and $gg\to b\anti b\hh,b\anti b\ha$ discovery modes
will provide an opportunity for determining the otherwise somewhat
elusive $\tanb$ parameter. The sensitivity to $\tanb$ depends
on the accuracy with which these cross sections can be measured
and the rate of their variation with $\tanb$.  The possibility
of determining $\tanb$ in this manner was noted
in Ref.~\cite{atlas074} (see remarks above
Table 34 of the referenced paper). However, the specific results
quoted there disagree with those obtained here, and seem to be in error
\cite{tanbremark}.
In what follows it is demonstrated
that a fairly simple global characterization of the $\tanb$
errors turns out to be possible.

\begin{figure}[htb]
\leavevmode
\begin{center}
\centerline{\psfig{file=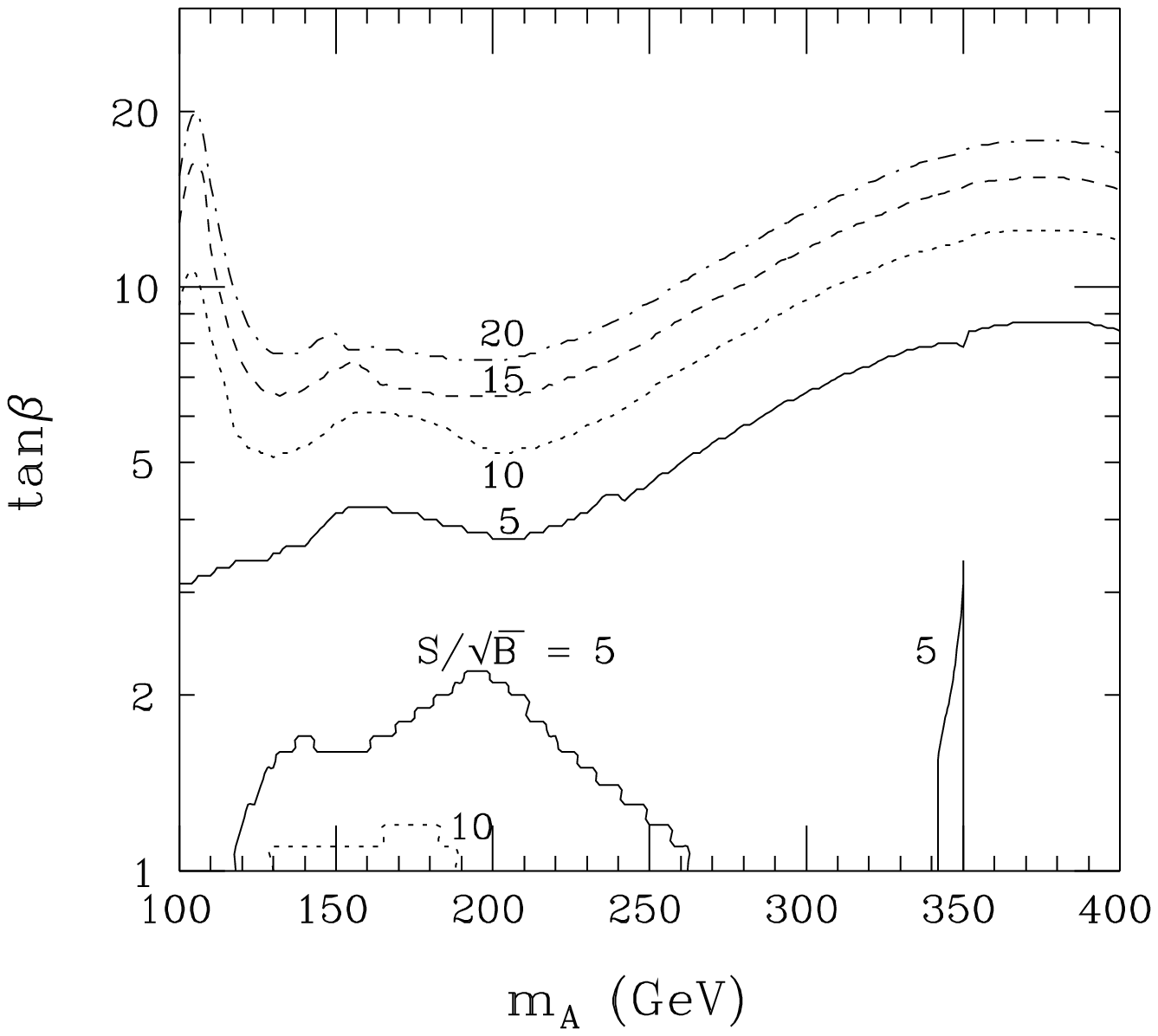,width=3.5in}}
\end{center}
\caption{The $\tanb$ values required for detection
of $\hh,\ha$ production with $\hh,\ha\to \tauptaum$
as a function of $\mha$ for $S/\protect\sqrt B=5,10,15,20$.}
\label{figtanb}
\end{figure}

We start with the signal and background results of Table 34
of Ref.~\cite{atlas074} for $\tanb=10$ and $\ha\to\tauptaum$ only.
There $S$ and $B$ are given for $L=100\fbi$ as a function of $\mha$, along
with the cross section employed before reduction by the efficiency
associated with cuts, particle identification and so forth. By comparing
$L\sigma$ (for $L=100\fbi$) 
with the tabulated $S$, we get the signal efficiency, $\eps_S$ for
each $\mha$. We then compute an effective statistical
significance for the combined $\hh$ and $\ha$ signals 
at $L=600\fbi$ as follows.
First we compute the total $\hh$ rate $S_H$
\begin{equation}
S_H=L\eps_S [\sigma(gg\to\hh)+\sigma(gg\to \hh b\anti b)]\br(\hh\to\tauptaum)\,.
\end{equation}
The analogous equation is used for the total $\ha$ rate $S_A$.
(The cross section times branching ratio for $\ha$
is found to be slightly larger, roughly by 10\%,
than quoted in Table 34 of Ref.~\cite{atlas074}.)
$B_H$ and $B_A$ are computed by scaling up the $L=100\fbi$ results
of Table 34 to $L=600\fbi$.
The net effective $S/\sqrt B$ for the combined $\hh$ and $\ha$
signals is computed according to the Ref.~\cite{atlas074} prescription:
\begin{equation}
\left[\left({S\over\sqrt B}\right)_{H}^2
+\left({S\over\sqrt B}\right)_{A}^2-2\eps_{HA}
\left({S\over\sqrt B}\right)_{H}\left({S\over\sqrt
B}\right)_{A}\right]^{1/2}\,,
\end{equation}
where $\eps_{HA}$ is a function of $r_M\equiv |\mhh-\mha|/\sigma_M$,
with $\sigma_M$ being the $\tauptaum$
mass resolution.  (At high luminosity we take $\sigma_M=21\gev$
--- see earlier SM LHC discussion for mass region M2.) The value
of $\eps_{HA}$ is: 0 (corresponding to no signal overlap) for $r_M>2$ ; 
$-0.33$ for $0.5\leq r_M\leq 2$; and $-1$ (\ie\ total overlap of the
signals) for $r_M<0.5$. The $\tanb$ values required for net $S/\sqrt B=5$,
10, 15 and 20 as a function of $\mha$ are shown in Fig.~\ref{figtanb}.
For large $\tanb$ values, $S/\sqrt B>5$ is always possible.  For
a limited range of $\mha$, $S/\sqrt B>5$ is also possible at low $\tanb$.

We next compute the error in the cross section determination as
\begin{equation}
{\Delta\sigma\over\sigma}=\left[ {S+B\over S^2}+(0.1)^2\right]^{1/2}\,,
\label{dsig}
\end{equation}
where the 0.1 is the presumed systematic uncertainty, assumed independent
of parameter choices. We have computed
the values of $\Delta\sigma/\sigma$ for the $\tanb$ values such that
$S/\sqrt B=5,10,15,20$ at $\mha=200\gev$ and $400\gev$.
For both $\mha$ values one finds
fractional errors of $[\Delta\sigma/\sigma]=0.22,0.14,0.12,0.11$, 
respectively. $\Delta\tanb$ can be approximately computed as
$\Delta\sigma[d\sigma/d\tanb]^{-1}$ or with greater precision by
searching for those values of $\tanb$ such that $\sigma$ changes by
$\Delta\sigma$; the results obtained in these two ways are virtually the same.
Dividing the absolute $\Delta\tanb$ error
by the $\tanb$ value required for the given $S/\sqrt B$,
one discovers that, for $S/\sqrt B\geq 10$,
the corresponding fractional errors in $\tanb$ are roughly 
independent of $\mha$: $\Delta\tanb/\tanb\sim 0.075,0.062,0.056$ for
$S/\sqrt B=10,15,20$ at both $\mha=200$ and $400\gev$.
Thus, when a $\geq 10\sigma$ signal can be detected in the $\tauptaum b\anti b$
final state channel a $\leq \pm 8\%$ determination of $\tanb$ will be possible,
a very useful level of accuracy. At $\tanb$ such that $ S/\sqrt B=5$,
there is more variation.  The full results for the high-$\tanb$
cases are summarized in Table~\ref{tanberrors}. At $\mha=200\gev$,
$S/\sqrt B=5$ is achieved at $\tanb\sim 2$ as well as at the higher $\tanb=3.5$
value at which we summarized results in Table~\ref{tanberrors}.
At $\tanb\sim 2$ and $\mha\sim 200\gev$, the percentage
error in the $\tanb$ determination is $\sim\pm 30\%$.

Of course, when there are two different $\tanb$ values yielding
the same signal rate (and also same $S/\sqrt B$), as at $\mha=200\gev$
above, we would be left with an ambiguity using the
totally inclusive procedures considered so far by ATLAS and CMS.
This ambiguity can be resolved by $b$-tagging.
For the high-$\tanb$ $S/\sqrt B=5,10,15,20$
contours, the signal rate is essentially entirely due to
the $gg\to\hh b\anti b,\ha b\anti b$ production mechanisms,
while on the low-$\tanb$ $S/\sqrt B=5$ and 10 contours, it is
the inclusive $gg\to \hh,\ha$ production mechanism
that dominates. Tagging
or anti-tagging $b$-quarks in the final state in association
with the $\tauptaum$ from $\hh,\ha$ decay would definitively
separate these mechanisms from one another and avoid any
ambiguity as to the correct $\tanb$ value.

\begin{table}[hbt]
\caption[fake]{We tabulate the percentage errors
at $\mha=200\gev$ and $400\gev$ 
for the $\hh,\ha\to\tauptaum$ signal and the corresponding
errors in the determination of $\tanb$ for the high-$\tanb$
contours such that $S/\sqrt B=5,10,15,20$, assuming
$L=600\fbi$ accumulated at the LHC.}
\footnotesize
\begin{center}
\begin{tabular}{|c|c|c|c|c|}
\hline
 Quantity & \multicolumn{4}{c|}{Errors} \\
\hline
\hline
$\mha$ & \multicolumn{2}{c|}{200 GeV} & \multicolumn{2}{c|}{400 GeV} \\
\hline
\ & $\Delta\sigma/\sigma$ & $\Delta\tanb/\tanb$ & $\Delta\sigma/\sigma$ & 
$\Delta\tanb/\tanb$  \\
\hline
$S/\sqrt B = 5$ & $\pm 20\%$ & $\pm 22\%$  & $\pm 22\%$ & $\pm 12\%$  \\
$S/\sqrt B = 10$ & $\pm 14\%$ & $\pm 7.8\%$ & $\pm 14\%$ & $\pm 7.4\%$ \\
$S/\sqrt B = 15$ & $\pm 12\%$ & $\pm 6.2\%$ & $\pm 12\%$ & $\pm 6.2\%$ \\  
$S/\sqrt B = 20$ & $\pm 11\%$ & $\pm 5.6\%$ & $\pm 11\%$ & $\pm 5.7\%$ \\
\hline
\end{tabular}
\end{center}
\label{tanberrors}
\end{table}

In the above analysis, we have implicitly assumed that 
$\br(\hh,\ha\to \tauptaum)$ will be either measured or calculable.  
More generally, conversion of 
measurements of $\sigma(\hh ,\ha )\br(\hh,\ha\to \tauptaum)$
to determinations of the cross sections and actual signal rate must include
systematic and/or statistical errors due 
to uncertainty in the $\tauptaum$ branching ratios.
Direct measurement of $\br(\hh,\ha\to \tauptaum)$ will require
NLC or FMC data. (See next subsection.) If only LHC data is available,
then the situation is more complicated, as we now describe.

At high enough $\tanb$, the enhancement of the 
$\hh,\ha$ couplings to $b\anti b$ and $\tauptaum$ 
in the MSSM will imply that these are the  only modes of importance 
and that they will be in the ratio\footnote{$\mb(\mha)$
is the running $b$-quark mass evaluated at scale $\mha\sim\mhh$.
For $\mha$ in the $150-400\gev$ range, $\mb^2(\mha)\sim 0.5 \mb^2(pole)$.}
$3\mb^2(\mha):m_{\tau}^2$. For such $\tanb$ values,
systematic uncertainty will be small. However, the $\tanb$
values of Fig.~\ref{figtanb} required for $S/\sqrt B=5$ and, to a lesser
extent $S/\sqrt B=10$, are not always
large enough to guarantee that other (\eg\ SUSY) decays of the $\hh,\ha$
can be neglected. (See next subsection for some examples.) 

Even at high $\tanb$ it would be very helpful to directly
measure the $\br(\tauptaum)/\br(b\anti b)$ ratio(s) as a confirmation
of the theoretical prediction.
To measure the ratio requires measuring the rates
for the $\hh b\anti b,\ha b\anti b\to
b\anti b b\anti b$ final states. This was considered in
Ref.~\cite{dgv}; this semi-theoretical study
found that $S/\sqrt B$ values
comparable to those in the $b\anti b \tauptaum$ final state are possible
if excellent $b$-tagging efficiency and purity are achieved at high luminosity.
Full detector simulation studies by the ATLAS and CMS experimental groups 
are underway.  Accuracy in the measurement of the $\hh b\anti b,\ha b\anti b\to
b\anti b b\anti b$ rates will be limited if $S/B$ is as small as typically
associated with $S/\sqrt B\sim 5$ signals in the study of Ref.~\cite{dgv}.
Precise statements must await the completion of ongoing work.

At lower $\tanb$ values, where
$gg\to \hh,\ha\to\tauptaum$  production mode(s) are dominant,
systematic uncertainties 
associated with imperfect knowledge of the $\tauptaum$
branching ratios will definitely be a major consideration.
At low $\tanb$, other decay channels (most notably $\hh\to \hl\hl$
and $\ha\to Z\hl$, not to mention SUSY pair channels
if kinematically allowed) are expected to be important
in the MSSM, but determining their magnitude will be very difficult 
without the NLC. Further,
isolation of the inclusive $gg\to \hh,\ha\to b\anti b$
production/decay mode is almost certainly impossible.

\subsection{Probing $\tanb$ and GUT-scale boundary conditions 
using $\hh\ha$ and $\hp\hm$ production at the NLC or FMC \protect\cite{gk}}

If supersymmetry is discovered, one of our primary goals will
be to fully test the model and determine the underlying GUT boundary conditions
at the GUT/Planck mass scale.  The heavy Higgs bosons of the model
are an important component in this program. First, detection of
the $\hh$, $\ha$ and $\hpm$ is required in order to verify the Higgs
sector content.  This may only be possible in the pair production modes
$\hh\ha$ and $\hp\hm$ at a $\epem$ or $\mupmum$ collider with
$\rts\gsim 2\mha$.  (Recall that the MSSM Higgs sector
structure requires $\mhh\sim\mha\sim\mhpm$ at higher masses.)
In Ref.~\cite{gk}, the influence of SUSY decays on our ability to
detect pair production is assessed and a strategy for using these and other
decays to probe the GUT boundary conditions is developed.
A related study has recently appeared in Ref.~\cite{fengmoroi}.

In Ref.~\cite{gk}, these issues are examined in the context of six not
terribly different GUT-scale boundary condition scenarios in
which there is universality for the soft-SUSY-breaking parameters
$\mhalf$, $m_0$ and $A_0$ associated with soft gaugino masses, soft scalar
masses and soft Yukawa coefficients, respectively \cite{susyref}.
After requiring that the electroweak symmetry breaking generated
as a result of parameter evolution yield the correct $Z$ boson mass,
the only other parameters required to fully specify a model 
in this universal-boundary-condition class are $\tanb$ and the
sign of the $\mu$ parameter (appearing in the superpotential $W\ni \mu \hat
H_1\hat H_2$). The six models considered in Ref.~\cite{gk} are
denoted \DM, \DP, \NSM, \NSP, \HSM, \HSP, where the superscript indicates
$\sign(\mu)$. Each is specified by a particular 
choice for $m_0:\mhalf:A_0$, thereby leaving
only $\mhalf$, in addition to $\tanb$,
as a free parameter in any given model. Pair production is
then considered in the context of each model as a function
of location in the kinematically and constraint allowed
portion of $(\mhalf,\tanb)$ parameter space.

Ref.~\cite{gk} finds that event rates for anticipated machine luminosities
are such $\hh\ha$ and $\hp\hm$ pair
production can be detected in final state modes where $\hh,\hh\to b\anti b$
or $t\anti t$ and $\hp\to t\anti b,\hm\to b\anti t$ even when the branching
ratios for SUSY decays are substantial. Further, the mass of the $\hh$
or $\ha$ can be determined with substantial accuracy using the fully
reconstructable all jet final states associated with these modes.
Perhaps of greatest ultimate importance, 
in much of the kinematically and phenomenologically allowed
parameter space Higgs branching ratios for 
a variety of different decay channels can be measured by ``tagging''
one member of the Higgs pair in a fully reconstructable all jet decay mode
and then searching for different types of final states in the decay of the 
second (recoiling) Higgs boson.

The power of Higgs pair observations for determining the GUT
boundary conditions is most simply illustrated by an example.
Let us suppose that the \DM\ model with $\mhalf=201.7\gev$ and $\tanb=7.5$
is nature's choice. This implies that $\mha=349.7\gev$ and
$\mcpmone=149.5\gev$. Experimentally, one would measure $\mha$ as above
and $\mcpmone$ (the lightest chargino) mass in the usual way and then
infer the required parameters for a given model. For the six models
the parameters are given in Table~\ref{mhalftanbtable}.
Note that if the correct GUT scenario can be ascertained experimentally, then
$\tanb$ and $\mhalf$ will be fixed.

\begin{table}[hbt]
\caption[fake]{We tabulate the values of $\mhalf$ (in GeV)
and $\tanb$ required in each of our six scenarios in order
that $\mha=349.7\gev$ and $\mcpmone=149.5\gev$.
Also given are the corresponding values of $\mhh$. Masses are in GeV.}
\begin{center}
\begin{tabular}{|c|c|c|c|c|c|c|}
\hline
 & \DM\ & \DP\ & \NSM\ & \NSP\ & \HSM\ & \HSP\ \\
\hline
\hline
$\mhalf$ & 201.7 & 174.4 & 210.6 & 168.2 & 203.9 & 180.0 \\
$\tanb$ & 7.50 & 2.94 & 3.24 & 2.04 & 12.06 & 3.83 \\
$\mhh$ & 350.3 & 355.8 & 353.9 & 359.0 & 350.1 & 353.2 \\
\hline
\end{tabular}
\end{center}
\label{mhalftanbtable}
\end{table}

Determination of the GUT scenario proceeds as follows.
Given the parameters required for the observed $\mha$
and $\mcpmone$ for each model, as tabulated in Table~\ref{mhalftanbtable},
the rates for different final states of the recoil (non-tagged) Higgs boson
in pair production can be computed. Those for the input
\DM\ model are used to determine the statistical accuracy
with which ratios of event numbers in different types of final states
can be measured.\footnote{We focus on ratios in order to
be less sensitive to systematic uncertainties in efficiencies \etc;
however, from Ref.~\cite{gk} it is clear that
absolute rates will also be useful in some instances.}
The ratios predicted in
the \DP, \NSM, \NSP, \HSM, and \HSP models will be different
from those predicted for the input \DM\ model.  Thus,
the statistical uncertainty predicted for the various ratios in the input \DM\
model can be used to compute the $\chisq$ by which the predictions of
the other models differ from the central values of the input \DM\ model.
The results for a selection of final state ratios are given
in Table~\ref{chisqtable}. The final states considered are:
$b\anti b$ and $t\anti t$ 
for the $\hh,\ha$; $\hl\hl$ (light Higgs pair, with $\hl\to b\anti b$) for
the $\hh$; $\hl\wp$ and $\tau^+\nu_\tau$ for the $\hp$ (or
the charge conjugates for the $\hm$); and SUSY modes
(experimentally easily identified by the presence of missing energy)
classified according to the number of charged leptons summed over
any number of jets (including 0). All branching ratios and reasonable
efficiencies are incorporated in the statistical errors employed
in constructing this table. The effective luminosity $\leff=80\fbi$ 
is equivalent to
an overall tagging and reconstruction efficiency for events
of $\eps=0.4$ at a total integrated luminosity of $L=200\fbi$.
Results presented are for $\rts=1\tev$.

\begin{table}[hbt]
\caption[fake]{We tabulate $\Delta\chi^2_i$ 
(relative to the \DM\ scenario) for the indicated branching
fraction ratios as a function of scenario,
assuming the measured $\mha$ and $\mcpmone$ values are $349.7\gev$
and $149.5\gev$, respectively. The SUSY channels have been resolved into 
final states involving a fixed number of leptons.  
The error used in calculating each $\Delta\chi^2_i$ is the approximate
$1\sigma$ error with which the given ratio could be measured
for $\leff=80\fbi$ at $\rts=1\tev$ {\it assuming that the \DM\
scenario is the correct one}.
}
\begin{center}
{\footnotesize
\begin{tabular}{|c|c|c|c|c|c|}
\hline
Ratio & \DP\ & \NSM\ & \NSP\ & \HSM\ & \HSP\ \\
\hline
\multicolumn{6}{|c|} {$\langle\hh,\ha\rangle$} \\
\hline
$ [0\ell][\geq0 j]/b\anti b,t\anti t$ 
 & 12878 & 1277 & 25243 & 0.77 & 10331 \\
$ [1\ell][\geq0 j]/b\anti b,t\anti t$ 
  & 13081 & 2.41 & 5130 & 3.6 & 4783 \\
$ [2\ell][\geq0j]/b\anti b,t\anti t$ 
  & 4543 & 5.12 & 92395 & 26.6 & 116 \\
$ \hl\hl/ b\anti b$  & 109 & 1130 & 1516 & 10.2 & 6.2 \\
\hline
\multicolumn{6}{|c|} {$\hp$} \\
\hline
$ [0\ell][\geq0j]/t\anti b$ 
 & 12.2 & 36.5 & 43.2 & 0.04 & 0.2 \\
$ [1\ell][\geq0j]/t\anti b$ 
 & 1.5 & 0.3 & 0.1 & 5.6 & 0.06  \\
$\hl W/ t\anti b$ 
 & 0.8 & 0.5 & 3.6 & 7.3 & 0.3 \\
$\tau\nu/ t\anti b$ 
 & 43.7 & 41.5 & 47.7 & 13.7 & 35.5 \\
\hline
$\sum_i\Delta\chi^2_i$ & 30669 & 2493 & 124379 & 68 & 15272 \\
\hline
\end{tabular}
}
\end{center}
\label{chisqtable}
\end{table}

From Table~\ref{chisqtable}
it is clear that the five alternative models can be discriminated
against at a high (often very high) level of confidence.  Further
subdivision of the SUSY final states into states containing
a certain number of jets yields even more discrimination power \cite{gk}.
Thus, not only will detection
of Higgs pair production in $\epem$ or $\mupmum$
collisions (at planned luminosities) be 
possible for most of the kinematically accessible
portion of parameter space in a typical GUT model, but also the detailed rates
for and ratios of different neutral and charged Higgs decay final states
will very strongly constrain the possible GUT-scale boundary condition
scenario and choice of parameters, \eg\ $\tanb$ and $\mhalf$, therein.

\subsection{Implications of LHC and NLC data
for $s$-channel discovery of the $\hh$ and $\ha$ at the FMC}

As we have already noted, 
colliders other than the FMC offer various mechanisms
to directly search for the $\ha,\hh$, but have significant limitations:
\begin{itemize}
\item There are regions in $(\mha,\tanb)$ parameter space at moderate
$\tan \beta$, $\mha\gsim 200\gev$ in which the $\hh,\ha$ cannot be detected
at the LHC.
\item At the NLC one can use the mode $\epem\to \zstar\to \hh\ha$,
but it is limited to $\mhh\sim \mha\lsim \sqrt{s}/2$.
\item A $\gamma \gamma$ collider could probe heavy Higgs up to masses of
$\mhh\sim \mha\sim 0.8\sqrt{s}$, but this would quite likely require
$L> 100{\fb}^{-1}$, especially if the Higgs bosons are at the upper
end of the  $\gamma \gamma$ collider energy spectrum~\cite{ghgamgam}.
\end{itemize}
In contrast, there is an excellent chance of being able to detect
the $\hh,\ha$ at a $\mupmum$ collider provided only that $\mha$ is smaller
than the maximal $\rts$ available. This could prove to be very important
given that GUT MSSM models usually predict $\mha\gsim 200\gev$.

A detailed study of $s$-channel production
of the $\hh,\ha$ has been made in Ref.~\cite{bbgh}, upon which
the ensuing discussion is based.
The signals become viable when $\tanb>1$
(as favored by GUT models) since the $\mupmum\hh$ and $\mupmum\ha$
couplings are proportional to $\tanb$. In particular, 
even though $\gamhh,\gamha$ are big (see Fig.~\ref{hwidths}) at high $\tanb$, 
due to large $b\anti b$ decay widths, $\br(\hh,\ha\to\mupmum)$
approaches a constant value that is large enough to imply
substantial cross sections $\sighhbar,\sighabar$. (We recall from
the earlier SM Higgs FMC discussion that for a general $\h$,
$\sighbar\propto \br(\h\to\mupmum)$
when the Gaussian beam spread $\sigrts$ is smaller than $\gamh$.)
The optimal strategy 
for $\hh,\ha$ detection and study at the FMC
depends upon the circumstances.

First, it could be that the $\hh$ and/or $\ha$ will already have been
discovered at the LHC.  With $L=300\fbi$ (ATLAS+CMS) of integrated           
luminosity, this would be the case if $\tanb\lsim 3$ or $\tanb$ 
is above an $\mha$-dependent lower bound (\eg\ $\tanb\gsim 10$ for $\mha\sim
400\gev$).\footnote{For $\tanb\lsim 3$, 
one makes use of modes such as $\hh\to\hl\hl\to
b\anti b \gam\gam$ and $\hh \to ZZ^{(*)}\to 4\ell$, when $\mhh\lsim 2\mt$,
or $\hh,\ha\to t\anti t$, when $\mhh,\mha\gsim 2\mt$.  At high $\tanb$,
the enhanced production rates for $b\anti b \hh,b\anti b\ha$ with
$\hh,\ha\to\taup\taum$ are employed.}
Even if the $\hh,\ha$ have not been detected,
strong constraints on $\mha$ are possible if precision measurements
of the properties of the $\hl$ (such as the $b\anti b/W\wstar$
and $c\anti c/b\anti b$ event rate ratios and the $(\mupmum\hl)^2$
coupling-squared) are made via $s$-channel production
at the FMC or in $\rts=500\gev$ running at the NLC, or
by combining these two types of data --- see earlier discussions
associated with the errors tabulated in
Tables~\ref{nlcerrors}, \ref{fmcerrors} and \ref{nlcfmcerrors}.
By limiting the $\rts$ scan for the $\hh$ and $\ha$
in the $s$-channel to the $\mha\sim\mhh$ 
mass region preferred by $\hl$ measurements,
we would greatly reduce the luminosity needed to
find the $\ha$ and $\hh$ via an $s$-channel scan as compared to that required
if $\mha$ is not constrained.

With such pre-knowledge
of $\mha$, it will be possible to detect and perform detailed
studies of the $\hh,\ha$ for essentially all $\tanb\geq 1$
provided only that $\mha\lsim \rts_{\rm max}$.\footnote{We
assume that a final ring optimized for maximal luminosity at $\rts\sim \mha$
would be constructed.} If $\tanb\lsim 3$, then
excellent resolution, $R\sim 0.01\%$, will be necessary for detection
since the $\ha$ and $\hh$
become relatively narrow for low $\tanb$ values (see Fig.~\ref{hwidths}).. 
For higher $\tanb$ values, $R\sim 0.1\%$ is adequate
for $\hh,\ha$ detection, but $R\sim 0.01\%$ would be required in order
to separate the rather overlapping $\hh$ and $\ha$ peaks (as a function
of $\rts$) from one another~\cite{bbgh}.

Even without pre-knowledge of $\mha$, 
there would be an excellent chance for discovery of the $\ha,\hh$
Higgs bosons in the $s$-channel at a $\mupmum$ collider if they
have not already been observed at the LHC.
This is because non-observation at the LHC implies 
that $\tanb\gsim 3$ while it is precisely for
$\tanb\gsim 2.5-3$ that detection of the $\ha,\hh$ is possible~\cite{bbgh}
in the mass range from 200 to 500 GeV via an $s$-channel scan in 
$\mupmum$ collisions. (The lower $\tanb$ reach given
assumes that $\ltot=200\fbi$ is devoted to the scan.
The detailed strategy outlined in Ref.~\cite{bbgh}, as to how
much luminosity to devote to different $\rts$ scan settings
in the $200-500\gev$ range, must be employed.)
That the LHC and the FMC are complementary in this
respect is a very crucial point. Together, the LHC and FMC
guarantee discovery of the $\ha,\hh$ after 3 to 4 years of
high luminosity operation each, provided $\mha\lsim 500\gev$.
Once $\mha,\mhh$ are known, very precise measurements of some of
the crucial properties of the $\hh,\ha$ 
(including a scan determination of their
total widths) become possible~\cite{bbgh}.

In the event that the NLC has not been constructed, it could be
that the first mode of operation of the FMC would be to optimize
for and accumulate luminosity at, say, $\rts=500\gev$. 
In this case, there is still a high probability
for detecting the $\hh,\ha$ if they have not been
observed at the LHC. First, if $\mha\sim\mhh\lsim \rts/2\sim 250\gev$ then
$\mupmum\to \hh\ha$ (and $\hp\hm$) pair production will be observed. Second,
although reduced in magnitude compared to an electron
collider, there is a long low-energy bremsstrahlung tail 
at a muon collider that provides a
self-scan over the full range of $\rts$ values below 
the nominal operating energy.
Observation of $\ha,\hh$ $s$-channel peaks in the $b\anti b$ mass 
($m_{b\anti b}$) distribution
created by this bremsstrahlung tail may be possible~\cite{bbgh}.
The region of the $(\mha,\tanb)$ parameter space plane for which
a peak is observable depends strongly on the $m_{b\anti b}$
resolution. For excellent $m_{b\anti b}$
resolution of order $\pm 5\gev$ and integrated luminosity
of $L=200\fbi$ at $\rts=500\gev$, 
the $\ha,\hh$ peak(s) are observable for $\tanb\gsim 4-5$
if $500\gev\geq\mha\geq 250\gev$.~\footnote{Required $\tanb$ values increase
dramatically as one moves into the $\mha\sim \mz$ zone,
but this region is covered by $\hh\ha$ pair production.}

Finally, if neither the LHC nor a FMC scan of the $\leq 500\gev$ region
has discovered the $\hh,\ha$, but supersymmetric particles and the $\hl$
have been observed, we would believe that the $\hh,\ha$ must exist
but have $\mha\sim\mhh\geq
500\gev$ . Analyses of the SUSY spectrum in the GUT context
and precision $\hl$ studies
might have yielded some prejudice for the probable $\mha$, and
an extension of the FMC energy up to the appropriate $\rts\sim \mha$ 
for $s$-channel discovery of the $\hh,\ha$
could be considered. However, at this point, even if we have
developed a favored range for $\mha$, it would probably
be most worthwhile to consider a machine with much higher $\rts$.
A popular reference $\mupmum$ collider design is one for $\sqrt{s}=4\tev$.
Ref.~\cite{gk} shows that such an energy with appropriately matched luminosity
would allow discovery of $\mupmum\to\ha\hh$ 
and $\hp\hm$ pair production,
via the $b\anti b$ or $t\anti t$ decay channels of the $\hh,\ha$
and $t\anti b,\anti t b$ decay channels of the $\hp,\hm$,
up to masses very close to $\mha\sim \mhh\sim \mhpm \sim
2\tev$, even if SUSY decays of the $\hh,\ha,\hpm$ are substantial. 
(This mass range certainly includes that expected in
any supersymmetric model that provides a solution to the
naturalness and hierarchy problems.) 
Detailed studies of the $\hh,\ha,\hpm$ of the type discussed
in the previous subsection would be possible once they were discovered.
An $\epem$ collider with $\rts\lsim 1.5-2\tev$ is also probably viable,
and could probe $\mha\sim\mhh\sim\mhpm\lsim \rts/2$~\cite{gk,fengmoroi}. 
In the absence
of a strong prejudice based on GUT boundary conditions, only the $2\tev$
option could be presumed (purely on the basis of naturalness)
to guarantee $\hh,\ha,\hpm$ discovery.

\subsection{Searching for a doubly-charged Higgs boson \protect\cite{glp}}

Doubly-charged Higgs bosons ($\dmm,\dpp$) appear in many 
extensions of the Standard Model Higgs sector,
such as left-right symmetric models, and can be relatively light,
the current bound being $\mdmm> 45\gev$ from LEP. 
They have received less attention than neutral and singly-charged Higgs
bosons because they can lead to phenomenological difficulties.
In particular, $\rho\equiv{\mw^2\over[\cos^2\theta_W\mz^2]}=1$ is not natural
unless any neutral Higgs boson that is part of the same Higgs multiplet
has zero vacuum expectation value.\footnote{Even if the 
model is constructed so that $\rho=1$
at tree-level, one loop corrections are infinite unless the vev is zero.}
Thus, models with zero vev are favored. In
left-right symmetric models, zero vev is natural for the 
neutral member of the Higgs triplet
associated with the left-handed sector; the right-handed sector 
neutral Higgs must have substantial vev and the associated
$\dmm_R$ phenomenology is very different.
Of course, representations can be chosen for which there is no
neutral member; \eg\ a $T=1/2,Y=-3$ representation contains 
only a $\dmm$ and a $\delm$.  Coupling constant unification 
should also be taken into account. It is amusing to
note that coupling constant unification occurs in the non-supersymmetric
Standard Model if a single $|Y|=2$ triplet representation is 
included in addition to the standard $|Y|=1$ doublet Higgs representation.
On the other hand, in the minimal supersymmetric extension of the SM,
inclusion of triplet Higgs field(s) destroys unification.
However, this can always be cured by introducing intermediate scale matter
so as to maintain unification, as done for example in 
left-right symmetric models. Thus, potential phenomenological
difficulties are not all that difficult to avoid, and
we should be on the look-out for signatures
of exotic Higgs representations, the clearest signal
being the existence of a doubly-charged Higgs boson.

An especially important feature of a $\dmm$ is the fact that for many
representation choices $\dmm\to \ell^-\ell^-$ couplings are allowed.
Indeed, in left-right symmetric models the corresponding
neutral field couplings give rise
to the see-saw mechanism and thereby naturally small neutrino masses.
Detection and study of a $\dmm$ provides important opportunities
for determining whether such couplings exist and how large they are.

We begin our discussion by considering the phenomenologically
natural models in which the vev of any neutral member of the 
multiplet containing the $\dmm$ is zero. This implies that
the $\dmm\to \wm\wm$ coupling is also zero. If the $\dmm$ also
couples to $\ell^-\ell^-$, then the resulting phenomenology
is very special and easily identified.  There are only
two production mechanisms: $\gam^\star,\zstar\to \dmm\dpp$
and $\ell^-\ell^-\to \dmm$. 
For the class of model being considered, it is also 
very possible that $\br(\dmm\to\ell^-\ell^-)\sim 1$ for $\ell=e$, $\mu$, or 
(most probably?) $\tau$. The only competing modes that might be present are
$\dmm\to \wm\delm$ and $\dmm\to\delm\delm$, where $\delm$ is a member of the same
SU(2)$_L$ multiplet as the $\dmm$.  Generally, $\mdelm$ is not very
different from $\mdmm$, and only the $\wm\delm$ mode has a significant likelihood
of being two-body allowed; in many models both the $\wm\delm$ and $\delm\delm$
modes can only proceed virtually.
Even when the $\wm\delm$ channel is two-body allowed, one finds
that $\ell^-\ell^-$ could be the dominant decay channel if the
$\dmm\to\ell^-\ell^-$ coupling is not too much lower than the current bound(s).
We note that decay of the $\dmm$ will occur inside the detector even if
the $\wm\delm$ decay mode is highly suppressed, so long
as the $\dmm\to\ell^-\ell^-$ coupling
is not extremely small (smaller than is preferred, for example,
for the $\dmm_L$ of a left-right symmetric model).

The $\ell^-\ell^-\to\dmm$ ($\ell=e$ or $\mu$) production mechanism
was studied in Ref.~\cite{gemem}. It can lead to detectable signals
down to remarkably small magnitudes of the $\dmm\to \ell^-\ell^-$
coupling --- far below current limits; for couplings near the current limits
very high production rates result, implying the possibility
of a $\dmm$ factory. However, for $\ell^-\ell^-\to \dmm$, 
one must have $\rts\sim \mdmm$. To avoid wasting time on a scan,
it is highly advantageous if $\mdmm$ is known ahead of time.
Thus, prior detection of $\dmm\dpp$ pair production would be of great value.
Discovery of a $\dmm$ with decays to $\ell^-\ell^-$
would in and of itself provide
a compelling motivation for building an $\ell^-\ell^-$ collider designed
for use as a $\dmm$ factory.

Pair production of $\dmm\dpp$ in $\epem$ or $\mupmum$ collisions 
requires only sufficient $\rts$.  Although no specific study was
performed, it seems very likely that discovery up to $\mdmm\lsim \rts/2$
would be possible.  However, the NLC and FMC are still more than a decade
away at best.

\begin{figure}[ht]
\leavevmode
\epsfxsize=3.250in
\hspace*{0.25in}
\epsffile{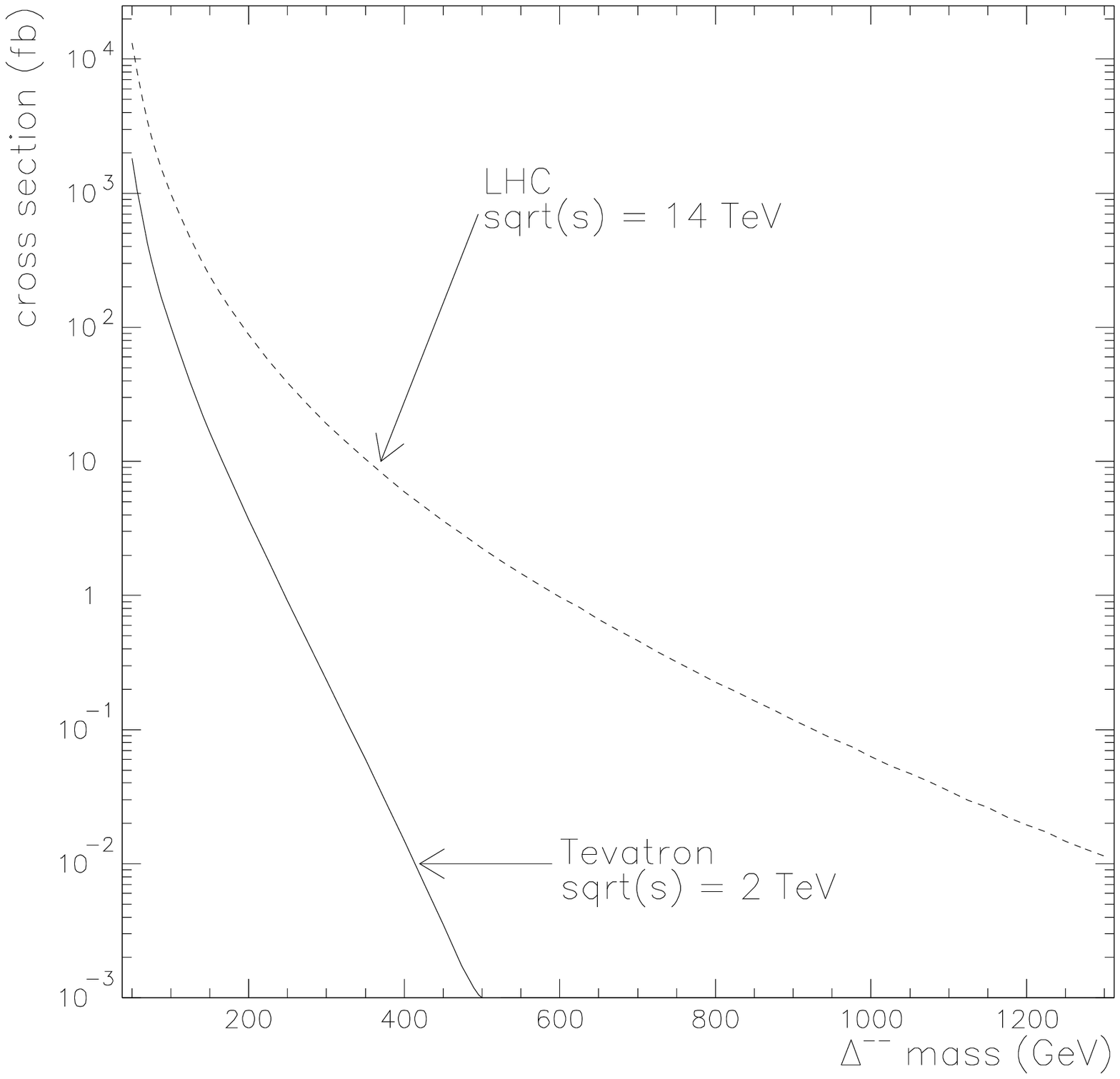}
\caption{$\dpp\dmm$ pair production
cross section as a function of $\dmm$ mass
for both the Tevatron and the LHC.}
\label{figxsec}
\end{figure}

In \cite{glp} discovery of $p\anti p\to \gam^\star,\zstar,\to \dmm\dpp$
with $\dmm\to\ell^-\ell^-$ and $\dpp\to\ell^+\ell^+$ is investigated.
The $\gam^\star,\zstar\to \dmm\dpp$ coupling is always present,
although slightly model-dependent; for definiteness a $T=1,T_3=-1,Y=-2$
$\dmm$ is considered (as found in several popular models). 
The Tevatron and LHC cross sections for $\dmm\dpp$ pair production
are plotted as a function of $\mdmm$ in Fig.~\ref{figxsec}.
Discovery limits for $\dmm\dpp$ are obtained assuming
that a single $\dmm\to\ell^-\ell^-$ decay channel is dominant with
$\ell=e$, $\mu$ or $\tau$.
A full Monte Carlo simulation is performed at Tevatron energies. Events
are generated in PYTHIA and passed through the Run I CDF detector 
simulation. For $\ell=e,\mu$ it is found that backgrounds are negligible
once a like-sign dilepton pair with high mass is required, and
it is purely a matter of having a handful of very clean events.
For $\ell=\tau$, the need to identify the $\tau$ by its decay
to an isolated lepton or hadron leads to a significant background level,
implying a smaller discovery reach.
Assuming $\br(\dmm\to\ell^-\ell^-)\sim 1$,
it is demonstrated that detection of the $\dmm$ at the Tevatron
(operating at $\rts=2\tev$ with $L=30\fbi$) will be possible for $\mdmm$ 
up to $300\gev$ for  $\ell=e$ or $\mu$ and  $180\gev$ for $\ell=\tau$.
These results should improve slightly if the greater coverage of the
TeV33 detector upgrades is incorporated.
The corresponding limits at the LHC are estimated by requiring
the same raw number of events before cuts and efficiencies as
needed at the Tevatron ($\sim 10$ for $\ell=e,\mu$ and $\sim 300$ for
$\ell=\tau$) yielding $\mdmm$ discovery up to roughly $925\gev$ ($1.1\tev$)
for $\ell=e,\mu$ and $475\gev$ ($600\gev$) for $\ell=\tau$,
assuming total integrated luminosity of $L=100\fbi$ ($L=300\fbi$).
For $\ell=e,\mu$, the reach of the 
LHC detectors will likely be even greater than
this, due to the improved lepton acceptance and resolution
anticipated over the current generation of hadron collider detectors.
For $\ell=\tau$, this simple extrapolation may not account
for a different signal-to-background ratio in 
$\tau$ selection at the LHC.  A full study is necessary to evaluate this.

Thus, if a $\dmm$ with moderate mass and the assumed
properties exists, discovery at TeV33 is not improbable; the LHC
allows discovery up to remarkably large masses.
Once found, the importance of pursuing $\ell^-\ell^-\to\dmm$ collisions
is easily argued \cite{gemem}. In particular,
it is very likely that the magnitude of the $\dmm\to\ell^-\ell^-$ coupling
can only be determined by doing so.
Indeed, observation of $\dmm\dpp$ pair production in only 
a single $\dmm\to \ell^-\ell^-$ channel provides
no information on the ${\ell\ell}$ coupling magnitude. 
(Of course, if more than one $\ell\ell$
channel is seen, ratios of the ${\ell\ell}$ couplings could be obtained.)
Only if the $\dmm\rta\delm\wm$ decay channel [for which
the partial width can be computed and compared to the
$\ell^-\ell^-$ partial width]
is also seen, can one get an estimate of the ${\ell\ell}$ 
coupling magnitude(s).
In contrast, an $e^-e^-$ ($\mu^-\mu^-$) collider would provide a 
direct measurement of the ${ee}$ (${\mu\mu}$) coupling.
This illustrates an important complementarity between the NLC or FMC and 
hadron colliders. Discovery of a $\dmm$ prior to
the construction and operation of the $\epem,\emem$ collider NLC complex
or the FMC analogue
would be very important in determining the energy range over which good 
luminosity and good energy resolution for $\emem$ or $\mu^-\mu^-$
collisions should be a priority.

Of course, the possibility that the $\dmm$ is part of a multiplet
whose neutral member has significant vev should not be ignored.
The $\dmm_R$ of the left-right symmetric model must fall into this class.
Such a $\dmm$ will have substantial $\wm\wm$ coupling.  There has
been substantial work on the related phenomenology \cite{dmmvev}.
An $\emem$ or $\mu^-\mu^-$ collider would be of particular value
in exploring such a $\dmm$.  In addition to the possibility of
direct $s$-channel production through the leptonic coupling, 
the non-zero $\wm\wm\to\dmm$ coupling will typically yield a substantial
cross section for $\emem~{\rm or}~\mu^-\mu^-\to \nu\nu\dmm$ production. 
Further, if $\mdmm\gsim 2\mw$, then $\dmm\to\wm\wm$ decays are
very likely to be dominant; detection of such a $\dmm$ at
a hadron collider might not be straightforward.  Thus, it could happen
that one would first discover the $\dmm$ in the $\wm\wm$-fusion mode,
at which point it would be important to turn to the $s$-channel
production probe of its possible $\emem$ or $\mu^-\mu^-$ couplings
by lowering the machine energy.

\section{Conclusions}

There have been two primary focuses in this report.
\begin{itemize}
\item 
We performed a first detailed study of
the accuracy with which the branching ratios,
couplings, total width and mass of a SM-like Higgs boson can be measured
in a model-independent way.
A number of new strategies and techniques were developed 
during the course of these studies.
A thorough evaluation
of the possibilities and errors at the LHC is still very much in progress.
Still, the results obtained to date indicate that many important
properties can be measured with substantial accuracy;
see Tables~\ref{2gamerrors}-\ref{m4errors}, Fig.~\ref{figdgam},
and Table~\ref{dmhsm}.
The simpler and cleaner environment at the NLC or FMC allowed us to perform a 
reasonably complete study at these machines, with very encouraging results;
see Tables~\ref{fmcsigbrerrors}-\ref{nlcfmcerrors}, Fig.~\ref{figdgam},
and Table~\ref{dmhsm}.
The errors quoted are those that would materialize after substantial
luminosity has been accumulated: $L=600\fbi$ at ATLAS+CMS at the LHC;
$L=200\fbi$ at the NLC (or FMC) in $\rts=500\gev$ running; and
$L=200\fbi$ in an $s$-channel scan at the FMC of the SM Higgs resonance peak.

One significant conclusion is the great desirability of being able
to accumulate $L=200\fbi$ both in $\rts=500\gev$ running and in
a FMC $s$-channel scan if the Higgs mass is below $2\mw$. 
This could be accomplished if both the NLC
and a low-energy FMC were constructed.

\item
We examined a number of issues and new ideas associated with extensions
of the simple one-doublet SM Higgs sector.  Of particular interest
were supersymmetric extensions of the SM, including the MSSM
and NMSSM. Contributions summarized here included:
\begin{enumerate}
\item
A demonstration of the substantial accuracy with which 
the $t\anti t\h$ and $ZZ\h$ couplings could be determined
for a general $\h$ of arbitrary CP nature at a $\rts=1\tev$ NLC.
\item A demonstration 
that in the next-to-minimal supersymmetric model (with Higgs
sector consisting of two Higgs doublets and one Higgs singlet), there
are regions of parameter space for which no Higgs boson can be found
at the LHC and/or LEP2, implying that in some circumstances Higgs
discovery would have to await the NLC or FMC.
\item A demonstration that at low $\tanb$, the $\ha$ of the MSSM 
could be detected at the LHC in the inclusive $\ha\to\gam\gam$ mode.
\item An assessment of the extent to which the observed
rates for $\hh,\ha\to\tauptaum$ at the LHC could be used to fix the elusive
$\tanb$ parameter.
\item A demonstration that $\epem\to \hh\ha,\hp\hm$ pair production at a
$\rts=1\tev$ NLC will be observable (even if SUSY decays of the
Higgs bosons are substantial)
and will provide powerful information for determining the correct 
boundary conditions at the unification scale.
\item A discussion of the implications of LHC and NLC data
for $s$-channel discovery of the $\hh$ and $\ha$ at the FMC, in particular
noting that for $(\mha,\tanb)$ parameters 
such that the $\hh,\ha$ cannot be seen at 
either the LHC or the NLC then they can 
be discovered up to $\mha\sim \rts$ at the FMC.
\item A demonstration that $pp\to \dmm\dpp$ (doubly-charged Higgs pair
production) will yield an observable signal up to surprisingly substantial
$\dmm$ masses at the Tevatron and up to very large masses at the LHC.
\end{enumerate}
\end{itemize}
Despite the very substantial progress summarized in this report, much work
remains to be done, both with respect to fine-tuning procedures for a SM-like
Higgs boson and with regard to improving old and finding new techniques
for discovering and studying exotic Higgs bosons.

\smallskip
\begin{center}
{\bf Acknowledgements}
\end{center}
\smallskip 
This work was supported in part by the U.S. Department of Energy
under Grant No.~DE-FG03-91ER40674 and by the National Science Foundation
under Grant No.~PHY94-07194.
Further support was provided by the Davis Institute for High Energy Physics.
JFG wishes to acknowledge the hospitality of the Institute for Theoretical
Physics where a portion of this report was written.
We thank R. Settles and P. Janot for their comments and suggestions.
%

\end{document}